\documentclass{jfm}
\usepackage{natbib}
\usepackage{upmath}
\usepackage[british]{babel}
\usepackage{amsmath,bm}
\usepackage{amssymb}
\usepackage{amsbsy}
\usepackage{amscd}
\usepackage{amstext}
\usepackage{tabularx}
\usepackage{float}
\usepackage{makeidx}
\usepackage{amsmath}
\usepackage{subfigure}
\usepackage{afterpage}
\usepackage[T1]{fontenc}
\usepackage[latin1]{inputenc}
\usepackage{multirow}
\usepackage{mathrsfs}
\usepackage{color}
\usepackage{hyperref}
\hypersetup{hypertex=true,
            colorlinks=false,
            linkcolor=blue,
            anchorcolor=blue,
            citecolor=blue}

\usepackage{graphicx} 
\usepackage{graphics,epsfig}
\usepackage{amsfonts}
\usepackage{psfrag}

\ifCUPmtlplainloaded \else
  \checkfont{eurm10}
  \iffontfound
    \IfFileExists{upmath.sty}
      {\typeout{^^JFound AMS Euler Roman fonts on the system,
                   using the 'upmath' package.^^J}%
       \usepackage{upmath}}
      {\typeout{^^JFound AMS Euler Roman fonts on the system, but you
                   dont seem to have the}%
       \typeout{'upmath' package installed. JFM.cls can take advantage
                 of these fonts,^^Jif you use 'upmath' package.^^J}%
      }
  \else
  \fi
\fi

\ifCUPmtlplainloaded \else
  \checkfont{msam10}
  \iffontfound
    \IfFileExists{amssymb.sty}
      {\typeout{^^JFound AMS Symbol fonts on the system, using the
                'amssymb' package.^^J}%
       \usepackage{amssymb}%

      }{}
  \fi
\fi

\ifCUPmtlplainloaded \else
  \IfFileExists{amsbsy.sty}
    {\typeout{^^JFound the 'amsbsy' package on the system, using it.^^J}%
     \usepackage{amsbsy}}
    {\providecommand\boldsymbol[1]{\mbox{\boldmath $##1$}}}
\fi

\newsavebox{\astrutbox}
\sbox{\astrutbox}{\rule[-5pt]{0pt}{20pt}}

\newcommand\xuda[1]{{\color{black} #1\color{black}}}

\title[Reinforcement-learning-based control of convectively-unstable flows]
{Reinforcement-learning-based control of convectively-unstable flows}

\author[D. Xu, M. Zhang]%
{
Da Xu, Mengqi Zhang
}

\affiliation{
Department of Mechanical Engineering, National University of Singapore, 9 Engineering Drive 1, 117575 Singapore 
  \\[\affilskip]
}

\date{\today}

\graphicspath{{figures/}}
\newlength\savewidth

\begin{document}
\maketitle

\begin{abstract}
This work reports the application of a model-free deep-reinforcement-learning-based (DRL) flow control strategy to suppress perturbations evolving in the 1-D linearised Kuramoto-Sivashinsky (KS) equation and 2-D boundary layer flows. The former is commonly used to model the disturbance developing in flat-plate boundary layer flows. These flow systems are convectively unstable, being able to amplify the upstream disturbance, and are thus difficult to control. The control action is implemented through a volumetric force at a fixed position and the control performance is evaluated by the reduction of perturbation amplitude downstream. We first demonstrate the effectiveness of the DRL-based control in the KS system subjected to a random upstream noise. The amplitude of perturbation monitored downstream is significantly reduced and the learnt policy is shown to be robust to both measurement and external noise. One of our focuses is to optimally place sensors in the DRL control using the gradient-free particle swarm optimisation algorithm. After the optimisation process for different numbers of sensors, a specific eight-sensor placement is found to yield the best control performance. The optimised sensor placement in the KS equation is applied directly to control 2-D Blasius boundary layer flows and can efficiently reduce the downstream perturbation energy. Via flow analyses, the control mechanism found by DRL is the opposition control.
Besides, it is found that when the flow instability information is embedded in the reward function of DRL to penalise the instability, the control performance can be further improved in this convectively-unstable flow.

\end{abstract}

\section{Introduction}
Active flow control is of wide interest due to its extensive industrial applications where the fluid motion is manipulated with energy-consuming controllers towards a desired target, such as the reduction of drag, the enhancement of heat transfer and delay of the transition from a laminar flow to turbulence \citep{brunton2015closed}. Depending on whether we have a model for describing the dynamics of the fluid motion, the control strategies can be categorised into model-based and model-free methods. The former assumes the controller's complete awareness of a model which can describe the flow behaviour accurately (e.g., Navier-Stokes equations). The linear system theory acts as the foundation for the model-based controllers, such as linear quadratic regulator and model predictive control. The model-based approach has been widely applied in the active flow control of academic flows \citep{kim2007linear,sipp2010dynamics,sipp2016linear}.

In real-world flow conditions, however, an accurate flow model is often unavailable. Even in the case where a model can be assumed, once the flow condition drastically changes, beyond the predictability of the model, the control performance will also deteriorate. Model-free techniques based on system identification method can tackle this issue and has enjoyed success to some extent in controlling flows. Nevertheless, limitations also exist for this method such as a large number of free parameters \citep{sturzebecher2003active,herve2012physics}. \xuda{Thus, more advanced model-free control methods based on machine learning have been put forward and studied to cope with the complex flow conditions \citep{gautier2015closed, duriez2017machine,rabault2019artificial,park2020machine}.} In this work, we examine the performance of a model-free deep-reinforcement-learning (DRL) algorithm in controlling convectively-unstable flows, subjected to random upstream noise. Our investigation will begin with the 1-D Kuramoto-Sivashinsky (KS) equation, which can be regarded as a reduced model of laminar boundary layer flows past a flat plate, and then extend to the controlling of the 2-D boundary layer flows. One of our focuses is to optimise the placement of sensors in the flow to maximise the efficiency of the DRL control. In the following, we will first review the recent works on DRL-based flow control and the optimisation of sensor placement in other control methods.

\subsection{DRL-based flow control}
With the advancement of machine learning (ML) technologies especially deep neural network (DNN) and the ever-increasing amount of data, DRL burgeons and has proven its power in solving complex decision-making problems in various applications, including robotics \citep{kober2013reinforcement}, game playing \citep{mnih2013playing}, flow control \citep{Lee1997}, etc. In particular, DRL refers to an automated algorithm that aims to maximise a reward function by evaluating the state of an environment that the DRL agent can interact with via sensors. It is particularly suitable for flow control due to their similar settings and is being actively researched in the field of active flow control \citep{rabault2019accelerating,brunton2020machine,brunton2021machine}.

Based on a Markov process model and the classical reinforcement learning algorithm, \cite{Gueniat2016} first proposed an experiment-oriented control approach and demonstrated its effectiveness in reducing the drag in a cylindrical wake flow. \cite{pivot2017continuous} presented a proof-of-concept application of DRL control strategy to the 2-D cylinder wake flow and achieved a 17\% reduction of drag by rotating the cylinder. \cite{koizumi2018feedback} adopted the DRL-based feedback control to reduce the fluctuation of the lift force acting on the cylinder due to the Karman vortex shedding via two synthetic jets. They compared the DRL-based control with traditional model-based control and found DRL achieved a better performance. A subsequent well-cited work which has roused much attention of the fluid community to DRL is due to  \cite{rabault2019artificial}. They applied the DRL-based control to stabilise the Karman vortex alley via two synthetic jets on the cylinder which was confined between two flat walls and analysed the control strategy by comparing the macroscopic flow features before and after the control. These pioneering works laid the foundation of the DRL-based active flow control and have sparked great interest of the fluid community in DRL.

Other technical improvement and parameter investigations have been achieved. \cite{rabault2019accelerating} proposed a multi-environment strategy to further accelerate the training process of the DRL agent. \cite{xu2020active} applied the DRL-based control to stabilise the vortex shedding of a primary cylinder via the counter-rotating small cylinder pair downstream the primary one. \cite{tang2020robust} presented a robust DRL control strategy to reduce the drag of cylinder using four synthetic jets. The DRL agent was trained with four different Reynolds numbers ($Re$) but was proved to be effective for any $Re$ between 60 and 400. \cite{paris2021robust} also studied a robust DRL control scheme to reduce the drag of cylinder via two synthetic jets. They focused on identifying the (sub-)optimal placement of the sensors in the cylinder wake flow from a pre-defined matrix of the sensors. \cite{ren2021applying} further extended the DRL-based control of the cylinder wake from the laminar regime to the weakly turbulent regime with $Re=1000$. \cite{li2022reinforcement} incorporated the physical knowledge from stability analyses into the DRL-based control of confined cylinder wakes, which facilitated the sensor placement and the reward function design in the DRL framework. Moreover, \xuda{\cite{castellanos2022machine} assessed both DRL-based control and linear genetic programming control (LGPC) for reducing the drag in a cylindrical wake flow at $Re = 100$. It was found that DRL was more robust to different initial conditions and noise contamination while LGPC was able to realise control with fewer sensors.} \cite{Pino2022} provided a detailed comparison between some global optimisation techniques and machine learning methods (LGPC and DRL) in different flow control problems to better understand how the DRL performs.

All the aforementioned works are implemented through numerical simulations. \cite{fan2020reinforcement} first experimentally demonstrated the effectiveness of DRL-based control for reducing the drag in turbulent cylinder wakes through the counter-rotation of a pair of small cylinders downstream the main one. \cite{shimomura2020closed} proved the viability of DRL-based control for reducing the flow separation around a NACA0015 airfoil in experiments. The control was realised through adjusting the burst frequency of a plasma actuator and the flow reattachment was achieved under the angle of attack of 12 and 15 degrees. For a more detailed description of the recent studies of DRL-based flow control in fluid mechanics, readers are referred to the latest review papers on this topic \citep{rabault2020deep,garnier2021review,viquerat2021review}. 

After reviewing the works on DRL applied to flow control, we would like to point out one important research direction that deserves to be further explored, which is to embed domain knowledge or flow physics including symmetry or equivariance in the DRL-based control strategy. It has been demonstrated in other scientific fields that ML methods embedded with domain knowledge or symmetry properties inherent in the physical system can outperform the vanilla ML methods \citep{Ling2016,Zhang2018,Karniadakis2021,Smidt2021,Bogatskiy2022}. This will not only enhance the sample efficiency, but also guarantee the physical property of the controlled result. In the DRL community, only few scattered attempts have been made \citep{Belus2019,Zeng2021,li2022reinforcement}, and more works need to be followed in this direction to fully unleash the power of DRL in controlling flows.


\subsection{Sensor placement optimisation}
Sensors probe the states of the dynamical system. The extracted information can be further used for a variety of purposes, such as classification, reconstruction or reduced-order modelling of a high-dimensional system through a sparse set of sensor signals \citep{brunton2016sparse,manohar2018data,loiseau2018sparse} and the state estimation of the large-scale partial differential equations \citep{khan2015computation,hu2016sensor}. Here we mainly focus on the field of active flow control, where sensors are used to collect information from the flow environment as feedback to the actuator. The effect of sensor placement on flow control performance has been investigated in boundary layer flows \citep{belson2013feedback} and cylinder wake flows \citep{akhtar2015using}. Identifying the optimal sensor placement is of great significance to the efficiency of the corresponding control scheme. 

Initially, some heuristic efforts were made to guide the sensor placement through modal analyses. For instance, \cite{strykowski1990formation} placed a second, much smaller cylinder in the wake region of the primary cylinder and recorded the specific placements of the second cylinder which led to an effective suppression of the vortex shedding. Later, \cite{giannetti2007structural} found that the specific placements determined by \cite{strykowski1990formation} in fact corresponded to the wavemaker region where the direct and the adjoint eigenmodes overlapped. \cite{aakervik2007optimal} and \cite{bagheri20091pply} proposed that in the framework of flow control, sensors should be placed in the region where the leading direct eigenmode has a large magnitude and actuators should be placed in the region where the leading adjoint eigenmode has a large magnitude when the adjoint modes and the global modes have a small overlap. Likewise, \cite{natarajan2016actuator} developed an extended method of structural sensitivity analysis to help place collocated sensor-actuator pairs for controlling flow instabilities in a high-subsonic diffuser. Such placements based on physical characteristics of the flow system may be appropriate for the globally-unstable flow where the whole flow system beats at a particular frequency and the major disturbances never leave a specific region (such flows are called absolutely-unstable, cf. \cite{Huerre1990}). 

However, for a convectively-unstable flow system with a large transient growth, such a method failed to predict the optimal placement, as demonstrated by \cite{CHEN2011}. They proposed to address the optimal placement issue using a mathematically rigorous method. They identified the $\mathscr{H}_2$ optimal controller in conjunction with the optimal sensor and actuator placement via the conjugate gradient method, in order to control perturbations evolving in spatially developing flows modeled by the linearised Ginzburg-Landau equation (LGLE). \cite{colburn2011gradient} improved the method by working directly with the covariance of the estimation error instead of the Fischer information matrix. \cite{chen2014fluid} further demonstrated the effectiveness of this method in the Orr-Sommerfeld/Squire equations for the optimal sensor and actuator placement. \cite{oehler2018sensor} analysed the trade-offs when placing sensors and actuators in the feedback flow control based on LGLE. \cite{manohar2018optimal} studied the optimal sensor and actuator placement issue using balanced model reduction with a greedy optimisation method. The effectiveness of this method was demonstrated in the $\mathscr{H}_2$ optimal control of LGLE, where a similar placement was achieved with that in \cite{CHEN2011} but with less runtime. More recently, \cite{sashittal2021Data} proposed a data-driven method to generate a linear reduced-order model of the flow dynamics and then optimised the sensor placement using adjoint-based method. \cite{jin2022optimal} explored the optimal sensor and actuator placement in the context of feedback control of vortex shedding using a gradient minimisation method and analysed the trade-offs when placing sensors. 

In most of the aforementioned studies, traditional model-based controllers were considered and gradient-based methods were adopted to solve the optimisation problem of the sensor placement, as the gradients of the control objective with respect to sensor positions can be calculated using explicit formulas involved in the model-based controllers. However, in the context of data-driven DRL-based flow control, such accurate gradient information is unavailable because of its model-free probing and controlling of the flow system in a trial-and-error manner. The sensor placement affects the DRL-based control performance in a nontrivial way. Thus, studies on the optimised sensor placement in DRL-based flow control are important but currently rare. In the DRL context, \cite{rabault2019artificial} attempted to use much fewer sensors of 5 and 11 to perform the same DRL training and found that the resulting control performance was far less satisfactory than that of the original placement of 151 sensors (see their appendix). \cite{paris2021robust} proposed a novel algorithm named S-PPO-CMA to obtain the (sub-)optimal sensor placement. Nevertheless, they selected the best sensor layout from some pre-defined sensors and the sensor position can not change in the optimisation process. \cite{ren2021applying} reported that an \textit{a posteriori} sensitivity analysis was helpful to improve the sensor layout but this method was implemented as a post-processing data analysis, unable to provide guidelines for the sensor placement \textit{a priori} (see also \cite{Pino2022}). \cite{li2022reinforcement} proposed a heuristic way to place the sensors in the wavemaker region in the confined cylindrical wake flow and the optimal layout was not determined theoretically in an optimisation problem. For optimisation problems where the gradient information is unavailable, it is natural to consider the non-gradient-based methods such as genetic algorithm (GA) and particle swarm optimisation (PSO) \citep{koziel2011computational}. For instance, \cite{mehrabian2007optimal} adopted a bio-inspired invasive weed optimisation algorithm to optimally place piezoelectric actuators on the smart fin for vibration control. \cite{yi2011optimal} utilised a generalised GA to find the optimal sensor placement in the structural health monitoring (SHM) of high-rise structures. \cite{blanloeuil2016particle} applied PSO algorithm to improve the sensor placement in an ultrasonic SHM system. The improved sensor placement enabled a better detection of multiple defects in the target area. \cite{wagiman2020new} adopted PSO algorithm to find an optimal light sensor placement of an indoor lighting control system. A 24.5\% energy saving was achieved with the optimal number and position of sensors. These works enlighten us on how to optimally distribute the sensors in the model-free DRL control.

\subsection{The current work}
\begin{figure}
  \centering  
  \subfigure[Absolutely unstable]{
  \includegraphics[width=0.465\textwidth]{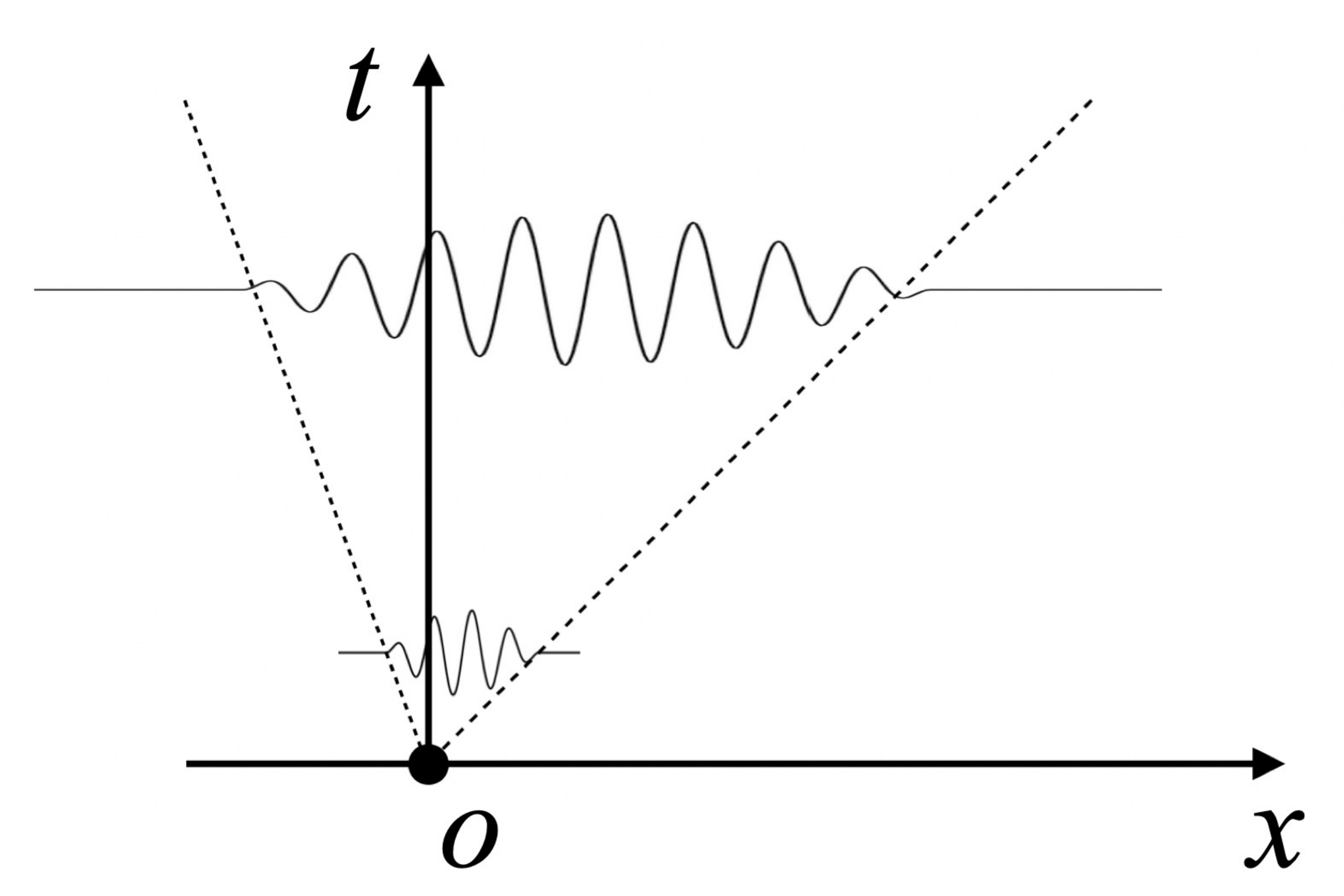}
  }
  \quad
  \subfigure[Convectively unstable]{
  \includegraphics[width=0.465\textwidth]{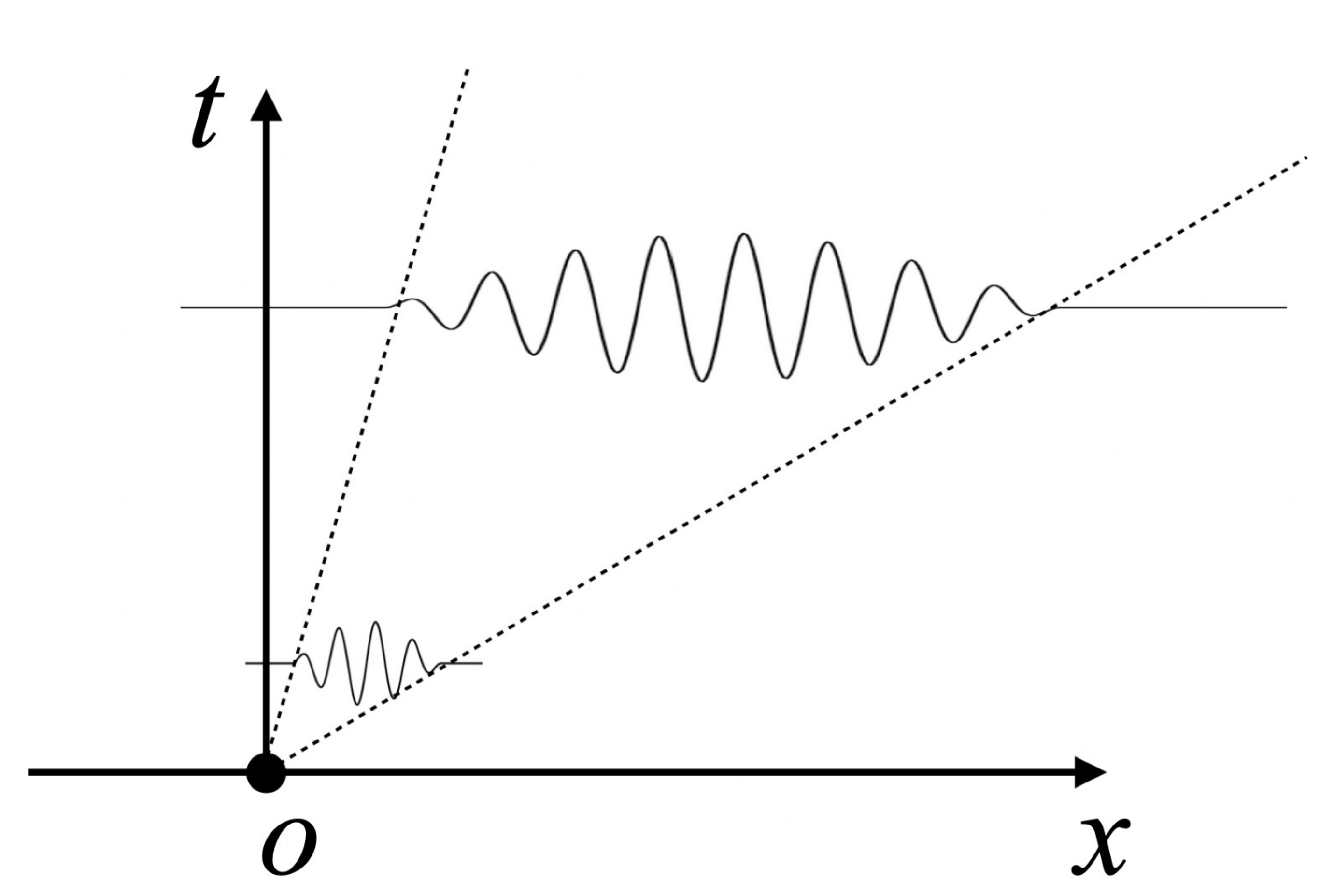}
  }
  \quad
  \caption{Schematic of absolute and convective instabilities. A localised infinitesimal perturbation can grow at a fixed location leading to (a) absolute instability, or decay at a fixed location but grow as convected downstream leading to (b) convective instability.}
  \label{figB}
\end{figure}

In the current work, we aim to study the performance of the model-free DRL-based strategy in controlling convectively-unstable flows. \xuda{The difference between absolute instability and convective instability is shown in figure \ref{figB} \citep{Huerre1990}. An absolutely unstable flow is featured by an intrinsic instability mechanism and is thus less sensitive to the external disturbance. A good example of this type of flow is the global onset of vortex shedding in cylindrical wake flows, which have been studied extensively in DRL-based flow control \citep{rabault2019artificial,fan2020reinforcement,paris2021robust,li2022reinforcement}. On the contrary, the convectively unstable flow acts as an noise amplifier and is able to selectively amplify the external disturbance, such as boundary layer flows and jets.} Compared to the absolutely-unstable flows, controlling the convectively-unstable flow subjected to unknown upstream disturbance is more challenging, representing a more stringent test of the control ability of DRL. \xuda{This type of flow is less studied in the context of DRL-based flow control.}

As a proof-of-concept study, we will in this work control the 1-D linearised Kuramoto-Sivashinsky (KS) equation and the 2-D boundary layer flows. \xuda{The KS equation is supplemented with an outflow boundary condition, without translational and reflection symmetries, and has been widely used as a reduced model of the disturbance developing in the laminar boundary layer flows.} We will investigate the sensor placement issue in the DRL-based control by formulating an optimisation problem in the KS equation based on the PSO method. Then the optimised sensor placement determined in the KS equation will be directly applied to the 2-D Blasius boundary layer flow solved by the complete Navier-Stokes (NS) equations, to evaluate the performance of DRL in suppressing the convective instability in a more realistic flow. \xuda{We were not able to directly optimise the sensor placement in the 2-D boundary layer flow using the PSO method because the computation in the latter case is exceedingly demanding. We thus circumvented this issue by resorting to the reduced-order 1-D KS equation.}

 

The paper is structured as follows. In Sec. \ref{problemformulation}, we introduce the flow control problem. In Sec. \ref{method}, we present the numerical method adopted in this work. The results on DRL-based flow control and the optimal sensor placement are reported in Sec. \ref{results}. Finally, in Sec. \ref{Conclusions}, we conclude the paper with some discussions. \xuda{In the appendices, we provide an explanation of the time delay issue in DRL-based control and a brief introduction to the classical model-based linear quadratic regulator (LQR). We also propose the stability-enhanced reward to further improve the control performance and investigate the DRL-based control of the nonlinear KS equation.}

\section{Problem formulation}\label{problemformulation}
\begin{figure}
  \centerline{\includegraphics[scale=0.3]{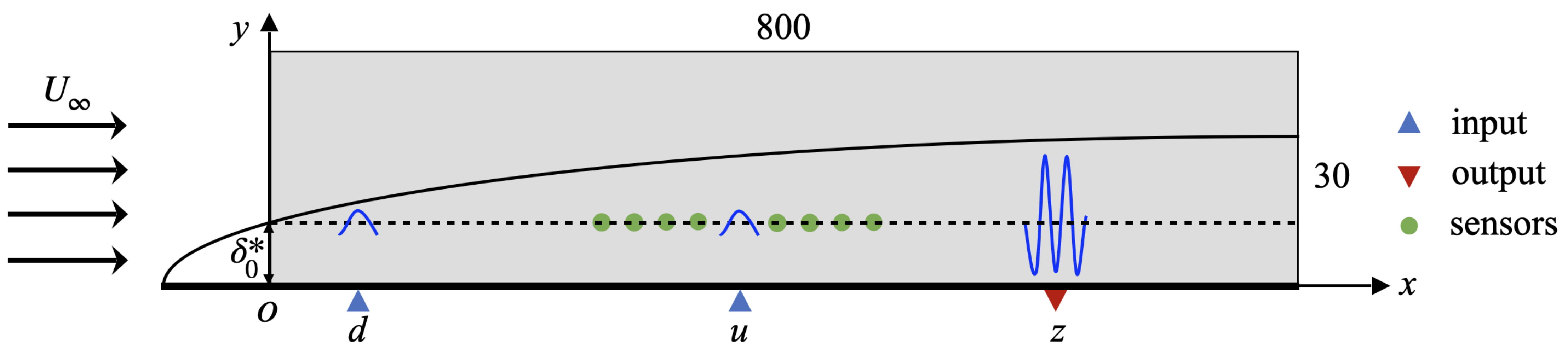}}
  \caption{Control setup for a 2-D flat-plate boundary layer flow. Random noise is introduced at point $d$ and a feedback control based on states collected by sensors is implemented at point $u$ to minimise the downstream perturbation measured at point $z$. $U_\infty$ is the uniform free-stream velocity and $\delta_0^*$ is the displacement thickness of the boundary layer at the inlet of computational domain shown by the grey box $\Omega = (0, 800) \times (0, 30)$ which is non-dimensionalised by $\delta_0^*$.}
\label{fig1}
\end{figure}

We investigate the control of perturbation evolving in a flat-plate boundary layer flow modelled by the linearised KS equation and the NS equations, as shown in figure \ref{fig1}. A localised Gaussian disturbance is introduced at point $d$. Due to the convective instability of the flow, the amplitude of the perturbation will grow exponentially with time while travelling downstream, if uncontrolled. The closed-loop control system is formed by introducing a spatially localised forcing at point $u$ as the control action, which is determined by the DRL agent based on the feedback signals collected by sensors (denoted by green points in figure \ref{fig1}). The control objective is to minimise the downstream perturbation measured by an output sensor at point $z$, trying to abate the exponential flow instability. 

\subsection{Governing equation of the 1-D Kuramoto-Sivashinsky equation}\label{KS control}
The original KS equation was first used to describe flame fronts in laminar flames \citep{kuramoto1976persistent,sivashinsky1977nonlinear} and \xuda{is one of the simplest nonlinear PDEs which exhibit spatiotemporal chaos \citep{Cvitanovic2010}. It has become a common toy problem in the studies of machine learning, such as the data-driven reduction of chaotic dynamics on an inertial manifold \citep{linot2020deep} and the DRL-based control of chaotic systems \citep{bucci2019control,Zeng2021}.}
Here, we investigate the 1-D linearised KS equation as a reduced-order-model representation of the disturbance developing in 2-D boundary layer flows, cf. \cite{fabbiane2014adaptive} for a detailed illustration of various model-based and adaptive control methods applied to the KS equation. Specifically, the linearised KS equation can describe the flow dynamics at a wall-normal position of $y=\delta_0^{*}$, where $\delta_0^*$ is the displacement thickness of the boundary layer at the inlet of the computational domain, as shown in figure \ref{fig1}.
The linearised step is briefly outlined below. First, the original non-dimensionalised KS equation reads
\begin{equation}
  \frac{\partial v}{\partial t}+v \frac{\partial v}{\partial x}=-\frac{1}{\mathcal{R}}\left(\mathcal{P} \frac{\partial^{2} v}{\partial x^{2}}+\frac{\partial^{4} v}{\partial x^{4}}\right),\quad x\in(0,L),
\label{eq2-1}
\end{equation}
where $v(x,t)$ represents the velocity field, $\mathcal R$ the Reynolds number (or $Re$ in boundary layer flows), $\mathcal P$ a coefficient balancing energy production and dissipation and finally $L$ the length of the 1-D domain. Since we study a fluid system that is close to a steady solution $V$ (a constant), we can decompose the velocity into a combination of two terms as
\begin{equation}
  v(x, t)=V+\varepsilon v^{\prime}(x, t)
\label{eq2-2}  
\end{equation}
where $v^{\prime}(x, t)$ is the perturbation velocity with $\varepsilon \ll 1$. Inserting Eq. \ref{eq2-2} into Eq. \ref{eq2-1} yields
\begin{equation}
  \frac{\partial v^{\prime}}{\partial t}=-V \frac{\partial v^{\prime}}{\partial x}-\frac{1}{\mathcal{R}}\left(\mathcal{P} \frac{\partial^{2} v^{\prime}}{\partial x^{2}}+\frac{\partial^{4} v^{\prime}}{\partial x^{4}}\right)-\varepsilon v^{\prime} \frac{\partial v^{\prime}}{\partial x}+f(x, t),\quad x\in(0,L).
\label{eq2-3}
\end{equation}
where the external forcing term $f(x,t)$ now appears on the right hand side for the introduction of disturbance and control terms.

When the perturbation is small enough, we can neglect the nonlinear term $- \varepsilon v^{\prime} \partial v^{\prime} / \partial x$ in Eq. \ref{eq2-3} and obtain the following linearised KS equation to model the dynamics of streamwise perturbation velocity evolving in flat-plate boundary layer flows
\begin{equation}
  \frac{\partial v^{\prime}}{\partial t}=-V \frac{\partial v^{\prime}}{\partial x}-\frac{1}{\mathcal{R}}\left(\mathcal{P} \frac{\partial^{2} v^{\prime}}{\partial x^{2}}+\frac{\partial^{4} v^{\prime}}{\partial x^{4}}\right)+f(x, t),\quad x\in(0,L)
\label{eq2-4}
\end{equation}
with the unperturbed boundary conditions
\begin{equation}
\text{Inflow:  } \left.v^{\prime}\right|_{x=0}=0,\left.\quad \frac{\partial v^{\prime}}{\partial x}\right|_{x=0}=0.  \ \ \ \text{Outflow:  }  \left.\frac{\partial^{3} v^{\prime}}{\partial x^{3}}\right|_{x=L}=0,\left.\quad \frac{\partial v^{\prime}}{\partial x}\right|_{x=L}=0.
\label{eq2-5}
\end{equation}
We adopt the same parameter settings as those in \cite{fabbiane2014adaptive}, i.e., $\mathcal{R} = 0.25$, $\mathcal{P} = 0.05$, $V = 0.4$ and $L = 800$, to model the 2-D boundary layer at $Re = 1000$.

The external forcing term $f(x, t)$ in Eq. \ref{eq2-4} includes both noise input and control input
\begin{equation}
f(x, t)=b_{d}(x) d(t)+b_{u}(x) u(t)
\label{eq2-7}
\end{equation}
where $b_{d}(x)$ and $b_{u}(x)$ represent the spatial distribution of noise and control inputs, respectively; $d(t)$ and $u(t)$ correspond to the temporal signals. The downstream output measured at point $z$ is expressed by \begin{equation}
z(t)=\int_{0}^{L} c_{z}(x) v^{\prime}(x, t) d x
\label{eq2-8}
\end{equation}
where $c_{z}(x)$ is the spatial support of the output sensor. In this work, all the three spatial supports $b_{d}(x)$, $b_{u}(x)$ and $c_{z}(x)$ take the form of a Gaussian function
\begin{subeqnarray}
g(x ; {x}_{n}, \sigma)=\frac{1}{\sigma} \exp \left(-\left(\frac{x-{x}_{n}}{\sigma}\right)^{2}\right) \qquad \qquad \qquad \\
\label{eq2-9}
b_{d}(x)=g\left(x ; {x}_{d}, \sigma_{d}\right), \quad b_{u}(x)=g\left(x ; {x}_{u}, \sigma_{u}\right), \quad
c_{z}(x)=g\left(x ; {x}_{z}, \sigma_{z}\right).
\label{eq2-10}
\end{subeqnarray}
We adopt the same spatial parameters used by \cite{fabbiane2014adaptive}, i.e., ${x}_{d} = 35$, ${x}_{u} = 400$, ${x}_{z} = 700$, and $\sigma_{d} = \sigma_{u} = \sigma_{z} = 4$. In addition, the disturbance input $d(t)$ in Eq. \ref{eq2-7} is modeled as a Gaussian white noise with a unit variance. In this work, we will use DRL to learn an effective control law, \textit{i.e.} how the control action $u(t)$ in Eq. \ref{eq2-7} varies with time, to suppress the perturbation downstream.

\subsection{Governing equations for the 2-D boundary layer}\label{Blasius control}
We will also test the DRL-based control in 2-D Blasius boundary layer flows, governed by the incompressible Navier-Stokes equations
\begin{equation}
\frac{\partial \boldsymbol{u}}{\partial t}+\boldsymbol{u} \cdot \nabla \boldsymbol{u}=-\nabla p+\frac{1}{R e} \nabla^{2} \boldsymbol{u}+\boldsymbol{f}, \quad \nabla \cdot \boldsymbol{u}=0
\label{eq2-11}
\end{equation}
where $\boldsymbol{u}=(u, v)^{T}$ is the velocity, $t$ the time, $p$ the pressure and $\boldsymbol{f}$ the external forcing term. $Re$ is the Reynolds number defined as $Re = U_\infty \delta_0^*/ \nu$, where $U_\infty$ is the free-stream velocity, $\delta_0^*$ the displacement thickness of the boundary layer at the inlet of the computational domain and $\nu$ the kinematic viscosity. Here we nondimensionalise the length and the velocity by $\delta_0^*$ and $U_\infty$, respectively, and use $Re = 1000$ in all the cases, which is consistent with that in the KS system. The computational domain is $\Omega = (0, L_x) \times (0, L_y) = (0, 800) \times (0, 30)$, denoted by the grey box in figure \ref{fig1}. The inlet velocity profile is obtained from the laminar base flow profile (Blasius solution). At the outlet of the domain, we impose the standard free-outflow condition with $\left(p \boldsymbol{I}-\frac{1}{R e} \nabla \boldsymbol{u}\right) \cdot \boldsymbol{n}=0$, where $\boldsymbol{I}$ is the identity tensor and $\boldsymbol{n}$ the outward normal vector. At the bottom wall, we apply the no-slip boundary condition and at the top of the domain the free-stream boundary condition with $u = U_\infty$ and $dv/dy = 0$.

\begin{figure}
  \centering  
  \subfigure[]{
  \includegraphics[width=0.465\textwidth]{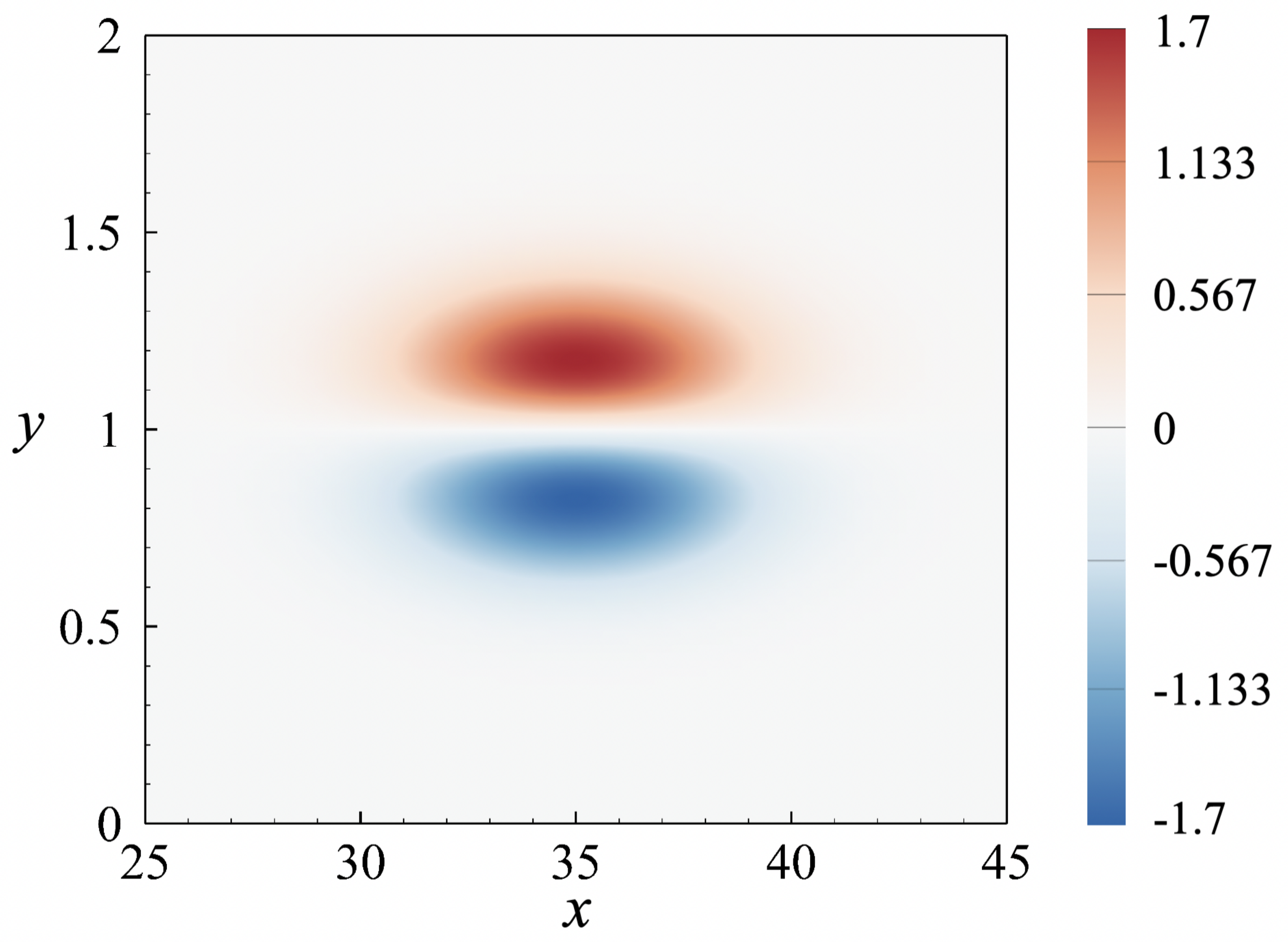}
  }
  \quad
  \subfigure[]{
  \includegraphics[width=0.465\textwidth]{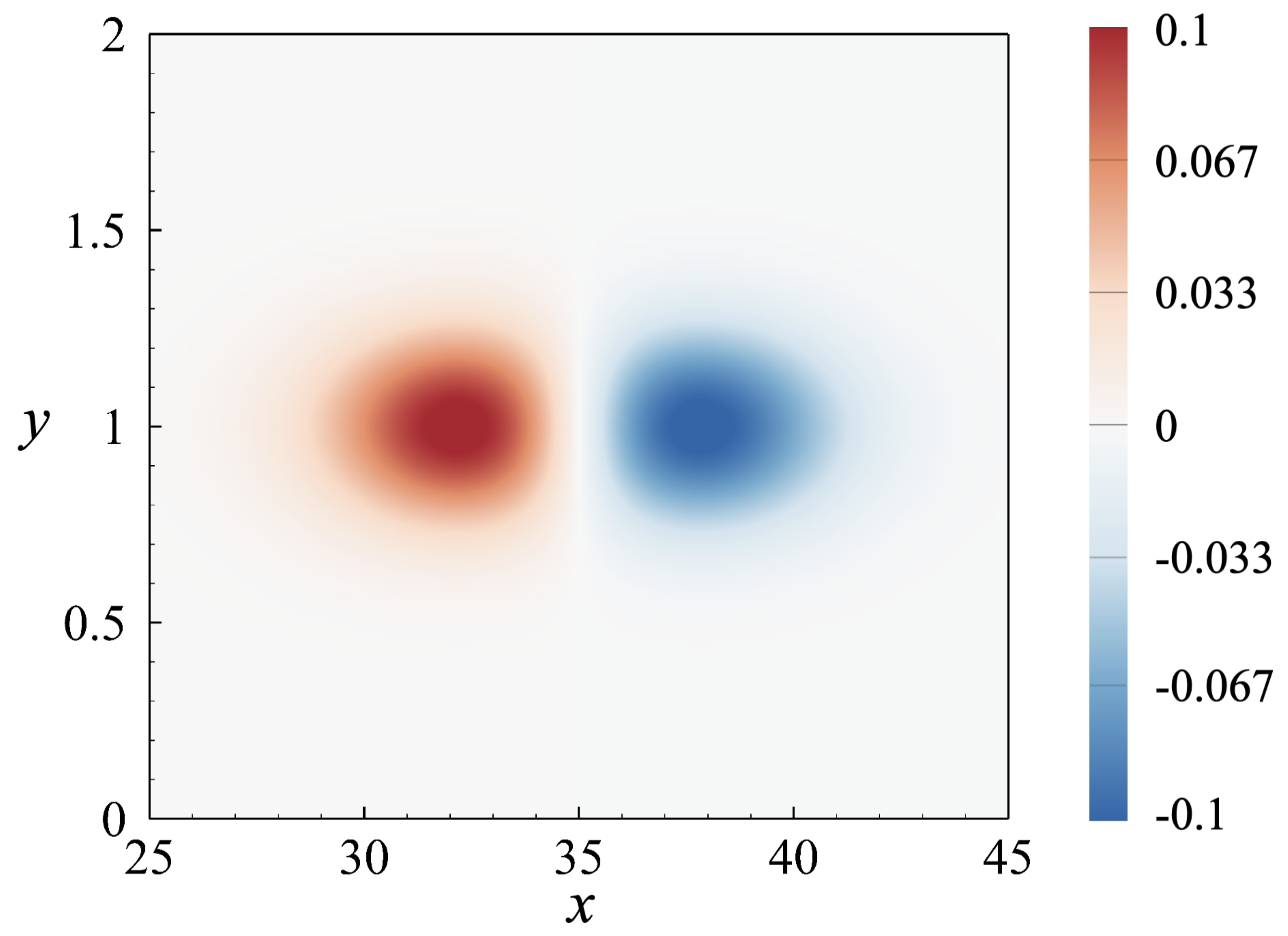}
  }
  \quad
  \caption{Contour plots of the 2-D Gaussian spatial distribution $\boldsymbol{b_{d}}(x, y)$. (a) Streamwise component; (b) wall-normal component.}
  \label{figA}
\end{figure}

The external forcing term $\boldsymbol{f}$ in the momentum equation includes both noise input $d(t)$ and control output $u(t)$ as below
\begin{equation}
\boldsymbol{f}=\boldsymbol{b_{d}}(x, y) d(t)+\boldsymbol{b_{u}}(x, y) u(t)
\label{eq2-12}
\end{equation}
where $\boldsymbol{b_{d}}$ and $\boldsymbol{b_{u}}$ are the corresponding spatial distribution functions, similar to those in Eq. \ref{eq2-7} except that the current supports assume 2-D Gaussian functions as below
\begin{equation}
\begin{aligned}
\boldsymbol{b}_{d}(x, y)=\left[\begin{array}{c}
\left(y-y_{d}\right) \sigma_{x} / \sigma_{y} \\
-\left(x-x_{d}\right) \sigma_{y} / \sigma_{x}
\end{array}\right] \exp \left(-\frac{\left(x-{x}_{d}\right)^{2}}{\sigma_{x}^{2}}-\frac{\left(y-{y}_{d}\right)^{2}}{\sigma_{y}^{2}}\right)\\
  \boldsymbol{b}_{u}(x, y)=\left[\begin{array}{c}
  \left(y-y_{u}\right) \sigma_{x} / \sigma_{y} \\
  -\left(x-x_{u}\right) \sigma_{y} / \sigma_{x}
  \end{array}\right] \exp \left(-\frac{\left(x-{x}_{u}\right)^{2}}{\sigma_{x}^{2}}-\frac{\left(y-{y}_{u}\right)^{2}}{\sigma_{y}^{2}}\right)
  \label{eq2-14}
\end{aligned}
\end{equation}
where $\sigma_x = 4$ and $\sigma_y = 1/4$ determine the size of the two spatial supports centred at $(x_d, y_d) = (35, 1)$ for noise input and $(x_u, y_u) = (400, 1)$ for control input, respectively. \xuda{As an illustration, the streamwise and wall-normal components of the spatial support $\boldsymbol{b_{d}}$ are presented in figure \ref{figA}, which are consistent with those in \cite{belson2013feedback}.}

The downstream output at point $z$ is a measurement of the local perturbation energy expressed by
\begin{equation}
z(t)=\sum_{i = 1}^{N}\left[\left(u^{i}(t)-U^{i}\right)^{2}+\left(v^{i}(t)-V^{i}\right)^{2}\right]
\label{eq2-15}
\end{equation}
where $N$ is the number of uniformly-distributed neighbouring points around point $z$ at (550, 1) and here we use a large $N = 78$; $u^i(t)$ and $v^i(t)$ are the instantaneous $x$ and $y$-direction velocity components at point $i$; $U^i$ and $V^i$ are the corresponding base flow velocities, i.e., a steady solution to the nonlinear Navier-Stokes equations. 

The upstream noise input $d(t)$ in Eq. \ref{eq2-12} is modeled as a Gaussian white noise with a zero mean and a standard deviation of $2 \times 10^{-4}$. Induced by this random disturbance input, the boundary layer flow considered here exhibits the behaviour of convective instability in the form of an exponentially growing Tollmien-Schlichting (TS) wave. The objective of the DRL control is to suppress the growth of the unstable TS wave by modulating $u(t)$ in Eq. \ref{eq2-12}. In figure \ref{fig1}, the input and output positions are specified and will remain unchanged in our investigation, but the sensor position will be improved in an optimisation problem, which will be discussed in Sec. \ref{Probe}.

\section{Numerical methods}\label{method}
\subsection{Numerical simulation of the 1-D KS equation}\label{NS}
We adopt the same numerical method used by \cite{fabbiane2014adaptive} to solve the 1-D linearised KS equation. The equation is discretised using a finite difference method on $n=400$ discretised grid points. The second- and fourth-order derivatives are discretised based on a centered five-node stencil while the convective term based on a one-node upwind scheme due to the convective feature. Boundary conditions in Eqs. \ref{eq2-5} are implemented using four ghost nodes outside the left and right boundaries. The spatial discretisation yields the following set of finite-dimensional state space equations (also referred to as \textit{plant} hereafter)
\begin{equation}
\begin{aligned}
\dot{\boldsymbol{v}}(t) &=\boldsymbol{A} \boldsymbol{v}(t)+\boldsymbol{B}_{d} d(t)+\boldsymbol{B}_{u} u(t) \\
z(t) &=\boldsymbol{C}_{z} \boldsymbol{v}(t)
\end{aligned}
\label{eq3-1}
\end{equation}
where $\boldsymbol{v}(t)\in R^n$ represents the discretised values at $n = 400$ equispaced nodes, $\dot{\boldsymbol{v}}(t)$ its time derivative and $\boldsymbol{A}$ the linear operator of the system. The input matrices $\boldsymbol{B}_{d}$ and $\boldsymbol{B}_{u}$, and output matrix $\boldsymbol{C}_{z}$ are obtained by evaluating Eq. \ref{eq2-10} at the nodes. The implicit Crank-Nicolson method is adopted for time-marching 
\begin{equation}
\boldsymbol{v}(t+\Delta t)=\boldsymbol{N}_{I}^{-1}\left[\boldsymbol{N}_{E} \boldsymbol{v}(t)+\Delta t\left(\boldsymbol{B}_{d} d(t)+\boldsymbol{B}_{u} u(t)\right)\right]
\label{eq3-2}
\end{equation}
where $\boldsymbol{N}_{I}=\boldsymbol{I}-\frac{\Delta t}{2} \boldsymbol{A}$, $\boldsymbol{N}_{E}=\boldsymbol{I}+\frac{\Delta t}{2} \boldsymbol{A}$ and $\Delta t$ is the time step chosen as $\Delta t = 1$ in the current work. 

We also investigate the dynamics of weakly nonlinear KS equation when the perturbation amplitude reaches a certain level, i.e., Eq. \ref{eq2-3} together with boundary conditions Eqs. \ref{eq2-5}, in Sec. \ref{Nonlinear}. Compared with the linear \textit{plant} described by Eq. \ref{eq3-1}, the weakly nonlinear \textit{plant} has an additional nonlinear term on the right hand side
\begin{equation}
\begin{aligned}
\dot{\boldsymbol{v}}(t) &=\boldsymbol{A} \boldsymbol{v}(t)+\boldsymbol{B}_{d} d(t)+\boldsymbol{B}_{u} u(t)+ \boldsymbol{N} (\boldsymbol{v}(t))\\
z(t) &=\boldsymbol{C}_{z} \boldsymbol{v}(t)
\end{aligned}
\label{eq3-2a}
\end{equation}
where $\boldsymbol{N}$ represents the nonlinear operator. In terms of the time marching of Eq. \ref{eq3-2a}, we adopt a third-order semi-implicit Runge-Kutta scheme proposed by \cite{kar2006semi} and also used by \cite{bucci2019control} and \cite{Zeng2021}, in which the nonlinear and the external forcing terms are time-marched explicitly while the linear term is marched implicitly using a trapezoidal rule.

It should be noted that the models given by $\boldsymbol{A}$, $\boldsymbol{B}_{u}$ and $\boldsymbol{C}_{z}$ are only for the purpose of numerical simulations but not for the controller design. DRL-based controller has no awareness of the existence of such models and it learns the control law from scratch through interacting with the environment and is thus model-free and data-driven, unlike the model-based controllers whose design depends specifically on such models.

\subsection{Direct numerical simulation of the 2-D Blasius boundary layer}\label{DNS}
Flow simulations are performed to solve Eqs. \ref{eq2-11} in the computational domain $\Omega$ using the open-source Nek5000 solver \citep{nek5000-web-page}. The spatial discretisation is implemented using the spectral element method, where the velocity space in each element is spanned by the $K$th-order Legendre polynomial interpolants based on the Gauss-Lobatto-Legendre (GLL) quadrature points. According to the mesh convergence study (not shown), we finally choose a specific mesh composed of 870 elements of the order $K = 7$, with the local refinement implemented close to the wall. In terms of the time integration, the two-step backward differentiation scheme is adopted in the unsteady simulations with a time step of $5 \times 10^{-2}$ unit times and the initial flow field is constructed from similarity solutions of the boundary layer equations. 

\subsection{Deep reinforcement learning}\label{RL}
\begin{figure}
  \centerline{\includegraphics[scale=0.32]{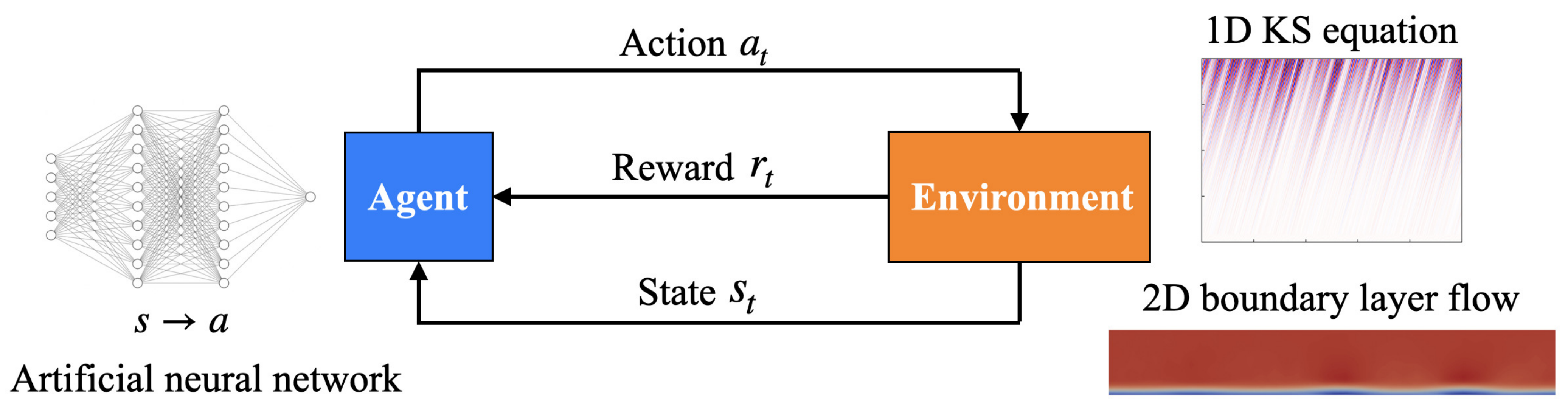}}
  \caption{The reinforcement learning framework in flow control. Agent: artificial neural network; environment: numerical simulation of 1-D KS equation or 2-D boundary layer flow; action: adjustment of the external forcing; reward: reduction of the downstream perturbation; state: streamwise velocities collected by sensors.}
\label{fig2}
\end{figure}

In the context of DRL-based method, the control agent learns a specific control policy via the interaction  with the environment.
The DRL framework in this work is shown in figure \ref{fig2}, where the agent is represented by an artificial neural network (ANN) and the environment corresponds to the numerical simulation of the 1-D KS equation or the 2-D boundary layer as explained above. In general, the DRL works as follows: first, the agent receives states $s_t$ from the environment; then, an action $a_t$ is determined based on the states and is exerted to the environment; finally, the environment returns a reward signal $r_t$ to evaluate the quality of the previous actions. This loop continues until the training process converges, i.e., the expected cumulative reward is maximised for each training episode. 

In the current setting of both KS equation and boundary layer flows, states refer to the streamwise velocities collected by sensors. Since the amplitude of velocity differs at different locations, states are normalised by their respective mean values and standard deviations before they are input to the agent. The number and position of sensors are determined from an optimisation process to be detailed in Sec. \ref{Probe}. Action corresponds to $u(t)$ in both Eq. \ref{eq2-7} and Eq. \ref{eq2-12}, with a predefined range of [-5, 5] and [-0.01, 0.01], respectively. Each control action is exerted to the environment for a duration of 30 numerical time steps before the next update. This operation is the so-called sticky action which is a common choice in the DRL-based flow control \citep{rabault2019artificial,rabault2019accelerating,beintema2020controlling}. Allowing quicker actions may be detrimental to the overall performance since the action needs some time to make an impact on the environment \citep{burda2018large}. The last crucial component in the DRL framework is reward which is directly related to the control objective. For the 1-D KS system, the reward is defined as the negative perturbation amplitude measured by the output sensor, i.e., the negative root-mean-square value of $z(t)$ in Eq. \ref{eq2-8}, while for the 2-D boundary layer flow, the reward is defined as the negative local perturbation energy measured around point $z$, i.e., $-z(t)$ in Eq. \ref{eq2-15}. However, due to the time delay effect in convective flows, the instantaneous reward is not a real response to the current action. To remedy the time mismatch, we store the simulation data first and retrieve them later after the action has reached the output sensor. This operation helps the agent perceive the dynamics of environment in a time-matched way and thus leads to a good training result. \xuda{For details on the time delay and how to remove it, please refer to Appendix \ref{A}}

In terms of the training algorithm for the agent, since the action space is continuous, we adopt a policy-based algorithm named deep deterministic policy gradient (DDPG) which is essentially a typical actor-critic network structure \citep{Sutton2018}. This method has also been used in other DRL works \citep{koizumi2018feedback,bucci2019control,Zeng2021,Pino2022,Kim2022}. The actor network $\mu_{\phi}(s)$, also known as the policy network, is parameterised by $\phi$ and outputs an action $a$ to be applied to the environment for a given state $s$. The aim of the actor is to find an optimal policy which maximises the expected cumulative reward, which is predicted by the critic network $Q_{\theta}(s, a)$ parameterised by $\theta$. The loss function $L(\theta)$ for updating the parameters of the critic reads
\begin{equation}
L(\theta)=\underset{\left(s, a, r, s^{\prime}\right) \sim \mathcal{D}}{\mathbb{E}}\left[\frac{1}{2}\left(\left[r(s, a)+\gamma Q_{\theta_{\text {targ }}}\left(s^{\prime}, \mu_{\phi_{\text {targ }}}\left(s^{\prime}\right)\right)\right]-Q_{\theta}(s, a)\right)^{2}\right]
\label{eq3-3}
\end{equation}
where $\mathbb{E}$ represents the expectation of all transition data $(s, a, r, s^{\prime})$ in sequence $\mathcal{D}$; $s, a$ and $r$ are the state, action and reward for the current step, respectively, and $s^{\prime}$ is the state for the next step. The right-hand-side of the equation calculates the typical error in RL, i.e., temporal-difference (TD) error, and $\gamma$ is the discount factor selected as 0.95 here. The \xuda{objective} function $L(\phi)$ for updating the parameters of the actor is obtained straightforwardly
\begin{equation}
L(\phi)=-\underset{s \sim \mathcal{D}}{\mathbb{E}}\left[Q_{\theta}\left(s, \mu_{\phi}(s)\right)\right]
\label{eq3-4}
\end{equation}
\xuda{It should be noted that although the actor aims to maximise the $Q$ value predicted by the critic, only gradient descent rather than gradient ascent is embedded in the adopted optimiser. Thus, a negative sign is introduced on the right-hand-side of Eq. \ref{eq3-4}.}

In addition, some technical tricks such as experience memory replay and a combination of evaluation net and target net are embedded in the DDPG algorithm,  in order to cut off the correlation among data and improve the stability of training. More technical details on DDPG can be found in \cite{silver2014deterministic} and \cite{lillicrap2015continuous}. \xuda{For other hyperparameters used in the current study, please refer to Appendix \ref{A}.}

\subsection{Particle swarm optimisation}\label{PSO}
As mentioned earlier, sensors are used to collect flow information from the environment as feedback to the DRL-based controller. Therefore, the sensor placement plays an important role in determining the final control performance. Some studies investigated the optimal sensor placement issue in the context of flow control using adjoint-based gradient descent method. In these studies, gradients of the objective function with respect to the sensor positions can be obtained exactly using explicit equations involved in the model-based controllers \citep{CHEN2011,belson2013feedback,oehler2018sensor,sashittal2021Data}. However, in our work, DRL-based control is model-free and such gradient information is unavailable. Therefore, we adopt a gradient-free method called particle swarm optimisation (PSO) to determine the optimal sensor placement in DRL.

PSO is a type of evolutionary optimisation methods which mimic the bird flock preying behaviour and was originally proposed by \cite{kennedy1995particle}. For an optimisation problem expressed by
\begin{equation}
y = f(x_1, x_2,...,x_D)
\label{eq3-5}
\end{equation}
where $y$ is the objective function and $x_1, x_2,...,x_D$ are $D$ design variables. PSO trains a swarm of particles which move around in the searching space bounded by the lower and upper boundaries and update their positions iteratively according to the following rule
\begin{subeqnarray}
v_{i d}^{k}&=& w v_{i d}^{k-1}+c_{1} r_{1}\left(p_{i d}^{best}-x_{i d}^{k-1}\right)+c_{2} r_{2}\left(s_{d}^{best}-x_{i d}^{k-1}\right) \label{eq3-6_1} \\
x_{i d}^{k}&=&x_{i d}^{k-1}+v_{i d}^{k-1}
\label{eq3-6}
\end{subeqnarray}
where $v_{i d}^{k}$ is the $d_{th}$-dimensional velocity of particle $i$ at the $k_{th}$ iteration and $x_{i d}^{k}$ is the corresponding position. The first term on the right hand side of Eq. \ref{eq3-6_1}(a) is the inertial term, representing its memory on the previous state with $w$ being the weight. The second term is called self-cognition, representing learning from its own experience, where $p_{i d}^{best}$ is the best $d_{th}$-dimensional position that the particle $i$ has ever found for the minimum $y$. The third term is called social cognition, representing learning from other particles, where $s_{d}^{best}$ is the best $d_{th}$-dimensional position among all the particles in the swarm. In Eq. \ref{eq3-6}(a), $c_1$, $c_2$ are scaling factors and $r_1$, $r_2$ are cognitive coefficients. After the iteration process converges, these particles will gather near a specific position in the search space and this is the optimised solution. 

\begin{figure}
  \centerline{\includegraphics[scale=0.3]{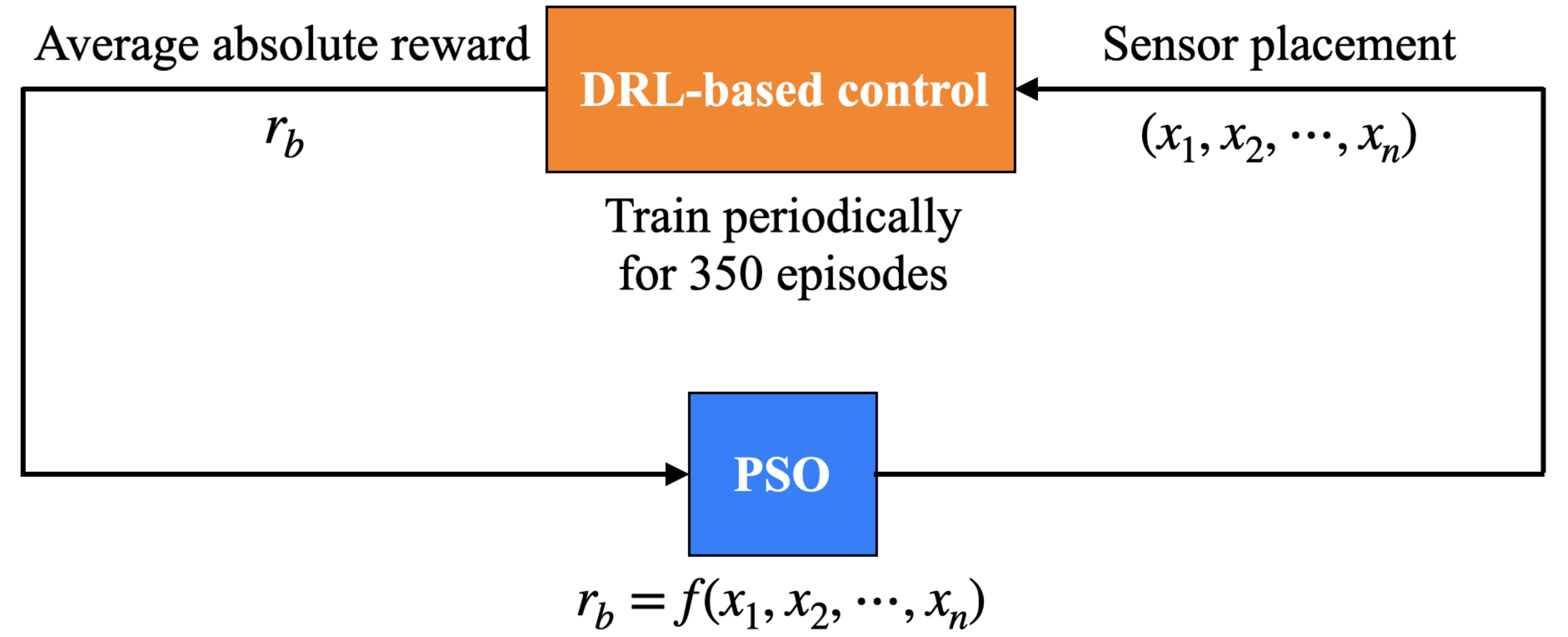}}
  \caption{Schematic diagram for sensor placement optimisation. Design variables are $n$ sensor positions $(x_1, x_2, ..., x_n)$. Objective function is the DRL-based control performance quantified by the average absolute value of reward $r_b$ from the 10 best episodes among all the 350 episodes.}
\label{fig3}
\end{figure}

For the optimal sensor placement in our case, design variables in Eq. \ref{eq3-5} are sensor positions $x_1, x_2, ... , x_n$ ($n$ is the number of sensors) and the objective function is related to the DRL-based control performance given the current sensor placement. As shown in figure \ref{fig3}, a specific sensor placement is input to the DRL-based control framework. \xuda{The DRL training is then executed for 350 episodes and this number will be explained in Sec. \ref{DRL}. For each episode, we keep a record of the absolute reward values for the action sequence and take the average of them as the objective function for this episode, denoted by $r_a$. In this sense, the quantity $r_a$ is representative of the average perturbation amplitude monitored at the downstream location $x = 700$ for one training episode. In addition, the upstream noise is random and varies for different training episodes, so we take the average of $r_a$ from the 10 best training episodes, denoted by $r_b$, as a good approximation to the test control performance.} Then $r_b$ is used as the objective function ($y$) in Eq. \ref{eq3-5} in the PSO algorithm based on the \textit{Pyswarm} package developed by \cite{miranda2018pyswarms}. The swarm size is selected as 50 and all the scaling factors are set default as 0.5. After a number of iterations, the algorithm converges and the optimal sensor placement is found. The following results in Sec. \ref{DRL} to \ref{robust} are based on the optimal sensor placement and the optimisation process is presented in Sec. \ref{Probe}. 

\section{Results and Discussion} \label{results}
\subsection{Dynamics of the 1-D linearised KS equation}\label{Dynamics}
\begin{figure}
  \centerline{\includegraphics[width=1.0\textwidth]{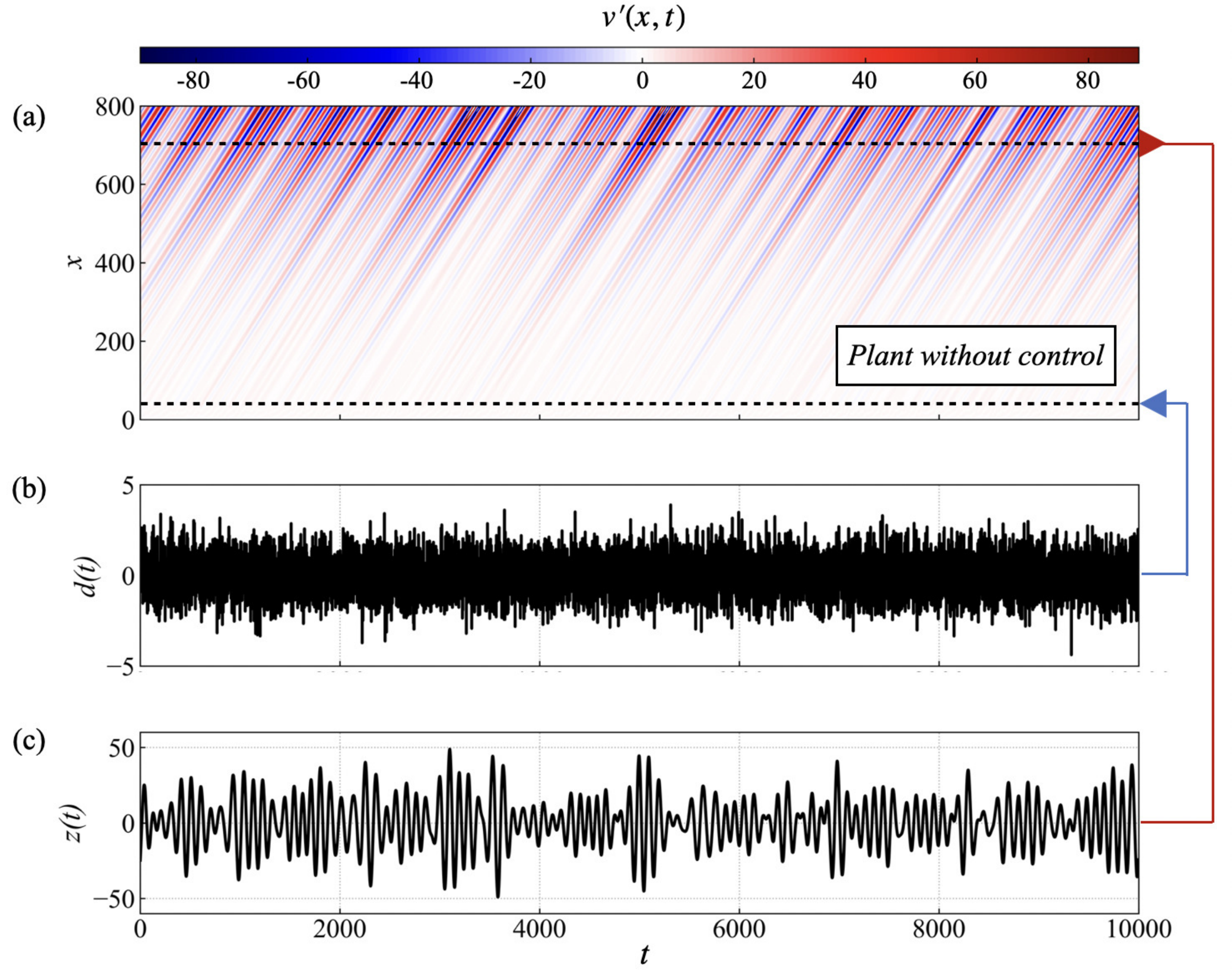}}
  \caption{Dynamics of the 1-D linearised KS equation when subject to a Gaussian white noise with a unit variance. (a) Spatial-temporal response of the system; (b) temporal signal of noise input $d(t)$ at $x = 35$; (c) output signal $z(t)$ measured by sensor at $x = 700$. The blue arrow represents input and red arrow represents output. }
\label{fig4}
\end{figure}

As introduced in Sec. \ref{problemformulation}, the 1-D linearised KS equation is an idealised equation for modelling the perturbation evolving in a Blasius boundary layer flow, with features such as non-normality, convective instability and a large time delay. Without control, the spatiotemporal dynamics of KS equation subjected to an upstream Gaussian white noise is presented in panel (a) of figure \ref{fig4}, where the perturbation is amplified significantly while travelling downstream and the pattern of parallel slashes demonstrates the presence of time delay. Panel \ref{fig4}(b) presents the temporal signal of noise with a unit variance and panel (c) displays the output signal at $x=700$. 

\begin{figure}
\centering  
\subfigure[Training history]{
\includegraphics[width=0.461\textwidth]{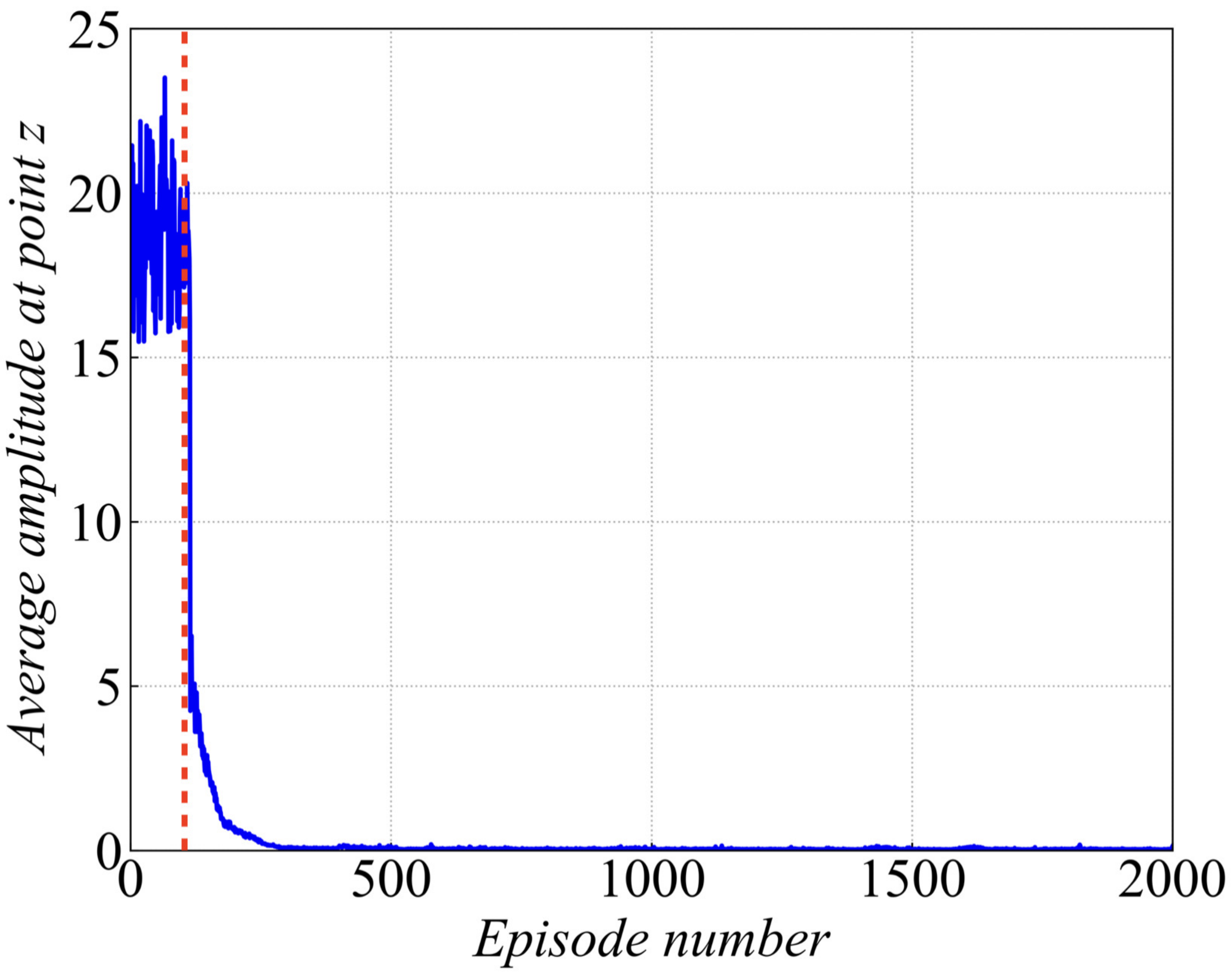}
}
\quad
\subfigure[DRL-based control performance]{
\includegraphics[width=0.474\textwidth]{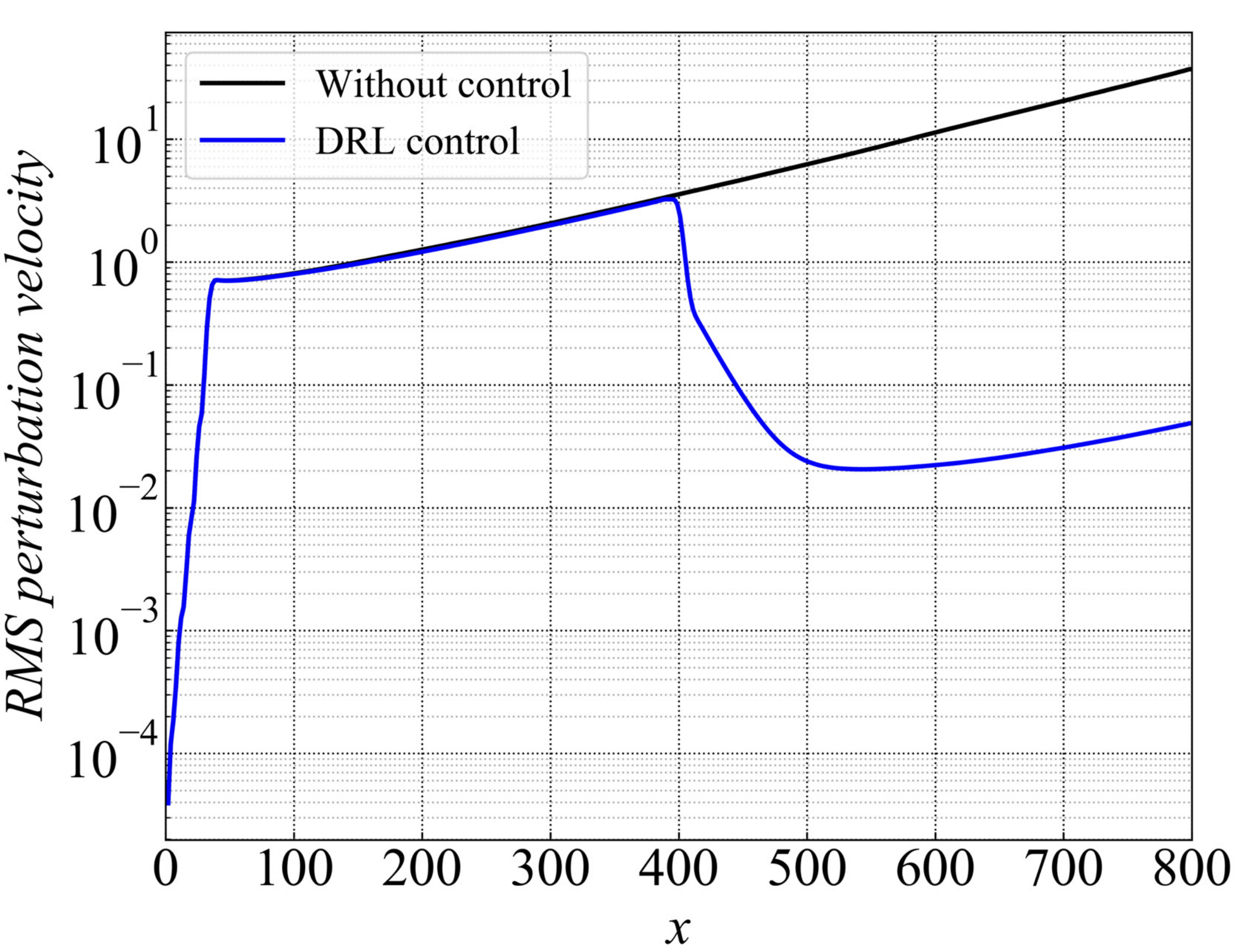}
}
\quad
\caption{DRL-based control for reducing the downstream perturbation evolving in the 1-D linearised KS equation. In panel (a), the average perturbation amplitude at point $z$ is recorded during the training process, \xuda{where the red dashed line denotes the time when the experience memory is full.} In panel (b), the root-mean-square (RMS) value of perturbation along the 1-D domain is plotted for both cases with and without control.}
\label{fig5}
\end{figure}

Due to the existence of linear instability in the flow, the amplitude of perturbation grows exponentially along the \textit{x} direction, which is verified by the black curve in figure \ref{fig5}(b), where the root-mean-square (RMS) value of perturbation calculated by
\begin{equation}
v^{\prime}(x)|_{rms}=\sqrt{\left(\overline{v^{\prime}(x)^{2}}-\overline{v^{\prime}(x)}^{2}\right)}
\end{equation}
is plotted along the 1-D domain and here $\overline{v^{\prime}(x)}$ represents the time-average of the perturbation velocity at a particular position $x$. In addition, by comparing figure \ref{fig4}(b) and \ref{fig4}(c), we find that the perturbation amplitude increases but some frequencies are filtered by the system. This is because when a white noise is introduced, a broadband of frequencies are excited. However, only frequencies corresponding to a certain range of wavelengths are unstable and amplified in the flow. This is the main trait of a convectively-unstable flow, plus the nonlinearity of the flow system, rendering it difficult for flow control. Such a flow is usually coined as a noise-amplifier in the hydrodynamic stability context \citep{Huerre1990}.

\subsection{DRL-based control of the KS system}\label{DRL}
In this section, we present the performance of DRL-based control on reducing the downstream perturbation evolving in the KS system. The training process is shown in figure \ref{fig5}(a) where the average perturbation amplitude at point $z$ ($x = 700$, see figure \ref{fig1}) is plotted for all the training episodes. In the initial stage of training, the perturbation amplitude is large and shows no sign to decrease, which corresponds to the filling process of experience memory in the DDPG algorithm. Once the memory is full (\xuda{denoted by the red dashed line in figure \ref{fig5}(a)}), the learning process begins and the amplitude decreases significantly to a value near zero and remains almost unchanged after 350 episodes, indicating that the policy network has converged and the optimal control policy has been learnt. Note that this explains the periodical training of 350 episodes when we implement the PSO algorithm, as mentioned in Sec. \ref{PSO}. Then, we test the learnt policy by applying it to the KS environment for 10000 time steps, with both the external disturbance input $d(t)$ and control input $u(t)$ turned on. 

\begin{figure}
  \centerline{\includegraphics[width=1.0\textwidth]{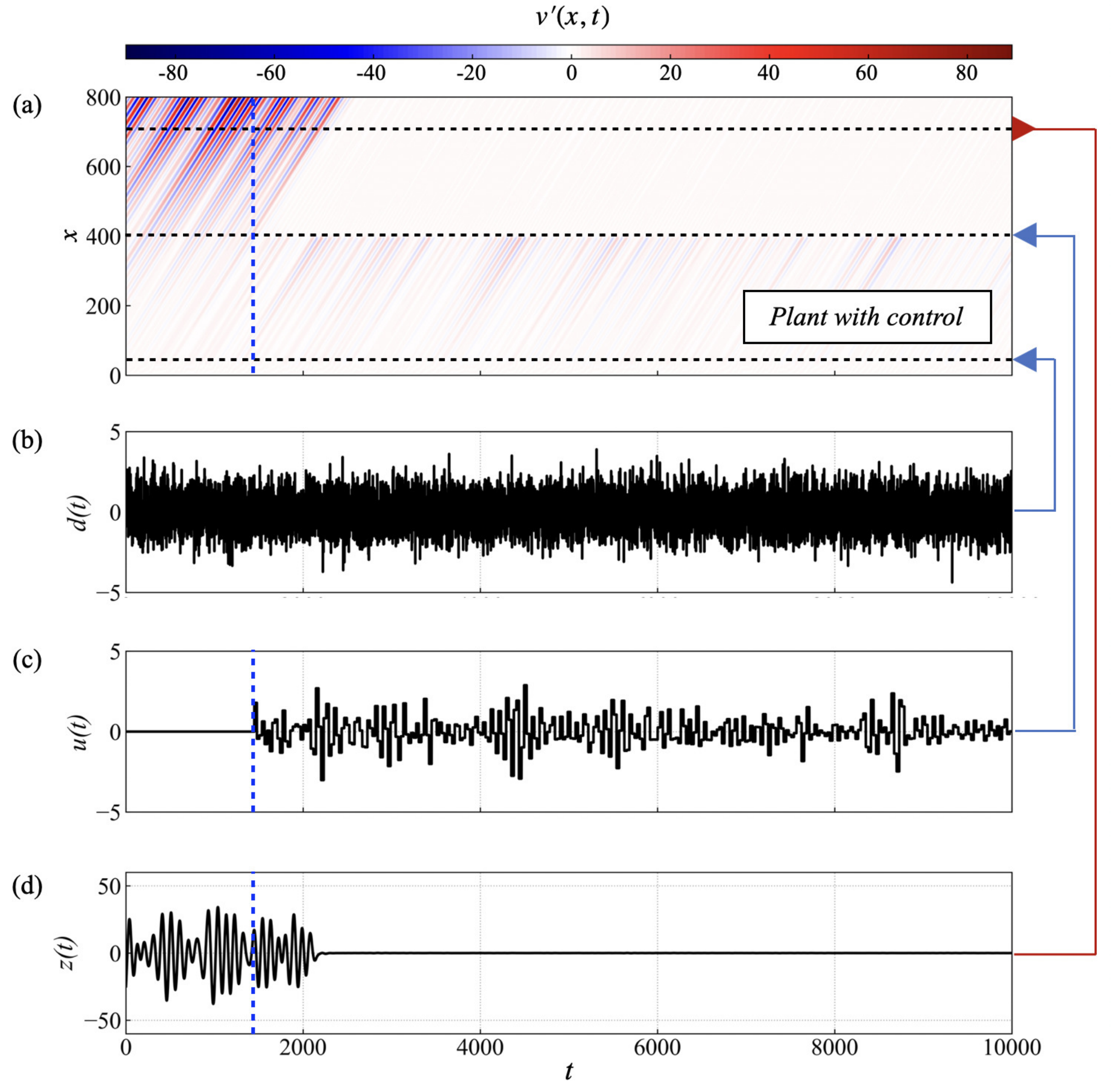}}
  \caption{Dynamics of the 1-D linearised KS equation with both the external disturbance and control inputs. (a) Spatial-temporal response of the system; (b) temporal signal of noise input $d(t)$ at $x = 35$; (c) temporal signal of control input $u(t)$ at $x = 400$ provided by the DRL agent; (d) output signal $z(t)$ measured by sensor at $x = 700$. The blue arrows represent inputs and red arrow represents output; the blue dashed line denotes the control starting time.}
\label{fig6}
\end{figure}

The spatiotemporal response of the controlled KS system is shown in figure \ref{fig6}(a) subjected to the noise input as shown in panel (b). Compared with that in figure \ref{fig4}(a), the downstream perturbation is significantly suppressed after the implementation of control action at $x = 400$ starting from $t = 1450$ (denoted by the blue dashed line). By observing figure \ref{fig6}(c) and \ref{fig6}(d), we find that there is a short time delay between the control starting time and the time when the downstream perturbation $z(t)$ starts to decline. This is due to the convective nature that the action needs some time to make an impact downstream. In addition, we also plot the RMS value of perturbation along the 1-D domain as shown by the blue curve in figure \ref{fig5}(b). In contrast to the originally exponential growth, with the DRL-based control, the perturbation amplitude is largely reduced downstream the control position ($x = 400$). For instance, the amplitude at $x = 700$ is reduced from about 20 to 0.03, which demonstrates the effectiveness of the DRL-based method in controlling the perturbation evolving in the 1-D KS system.

\subsection{Comparison between DRL and model-based LQR}\label{LQR}
In order to better evaluate the performance of DRL applied to the 1-D KS equation, we compare the DRL controlled results with those of the classical linear quadratic regulator (LQR) method. The LQR method for controlling the KS equation has been detailedly explained in \cite{fabbiane2014adaptive}. \xuda{For the basics of LQR, the reader may consult Appendix \ref{B}.}

\begin{figure}
\centering  
\subfigure[Control performance of ideal LQR and of DRL with action bound of ${[-5,5]}$]{
\includegraphics[width=1.0\textwidth]{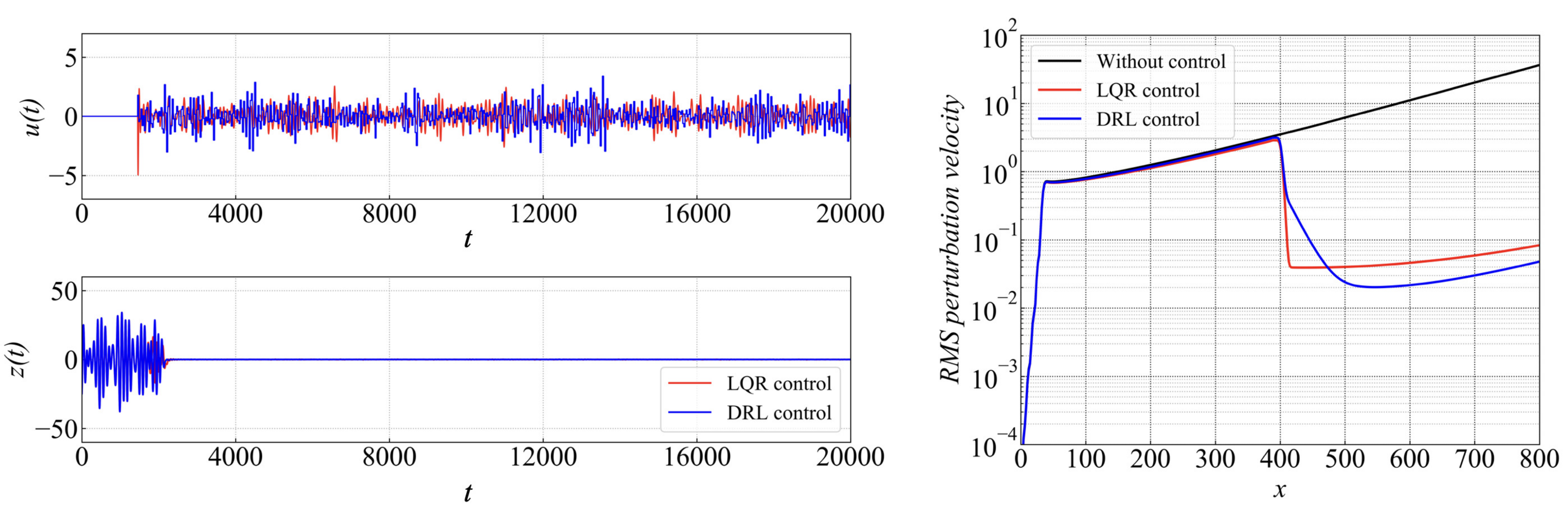}
}
\quad
\subfigure[Control performance of LQR and DRL both with action bound of ${[-3,3]}$]{
\includegraphics[width=1.0\textwidth]{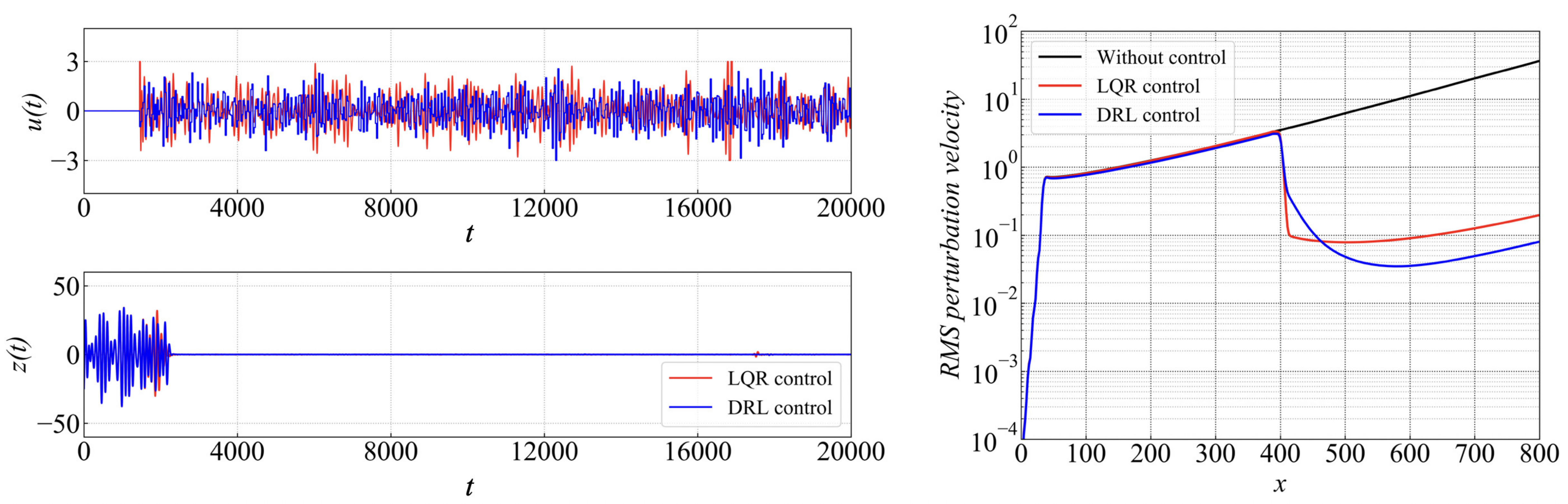}
}
\quad
\subfigure[Control performance of LQR and DRL both with action bound of ${[-2,2]}$]{
\includegraphics[width=1.0\textwidth]{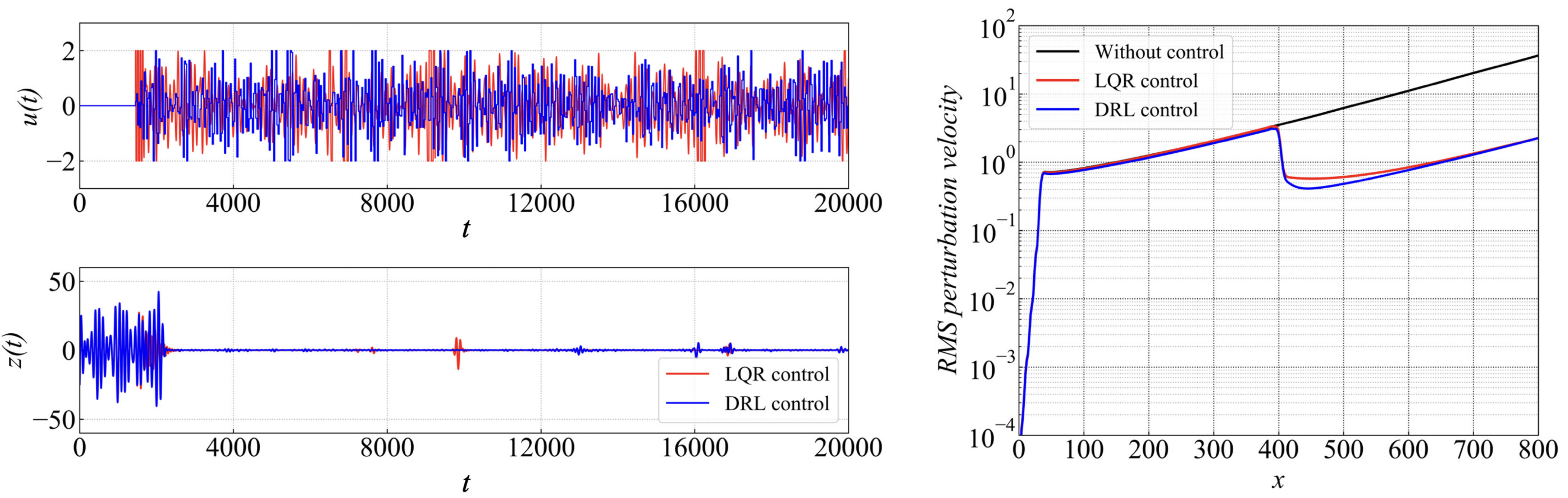}
}
\quad
\subfigure[Control performance of LQR and DRL both with action bound of ${[-1,1]}$]{
\includegraphics[width=1.0\textwidth]{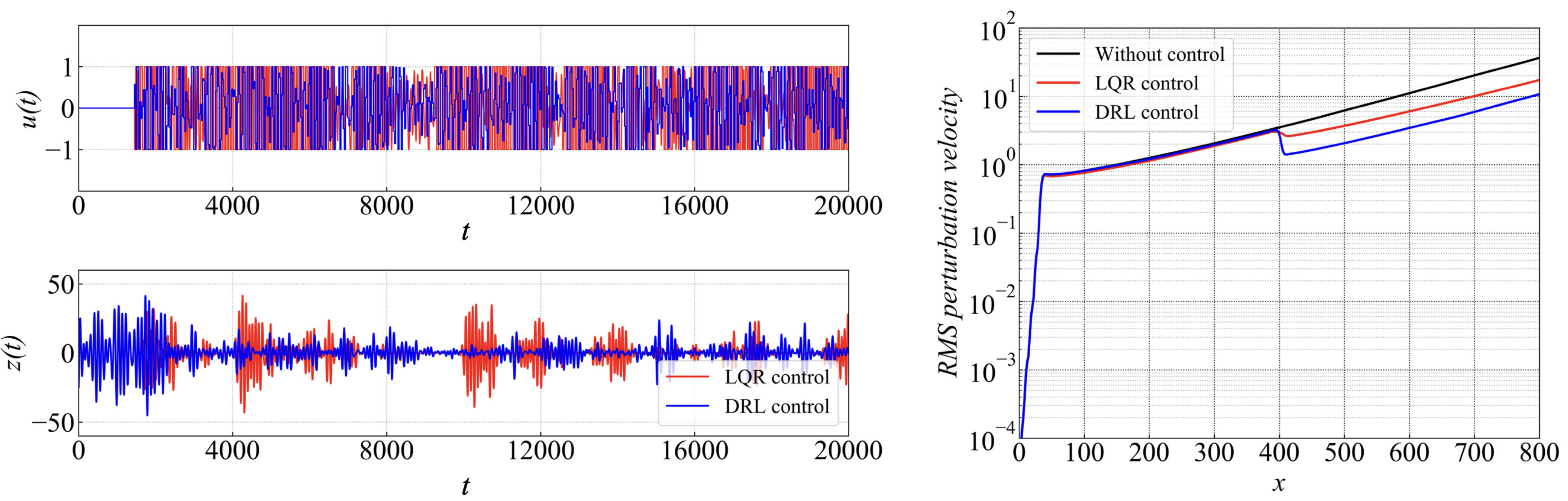}
}
\quad
\caption{Comparison between DRL-based control and LQR control when applied to the 1-D linearised KS system with different action bounds. In each panel, the top left figure presents the time-variation curve of the control action $u(t)$ and the bottom left presents the corresponding sensor output $z(t)$; the right figure shows the RMS value of perturbation velocity along the 1-D domain.}
\label{fig19}
\end{figure}

Ideally, if there are no action bounds imposed on LQR, it will generate the optimal control performance as shown in red curves in figure \ref{fig19}(a) and we call it the ideal LQR control. It is shown that the perturbation amplitude downstream the actuator position is dramatically reduced and the corresponding action happens to be in the range of [-5, 5]. Thus, for a consistent comparison of LQR and DRL, we apply the same action bound of [-5, 5] to DRL-based control and present the resulting control performance as blue curves. It is found that the DRL-based control is slightly better than the LQR control in reducing the downstream perturbation as shown in the right panel of figure \ref{fig19}(a). Despite the fact that in the near downstream region from $x = 400$ to $x = 470$, the amplitude of DRL-based control is higher than that of LQR control, the reduction of perturbation is more significant in the DRL control in a further downstream region. 

In real control applications, one needs to consider the saturation effect of the actuators and this is realised by imposing the action bound \citep{grundmann2008active,corke2010dielectric}. Here we adopt the same method used by \cite{fabbiane2014adaptive} to bound the LQR control action as below
\begin{equation}
u_{L Q R}= \begin{cases}u_{L Q R} & \text { if } \bar{u}_{\min }<u_{L Q R}<\bar{u}_{\max }, \\ \bar{u}_{\min } & \text { if } \bar{u}_{\min } \geqslant u_{L Q R}, \\ \bar{u}_{\max } & \text { if } \bar{u}_{\max } \leqslant u_{L Q R}.\end{cases}
\label{eq4-1}
\end{equation}
We first impose the same action bound of [-3, 3] to both LQR and DRL-based control and compare their control performance in figure \ref{fig19}(b). When the saturation function is applied to the LQR control signal, the controller becomes sub-optimal and its performance to reduce the perturbation becomes worse. It is observed that at about $t = 17000$, the action is saturated and thus there appears a small spike later in the output signal $z(t)$ as shown in the bottom left subplot of figure \ref{fig19}(b). On the other hand, the DRL-based control also performs slightly worse with the smaller action range but it still outperforms LQR control with the same action bound. Next, we further narrow down the action range to [-2, 2] and this time both LQR and DRL-based control deteriorate due to the saturation of action and their control performances are at the similar level, as shown in figure \ref{fig19}(c). Finally, we further limit the action range to [-1, 1] and compare the corresponding control performance of the two methods in figure \ref{fig19}(d). In this condition, LQR controller deteriorates severely and the reduction of downstream perturbation is rather limited. In contrast, DRL-based control performs better with a relatively clear trend of reduction of perturbation downstream the actuator position, although the extent of reduction is not significant compared to the previous cases with larger action bounds as shown in figure \ref{fig19}(a-c). We would like to emphasise that compared to the model-based LQR method, the DRL-based method is totally model-free, which means that the control law is learnt with no awareness of the assumed models of the \textit{plant} as mentioned in Sec. \ref{NS}. In addition, the control action is determined based solely on states collected by some sensors from the flow field, instead of the full knowledge of the \textit{plant} as required by LQR control. 

\begin{figure}
\centering  
\subfigure[Control performance comparison]{
\includegraphics[width=0.467\textwidth]{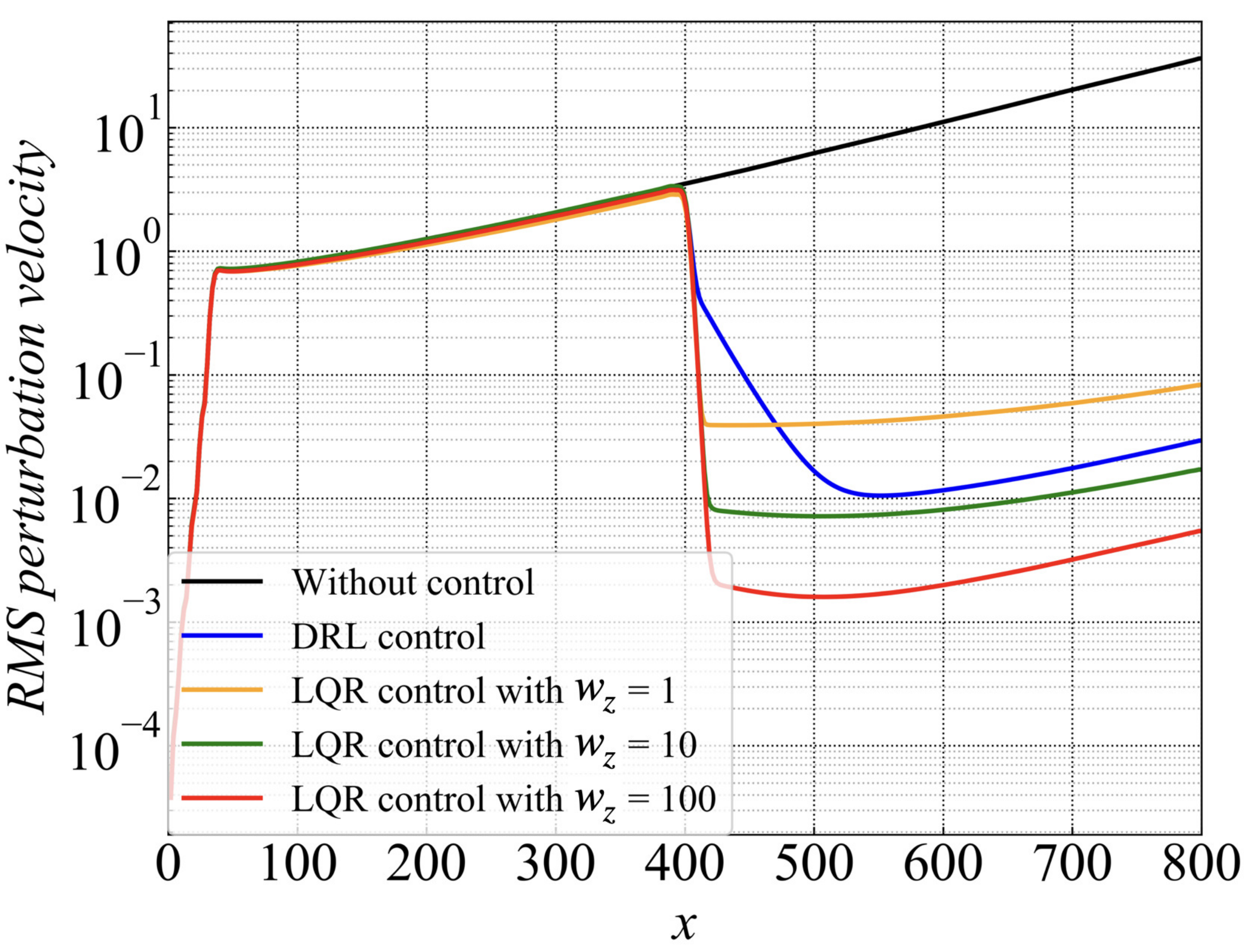}
}
\quad
\subfigure[Time variation of control action]{
\includegraphics[width=0.469\textwidth]{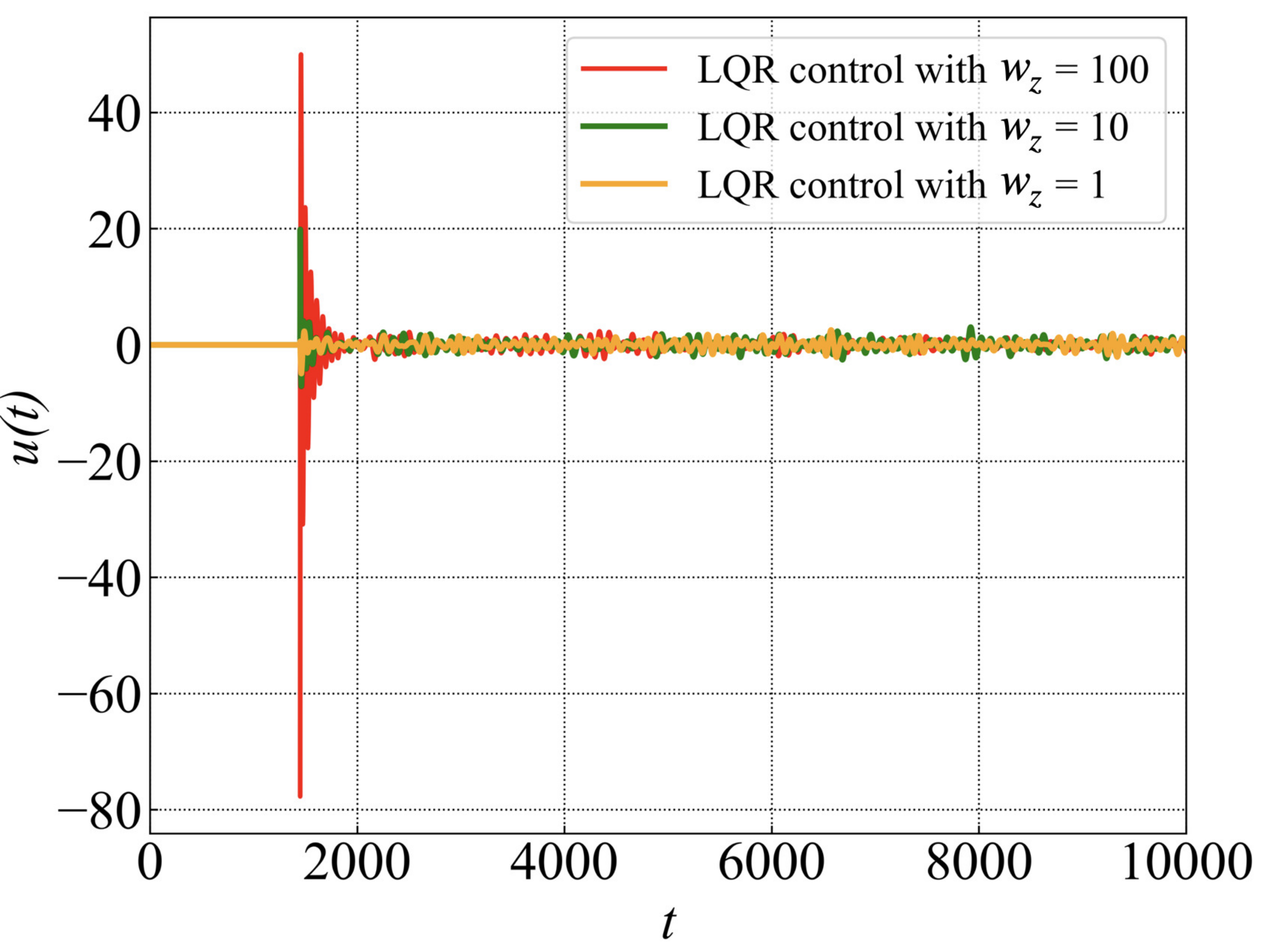}
}
\quad
\caption{Comparison of LQR control performance with different values of weight $w_z$. In panel (a), the RMS value of perturbation velocity along the 1-D domain is plotted. In panel (b), the time-variation curves of control action $u(t)$ are presented with the control starting at $t = 1450$.}
\label{fig20}
\end{figure}

The current LQR controller is based on a specific selection of weights, i.e., $w_z = w_u = 1$ as given in Eq. \ref{eq-A2}. This selection was adopted by \cite{fabbiane2014adaptive}, indicating that equal weight has been assigned to the two parts, i.e., sensor output $z(t)$ and control input $u(t)$. These parameters influence the control performance of LQR. For example, the LQR performance can be improved by increasing the weight $w_z$ from 1 to 10 and 100 with a fixed $w_u$ at 1 (which means that we focus more on the output part). The results are shown in figure \ref{fig20}(a). However, the ensuing problem is that with the increase of $w_z$, the magnitude of control action will experience a rapid increase at the very initial stage of the control process as shown in figure \ref{fig20}(b), which is unavoidable because larger weight is considered to force the output to the target value as soon as possible and this is realised by a larger action in the beginning. In realistic applications, the choice of weights is a trade-off between the control performance and the available range of control action. 

In addition to the control performance, the cost of energy effort is another important factor in active flow control. The cost of energy effort in our context can be quantified by the average magnitude of action during the control process. We compare this aspect for the two methods in Table \ref{table2}, which shows that the average magnitude of action for the two methods are in general comparable to each other, except that in the case of the smallest action bound of $[-1,1]$, the averaged action of DRL is about 20\% less than that of the LQR.


\begin{table}
  \begin{center}
\def~{\hphantom{0}}
  \begin{tabular}{cccc}
      $\ Control  \ method \ $ & $\ Action \ bound \ $ & $ \ Average \ magnitude \ of \ action \ $ \\[3pt]
             & no & 0.6843\\
             & [-3.0, 3.0] & 0.7319\\
       LQR   & [-2.0, 2.0] & 0.7092\\
             & [-1.0, 1.0] & 0.8139\\
       \hline
             & [-5.0, 5.0] & 0.7132\\
             & [-3.0, 3.0] & 0.7265\\
       DRL   & [-2.0, 2.0] & 0.7176\\
             & [-1.0, 1.0] & 0.6533\\
  \end{tabular}
  \caption{Average magnitude of action for different methods with different action bounds.}
  \label{table2}
  \end{center}
\end{table}

\subsection{Robustness of the DRL policy in controlling convectively-unstable flows}\label{robust}
\begin{figure}
  \centerline{\includegraphics[width=1.0\textwidth]{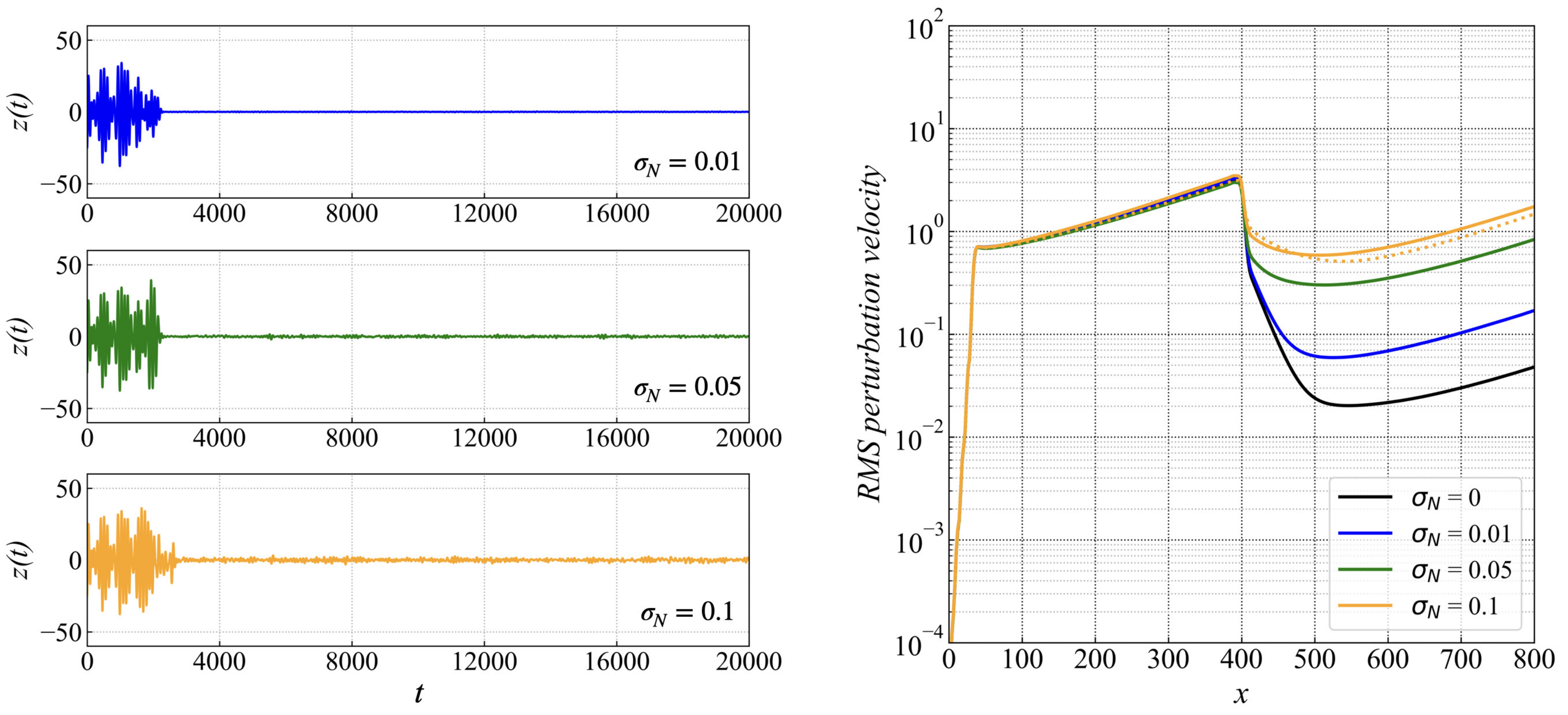}}
  \caption{DRL-based control performance with states contaminated by a Gaussian noise of three different standard deviations $\sigma_N$. In the left panel, the sensor output $z(t)$ is plotted for the three cases. In the right panel, the RMS value of perturbation along the 1-D domain is plotted for the three cases. Besides, the black curve for control with uncontaminated states and the orange dashed curve for control with the agent trained with $\sigma_N = 0.1$ are also shown for comparison.}
\label{fig7}
\end{figure}

In this section, we test the robustness of the learnt control policy to two types of noise, i.e., measurement noise and external noise. In realistic conditions, the states collected by sensors are always subjected to noise and we call it measurement noise. We consider this effect by adding a Gaussian noise with standard deviations $\sigma_N$ of 0.01, 0.05 and 0.1, respectively, to the original states and test the robustness of the learnt policy which was trained with $\sigma_N=0$ to the measurement noise of different levels. As shown in the left panel of figure \ref{fig7}, with the increase of noise level, the overall result is still satisfactory except that the DRL-based control performance deteriorates in the sense that there are some small oscillations being monitored in the sensor output $z(t)$. We plot the corresponding RMS value of perturbation along the 1-D domain in the right panel, where the control performance with uncontaminated states, i.e., $\sigma_N = 0$, is also shown by the black curve for comparison. With the noise level increasing to 0.1 (accounting for about 10\% of the originally normalised state signals), the controlled amplitude of $z(t)$ at $x = 700$ increases from 0.03 to 1.0. Despite the slight deterioration of the control performance with the increased noise level, the current policy is still effective in reducing the downstream perturbation (recall that the uncontrolled amplitude at $x=700$ is 20). We have also attempted to train a DRL agent with $\sigma_N = 0.1$ and test it in the same condition. The resulting control performance is depicted by the orange dashed curve in the right panel of figure \ref{fig7} and it can be seen that the result is close to the orange solid curve which was obtained from the agent trained with $\sigma_N = 0$ but tested with $\sigma_N = 0.1$. Therefore, the DRL agent is robust in the sense that, once trained, it can effectively control a new flow situation with some additional noise. \xuda{This robustness property benefits from the closed-loop nature of the control system and the decorrelation of noise among state observations, as explained by \cite{paris2021robust}. More specifically, the action error caused by measurement noise does not accumulate over time since the feedback state is able to rectify the previous erroneous action and prevent it from deviating too far.}

\begin{figure}
  \centerline{\includegraphics[width=1.0\textwidth]{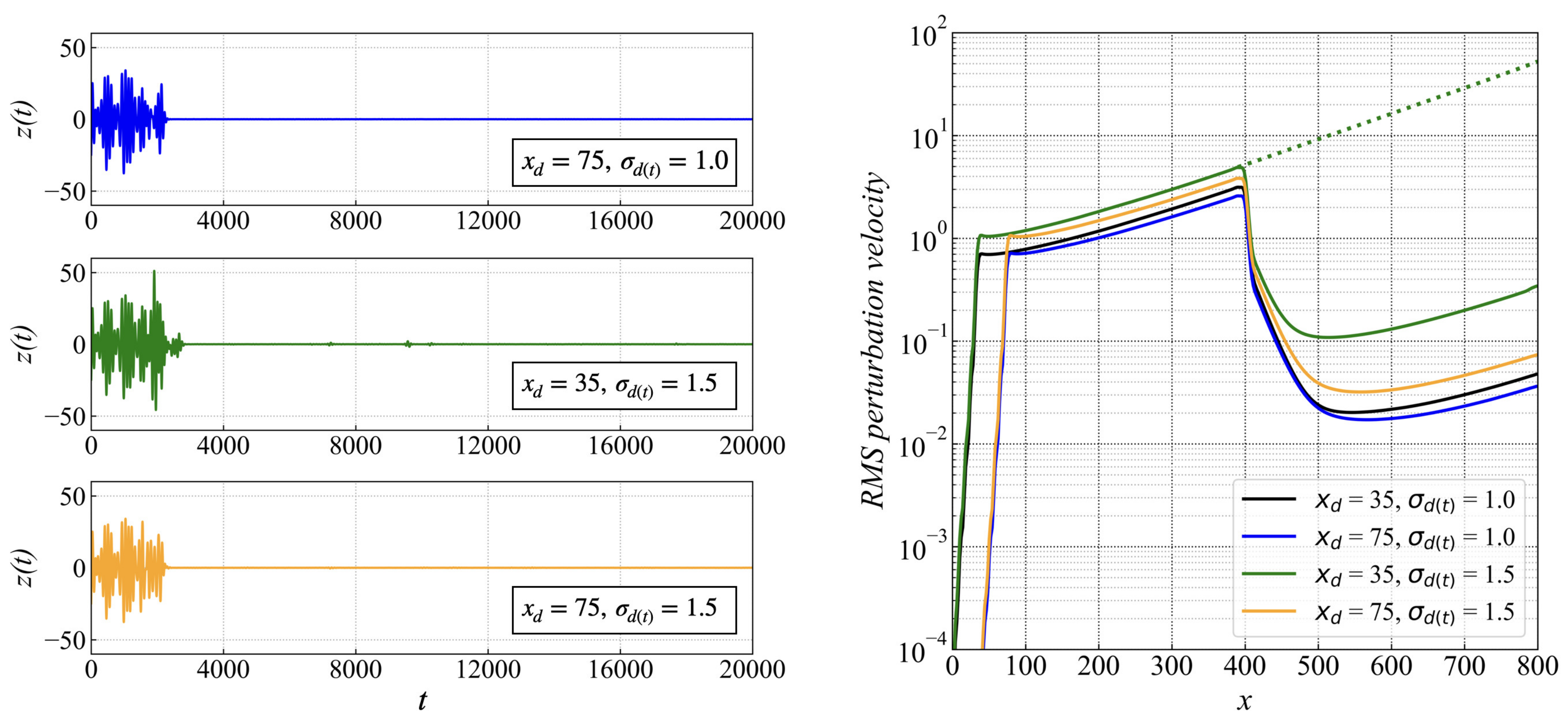}}
  \caption{DRL-based control performance under three different upstream noise conditions: a Gaussian noise input at $x_d$ with a standard deviation of $\sigma_{d(t)}$. In the left panel, the sensor output $z(t)$ is plotted for the three cases. In the right panel, the RMS value of perturbation along the 1-D domain is plotted for the three cases and that for control testing with the same environment as training, i.e., $x_d = 35$ and $\sigma_{d(t)} = 1.0$ is also given by the black curve for comparison.}
\label{fig8}
\end{figure}

The second type of noise is the external noise. The current policy was trained under a fixed external noise condition, i.e., a white noise input at $x_d = 35$ with a unit standard deviation $\sigma_{d(t)} = 1.0$. We want to test whether the current policy is still valid in new external noise situations. For instance, we can move the noise position from $x_d = 35$ to $x_d = 75$ with the noise level unchanged or increase the noise level from $\sigma_{d(t)} = 1.0$ to $\sigma_{d(t)} = 1.5$ with the noise position unchanged, or change both of them. We test the control policy in such new environments and plot the obtained control performance in figure \ref{fig8}. Due to the increase of noise level and/or the shift of noise position, the perturbation field upstream the control position is only trivially different from each other, as shown in the right panel. Once the DRL-based control is applied, the downstream perturbation is significantly reduced, as demonstrated in both the left and right panels. This robustness property benefits from the state normalisation process as mentioned in Sec. \ref{RL}, which has also been demonstrated in \cite{paris2021robust}. Among the three testing cases, the second one ($x_d = 35$ and $\sigma_{d(t)} = 1.5$) depicted by the green curve leads to a relatively large downstream perturbation. This is due to the fact that the uncontrolled perturbation level increases (depicted by the green dashed curve in the right panel) and sometimes is out of the current action range of [-5, 5], which will lead to the small spikes as shown in the middle figure of the left panel. 

In the end, we would like to mention that in the confined cylinder wake flow as studied by \cite{rabault2019artificial,paris2021robust,li2022reinforcement}, the nature of flow instability is locally absolute, cf. \cite{Monkewitz1987} and \cite{giannetti2007structural} for the general theory of flow instability in unconfined wake flows or see the flow stability and sensitivity analyses in \cite{li2022reinforcement} for the confined wake flow. In such a flow, the noise level is only secondary compared to the primary absolute instability mechanism accounting for the perturbation amplification, cf. \cite{Huerre1990} for a classical review of the absolute and convective instabilities. On the other hand, in the case of KS equation or boundary layer flows, the nature of instability is convective, for which the perturbation amplification depends more critically on the upstream noise. Our numerical demonstration of the DRL control in the convectively-unstable flow is a stricter test of its robustness and proves its applicability in such flows.

\subsection{Optimisation of sensor placement}\label{Probe}
All the above results are obtained based on the optimal sensor placement. In this section, we explain how the optimal sensor placement has been identified. As mentioned in Sec. \ref{PSO}, we adopt the PSO method to find the optimal sensor placement. Before the searching for the optimal sensor positions, we need to specify lower and upper bounds for the searching space. In the current work, we investigate different cases with 1, 2, 4, 6, 8 and 10 sensors located in $x\in(370,430)$. This range is determined through our preliminary tests. We have attempted to add more sensors outside the current range and found that they had unimportant effects on the final control performance.

\begin{figure}
\centering  
\subfigure[Optimisation process using PSO]{
\includegraphics[width=0.461\textwidth]{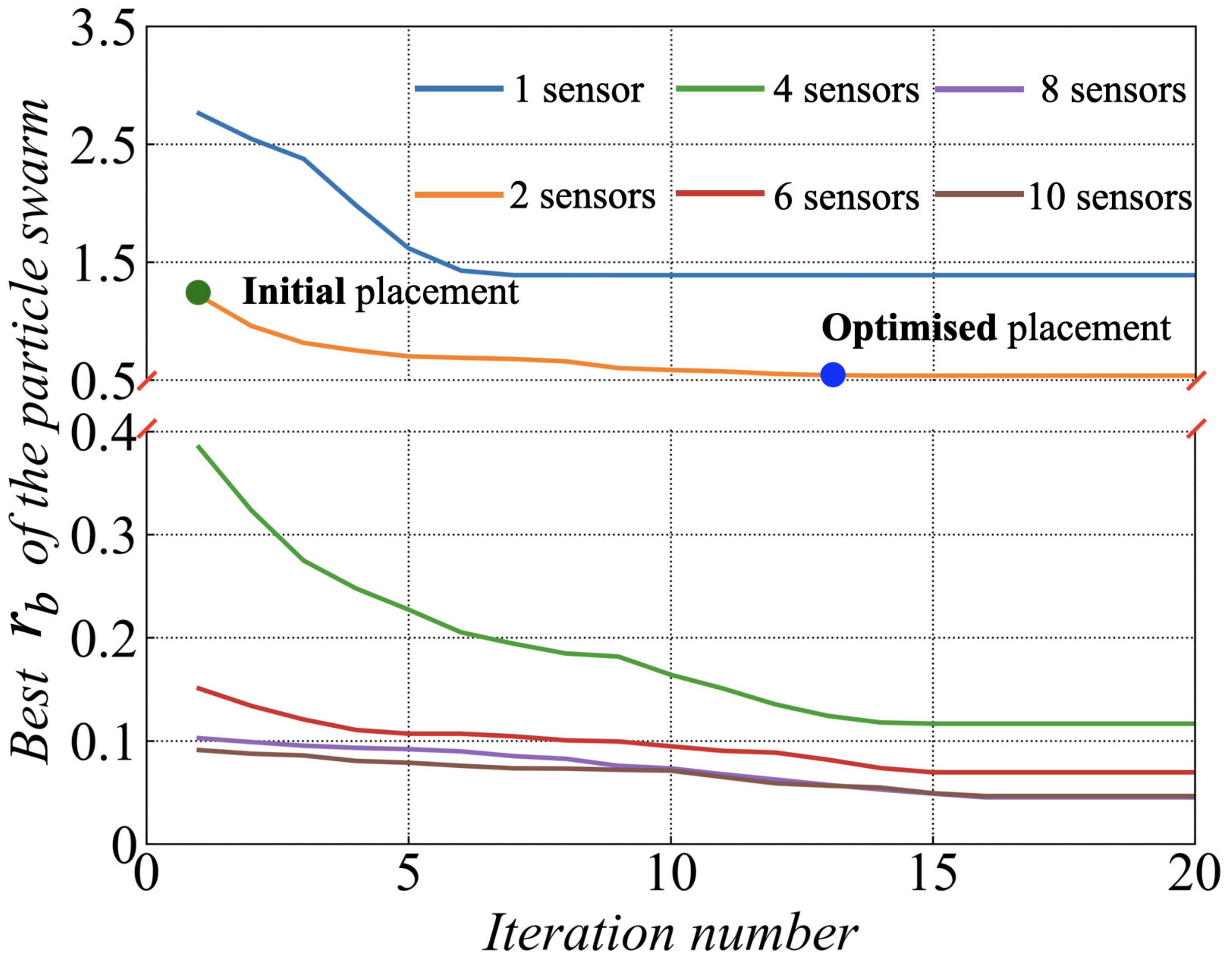}
}
\quad
\subfigure[Control performance comparison]{
\includegraphics[width=0.475\textwidth]{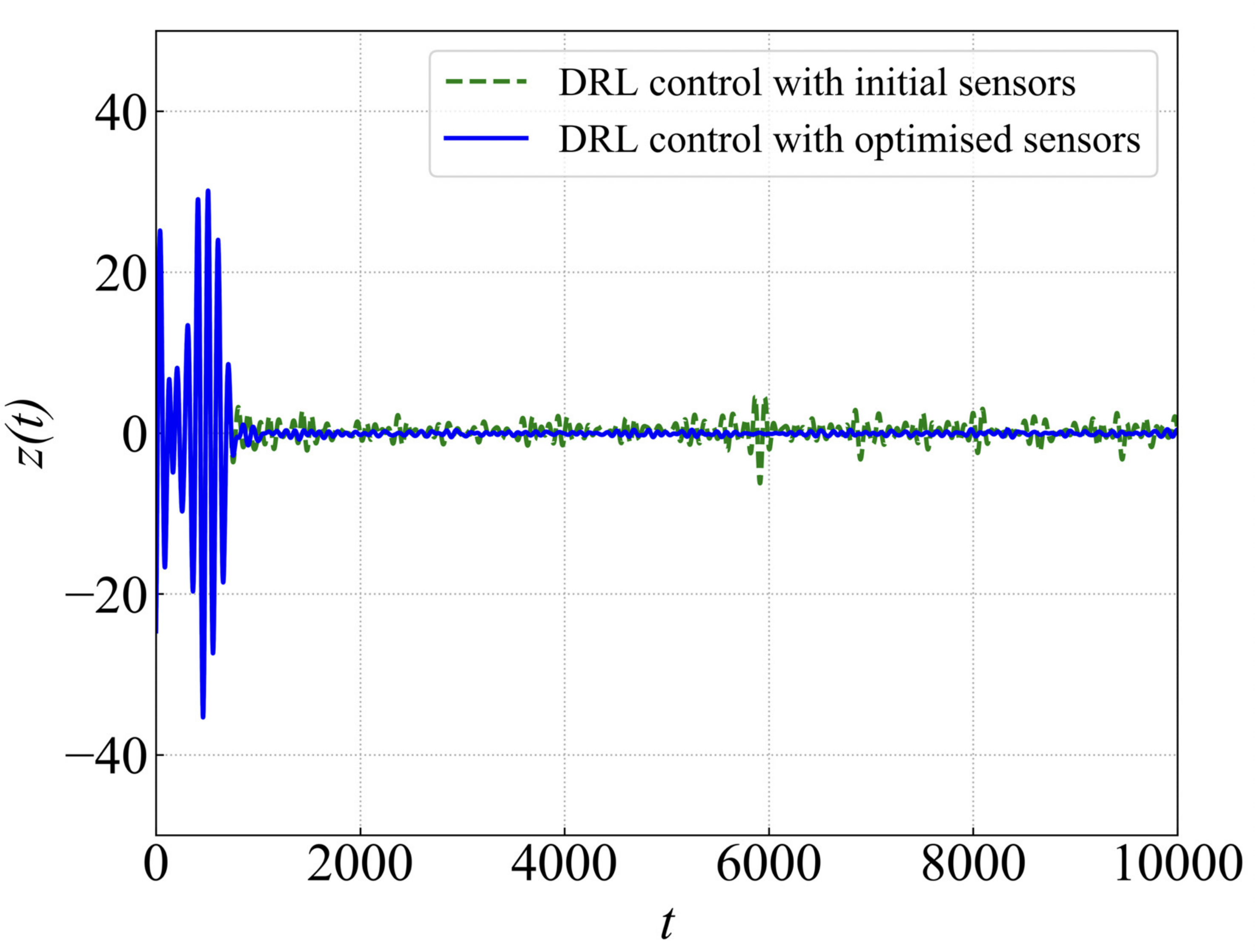}
}
\quad
\caption{Optimal sensor placement in DRL-based flow control. In panel (a), the optimisation process using PSO is presented for cases with different numbers of sensors, and finally the optimal placement is found which leads to the lowest perturbation amplitude $r_b$. In panel (b), the control performance comparison between the initial placement and the optimised one is plotted (taking 2 sensors as an example).}
\label{fig9}
\end{figure}

The optimisation processes for all the considered cases are presented in figure \ref{fig9}(a), where the trend of these curves are similar. Here we take the case with two sensors as an example to illustrate this process. The initial placement of the sensors is random. After the first iteration, the algorithm returns an initial sensor placement (denoted by the green point) that all the particles in the swarm have ever found, which corresponds to the lowest objective function $r_b$ so far, as defined in Sec. \ref{PSO}. Then, based on the initial placement, the particles attempt to find a new placement which leads to a lower $r_b$ in the next iteration. Finally, after a number of iterations, the searching process converges and the optimal placement is identified, denoted by the blue point. To demonstrate the effectiveness of the optimisation, we compare the control performance given by the initial sensor placement and the optimised one. As shown in figure \ref{fig9}(b), the optimised placement indeed results in a better control performance, i.e., a lower perturbation amplitude at the downstream position $x_z = 700$ than that of the unoptimised one. 

\xuda{The optimised sensor layouts with different numbers of sensors are presented in figure \ref{fig10}(a), where circle points represent the optimised sensor placements (OSP), named as OSP plus the number of sensors, and the actuator position is also given as a reference. As suggested by one of the reviewers, we will test whether only using sensors upstream the actuator is sufficient for control. To do so, we consider three newly added sensor placements, as shown by crosses in the figure, among which SP1 is a four-sensor layout with the downstream sensors removed from OSP8; SP2 is a five-sensor layout with the downstream sensors removed from OSP10; SP3 is the uniformly-spaced four-sensor layout upstream the actuator which is scaled by the parallel slash pattern as shown in figure \ref{fig4}(a). 

\begin{figure}
  \centering  
  \subfigure[Optimised sensor placement]{
  \includegraphics[width=0.461\textwidth]{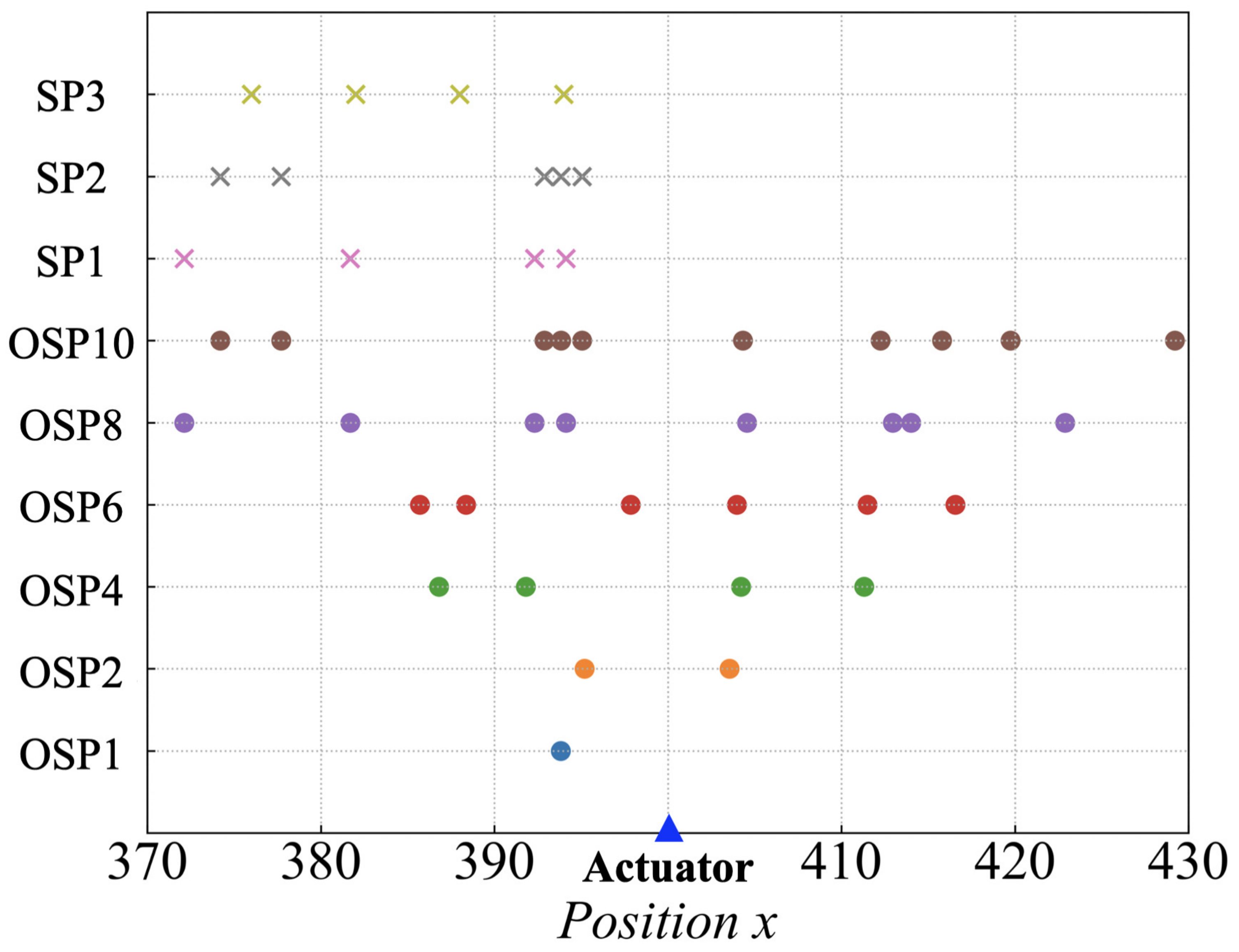}
  }
  \quad
  \subfigure[DRL-based control performance]{
  \includegraphics[width=0.474\textwidth]{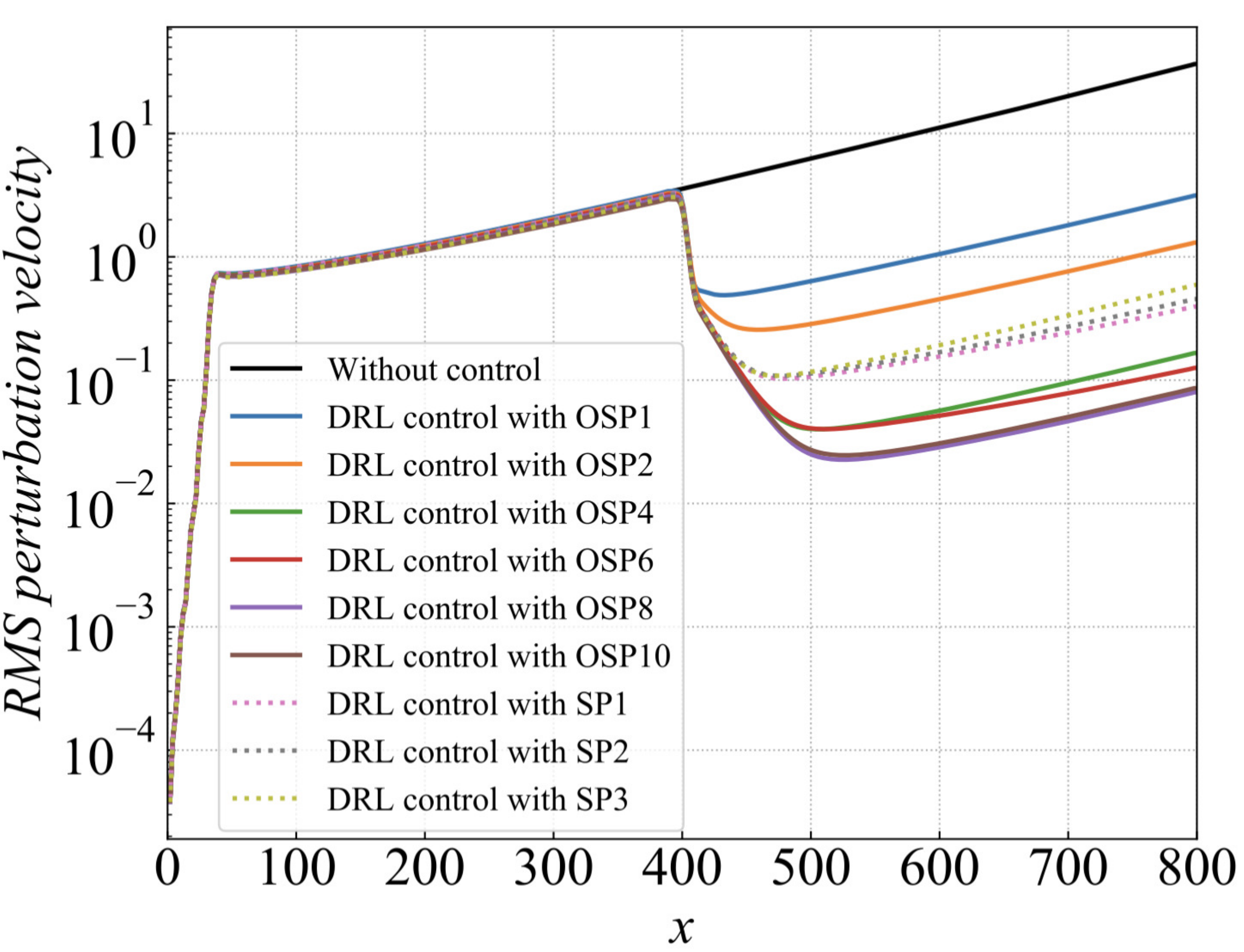}
  }
  \quad
  \caption{Optimal sensor placement and the effect of the number of sensors. In panel (a), the optimal placement is presented for cases with different numbers of sensors. In addition, three specific layouts with all sensors placed upstream are also considered. In panels (b), we compare the corresponding control performance with different sensor placements.}
  \label{fig10}
\end{figure}

The corresponding control performance is presented in figure \ref{fig10}(b), where the RMS value of perturbation along the 1-D domain is plotted. It is shown that if only one sensor is used, it is better to place it upstream the actuator than downstream to help the agent detect the upcoming disturbance. Nevertheless, due to the implementation of sticky actions where the control action keeps constant every 30 time steps, a single upstream sensor can only inform the agent of a single upcoming parallel slash in a packet of slashes that crosses the actuator over the duration that the action is ``stuck''. As the slashes are generated by random noise, it is not possible for the agent to choose an action that accounts for the packet of slashes based on a single measurement. With more sensors placed upstream, it is more likely that the agent will be fully informed to select the best sticky action to suppress the packet. This hypothesis is verified by the fact that the performance of the three layouts SP1, SP2 and SP3 (crosses in the figure) is better than that of OSP1 and OSP2 (one upstream and one downstream). However, the four-sensor layout with all sensors placed upstream is inferior to OSP4 (two upstream and two downstream), which indicates that the downstream sensors are also necessary for a better control performance. This is determined by the inherent feature of DRL framework, where the agent interacts with the environment in a closed-loop. After the agent imposes an action to the environment, the environment needs to return states as feedback which determines the next action. Overall, with the number of sensors increasing from 1 to 8, the resulting control performance is continuously improved as expected since more sensors will provide a more complete measurement of the flow environment. However, in the current case, as the number of sensors is further increased from 8 to 10, no further improvement is observed, which indicates that the information given by the two additional sensors is redundant. That is the reason why we adopt eight sensors in Sec. \ref{RL} and the placement of them is depicted by the purple points in figure \ref{fig10}(a).

Some discussions on the role of sensors in DRL control are in order. In the model-free DRL framework, sensors are used to collect states which not only provide a measurement of the flow environment but also reflect the effect of the action after it is taken. In this context, a proper number of sensors should be placed both upstream and downstream the actuator. In contrast, in traditional model-based control methods such as the linear quadratic Gaussian regulator, the sensor is used to estimate the full states of the \textit{plant} via the Kalman filter and then the control action is calculated based on the estimated states. In this situation, one sensor can usually work well since this demonstration does not implement sticky actions and is able to observe each slash individually that passes through and actuate accordingly. For instance, \cite{belson2013feedback} investigated the effects of different types and positions of sensor and actuator on the model-based control of boundary layer flows and found a specific sensor-actuator pair with good control performance and robustness.}


\subsection{DRL-based control performance in 2-D boundary layer flows}\label{DRL-BL}
We have demonstrated that DRL-based method is effective in controlling the convectively-unstable perturbation evolving in the 1-D KS system. Since the KS equation is a reduced model for the dynamics of the streamwise perturbation velocity evolving in the boundary layer flow at the wall-normal position $y=1$, it is reasonable to expect that the optimised sensor placement that we have derived in the KS system can be translated to the control of 2-D Blasius boundary layer flows. We note that the nonlinear NS equations are the governing equations in this case.

\begin{figure}
\centering  
\subfigure[Training history]{
\includegraphics[width=0.472\textwidth]{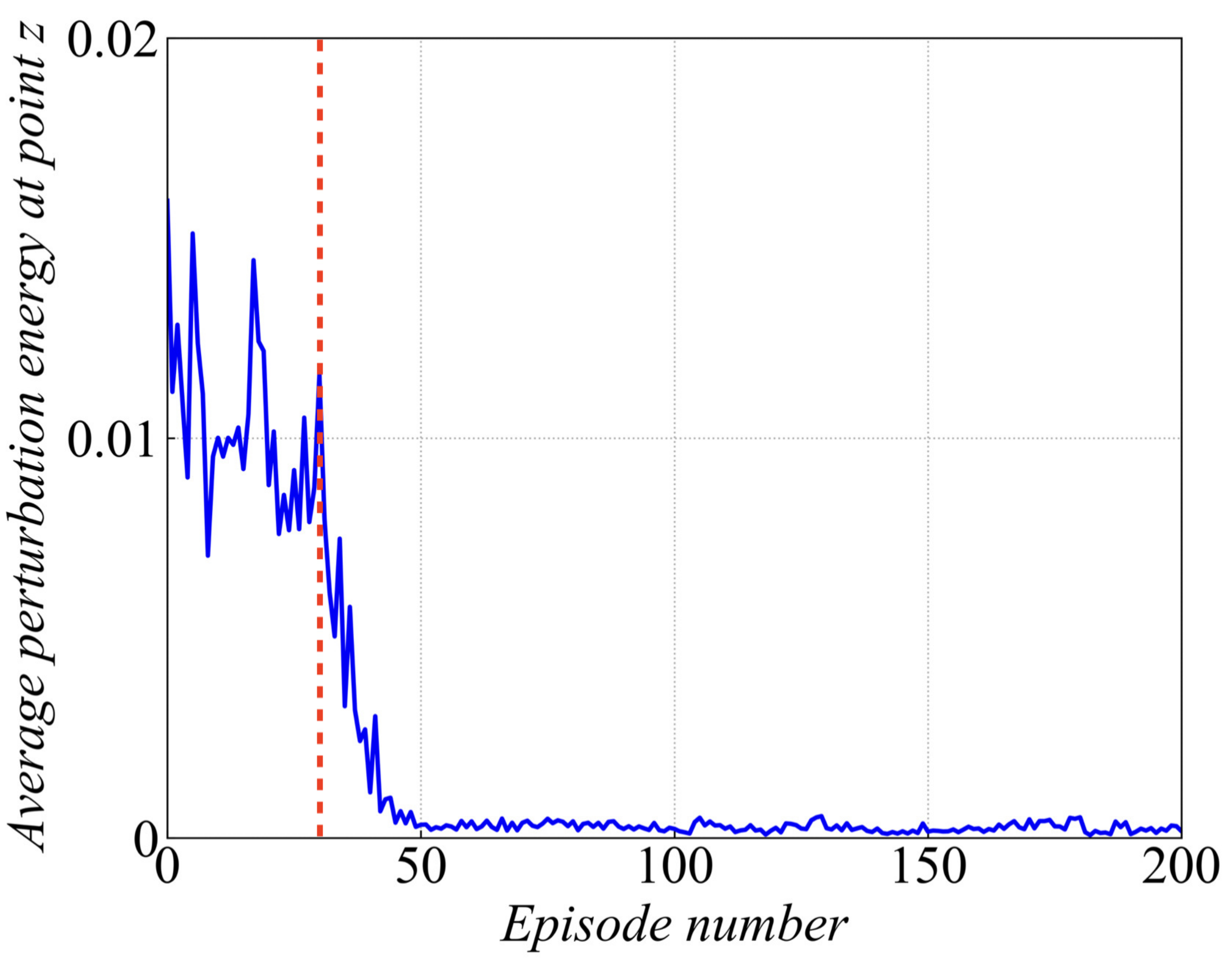}
}
\quad
\subfigure[Test process]{
\includegraphics[width=0.462\textwidth]{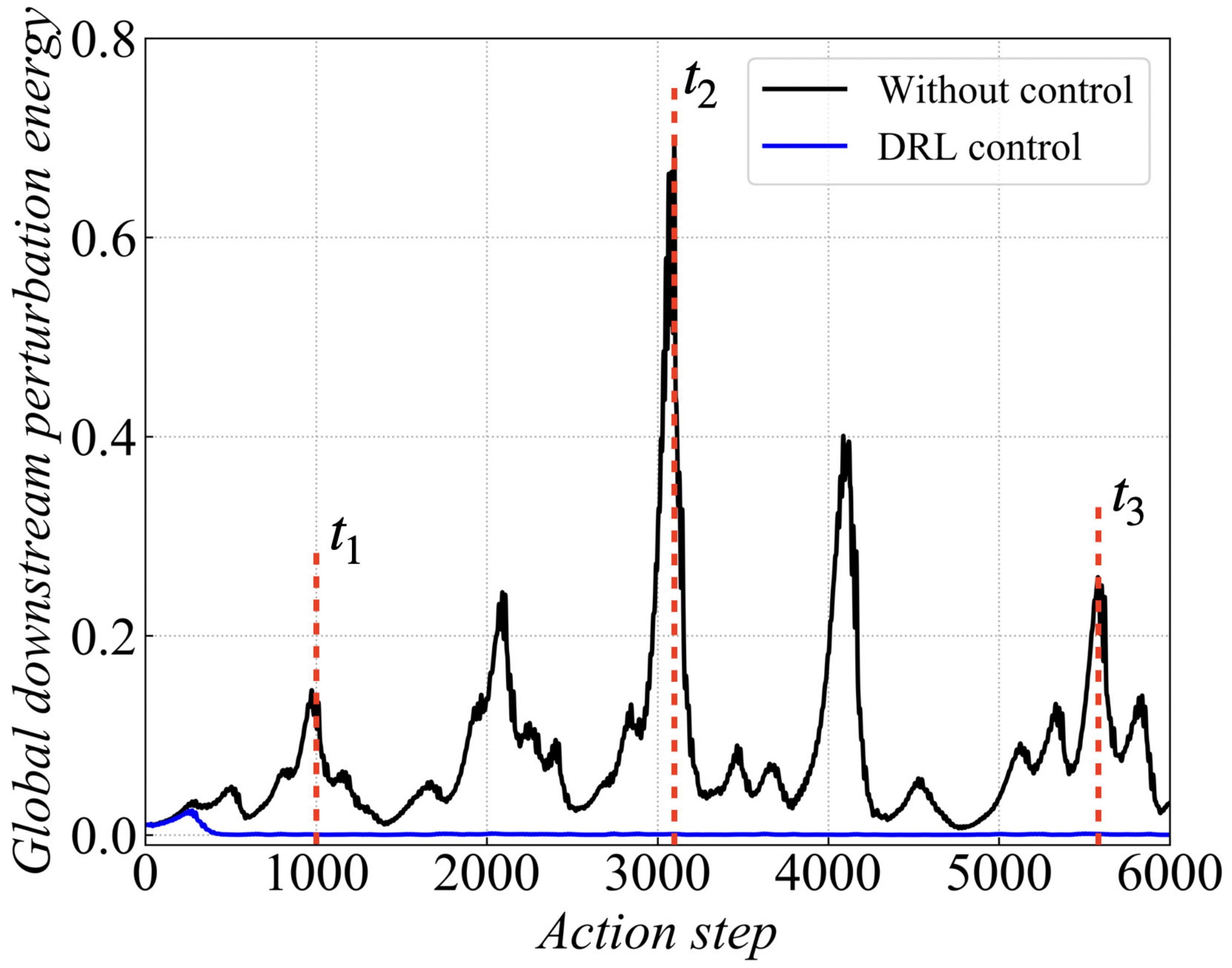}
}
\quad
\subfigure[Effect of the number of sensors]{
\includegraphics[width=0.465\textwidth]{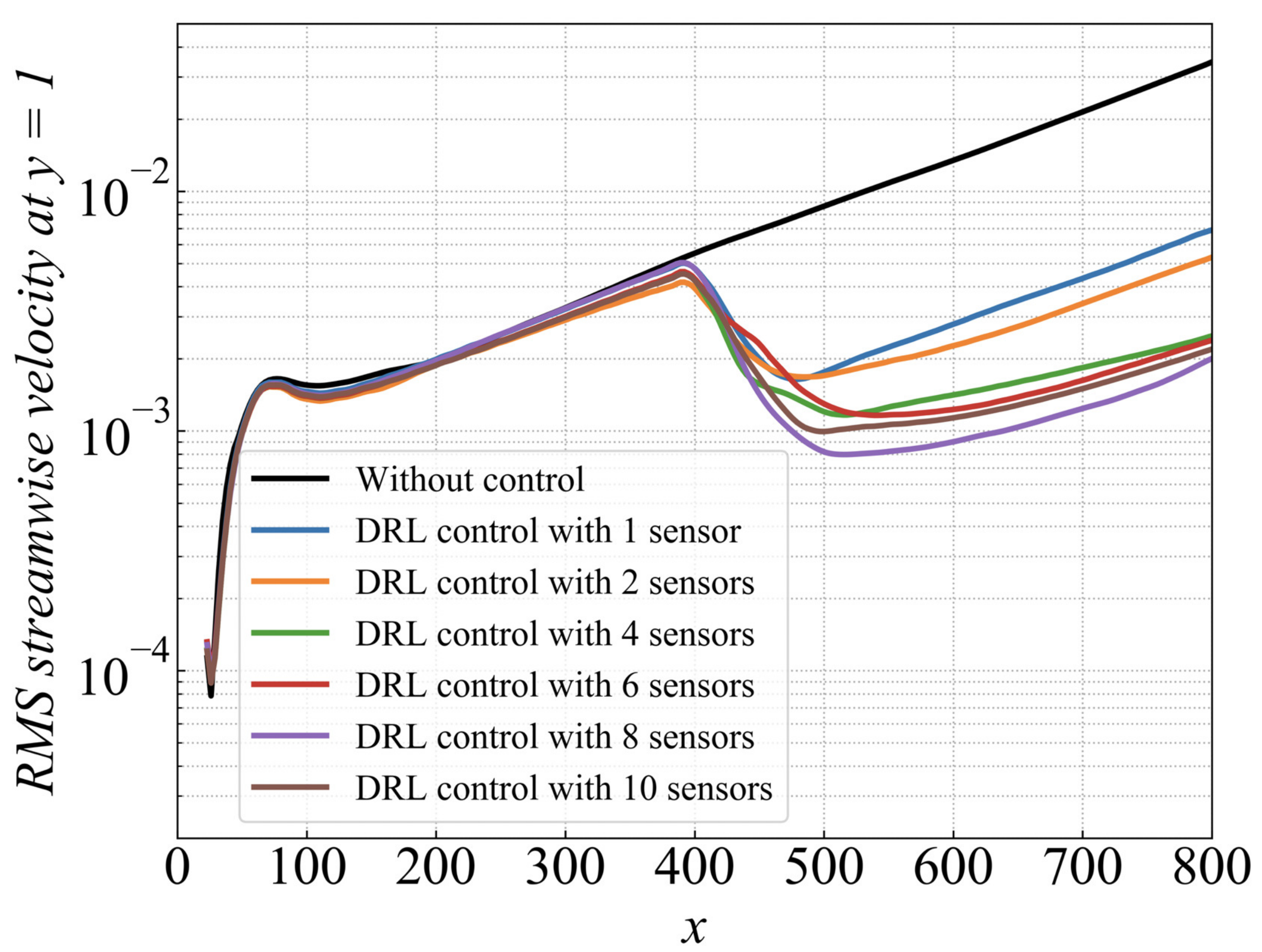}
}
\quad
\subfigure[Effect of the amplitude of disturbance]{
\includegraphics[width=0.465\textwidth]{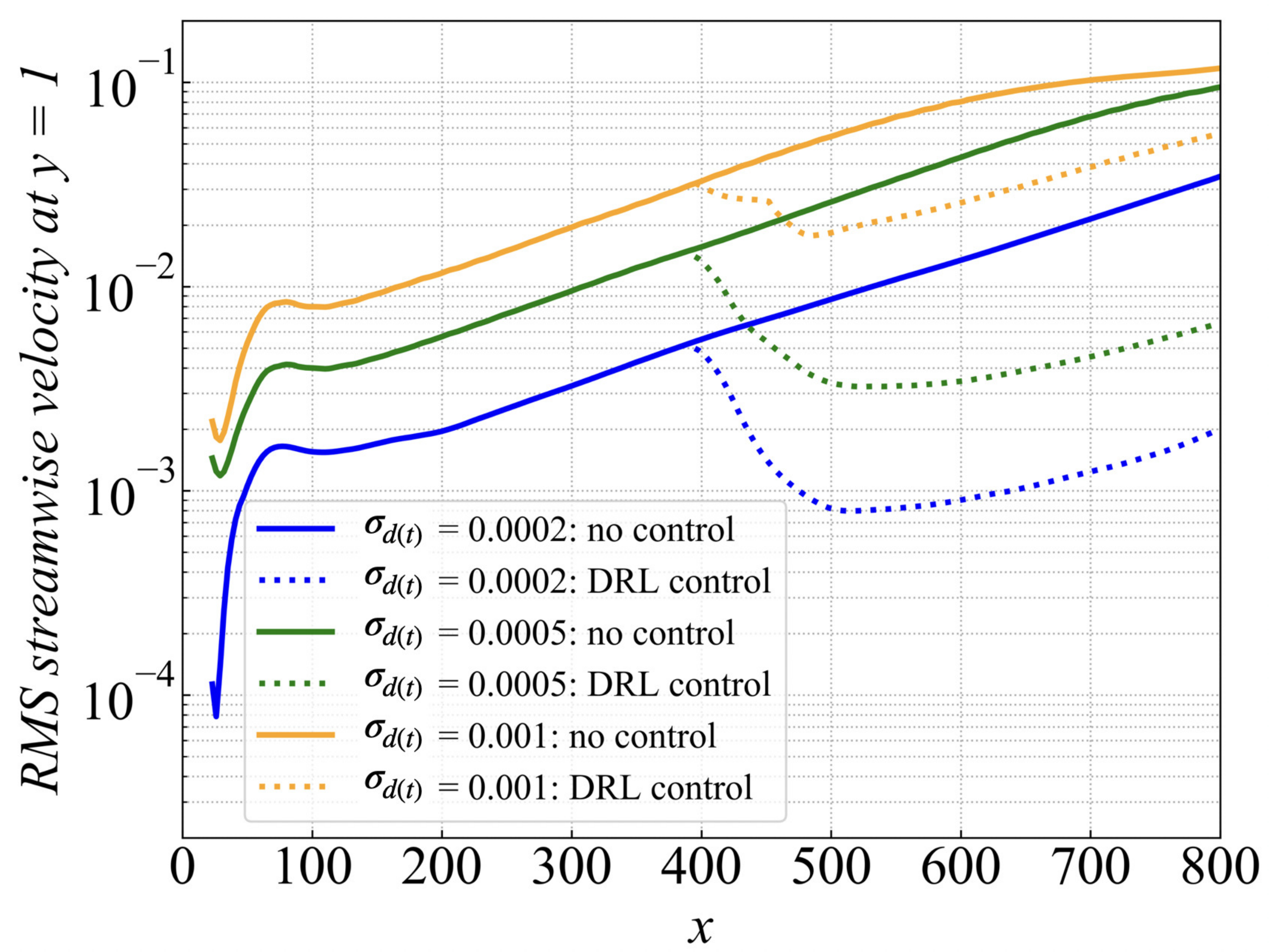}
}
\quad
\caption{DRL-based control for the 2-D Blasius boundary layer flow, subjected to an upstream disturbance input $d(t)$. In panel (a), the average local perturbation energy around point $z$ is recorded during the training process. In panel (b), the global perturbation energy downstream the actuator is recorded during the test process. In panel (c), the RMS value of streamwise velocity along $y=1$ is plotted for DRL-based control with different numbers of sensors. In panel (d), the RMS value of streamwise velocity along $y=1$ is plotted for DRL-based control with disturbance input of different amplitudes $\sigma_{d(t)}$. In addition, curves for cases without control are also shown as a reference.}
\label{fig13}
\end{figure}

Here we adopt the same sensor placements with different numbers of sensors as shown in figure \ref{fig10}(a) in the DRL training for controlling boundary layer flows. A typical training process is shown in figure \ref{fig13}(a), where \xuda{the red dashed line again denotes the time when the experience memory is full and then} the absolute value of the average reward per episode, i.e., the local perturbation energy around point $z$ as defined in Eq. \ref{eq2-15}, generally decreases until convergence after about 50 episodes. Note that the reward is defined as the negative local perturbation energy; nevertheless, the global perturbation level in the whole flow field decreases. As shown in figure \ref{fig13}(b), we test the learnt control policy by applying it to the boundary layer flow for 6000 action steps which are 10 times longer than the training episode, with both the disturbance input $d(t)$ and control input $u(t)$ activated. It is found that the global perturbation energy downstream the actuator is significantly reduced and maintained at a level close to zero throughout the test process. Here the global downstream perturbation energy $e(t)$ is calculated as below
\begin{equation}
e(t)=\sum_{i = 1}^{M}\left[\left(u^{i}(t)-U^{i}\right)^{2}+\left(v^{i}(t)-V^{i}\right)^{2}\right]
\label{eq4-aa}
\end{equation}
where $M$ is the number of uniformly-distributed, customised grid points in the downstream domain of $(410, 800) \times (0, 30)$ and here $M=2052$; $u^i(t)$ and $v^i(t)$ are the instantaneous $x$- and $y$-direction velocity components at point $i$; $U^i$ and $V^i$ are the corresponding base flow velocities.

We then extract the RMS value of the streamwise velocity along $y=1$ and plot it in figure \ref{fig13}(c), where the black curve is for the case without control and other coloured curves represent DRL-based control with different numbers of sensors, respectively. Similar to that in the KS system (cf. figure \ref{fig10}(b)), when no control is implemented, the random noise introduced upstream exhibits an exponential growth while convecting downstream. Such behaviour of linear instability is observed when the external disturbance $d(t)$ is small enough with a standard deviation of $2 \times 10^{-4}$ used here. With the control activated, the perturbation amplitude downstream the actuator is significantly reduced. Besides, the effect of the number of sensors on the control performance is also similar to that in KS system. With the number of sensors increasing from 1 to 8, the resulting control performance is continuously improved, while with a further increase to 10, the control performance slightly degrades, similar to the situation in the KS equation. 

Next, the amplitude of the input disturbance $d(t)$ is increased to investigate the effect of nonlinearity on the DRL-based control performance. \xuda{Each disturbance level represents an agent trained in that specific system, which means the training conditions are the same as the test conditions.} With the standard deviation $\sigma_{d(t)}$ increasing from $2 \times 10^{-4}$ to  $5 \times 10^{-4}$ and $1 \times 10^{-3}$, the nonlinear behaviour gradually starts to play a role in the dynamics, as shown by the solid curves in figure \ref{fig13}(d) where the linear exponential growth is lost in the downstream region. The nonlinearity in turn slightly degrades the DRL-based control performance as shown by the dashed curves in figure \ref{fig13}(d). \xuda{The comparison of flow contours of streamwise perturbation velocity at $y = 1$ with and without control for the three input disturbance levels are shown in figures \ref{figa1}-\ref{figa3}, respectively. In each figure, panel (a) represents the flow field triggered solely by an upstream disturbance input $d(t)$ and panel (b) displays that with both $d(t)$ and DRL-based control $u(t)$ being activated. It is shown that in all the three cases, the downstream perturbation is suppressed to some extent with DRL control but the degree of effectiveness differs. With the increase of disturbance amplitude, the required magnitude of control action $u(t)$ increases (still within the predefined range of $[-0.01, 0.01]$) but the resultant control performance deteriorates by comparing contours in panel (b) among three figures, and also by comparing the recorded output signal $z(t)$, i.e., the localised perturbation energy around (550,1), among figures \ref{figa1}(c), \ref{figa2}(c) and \ref{figa3}(d).

\begin{figure}
  \centerline{\includegraphics[width=1.0\textwidth]{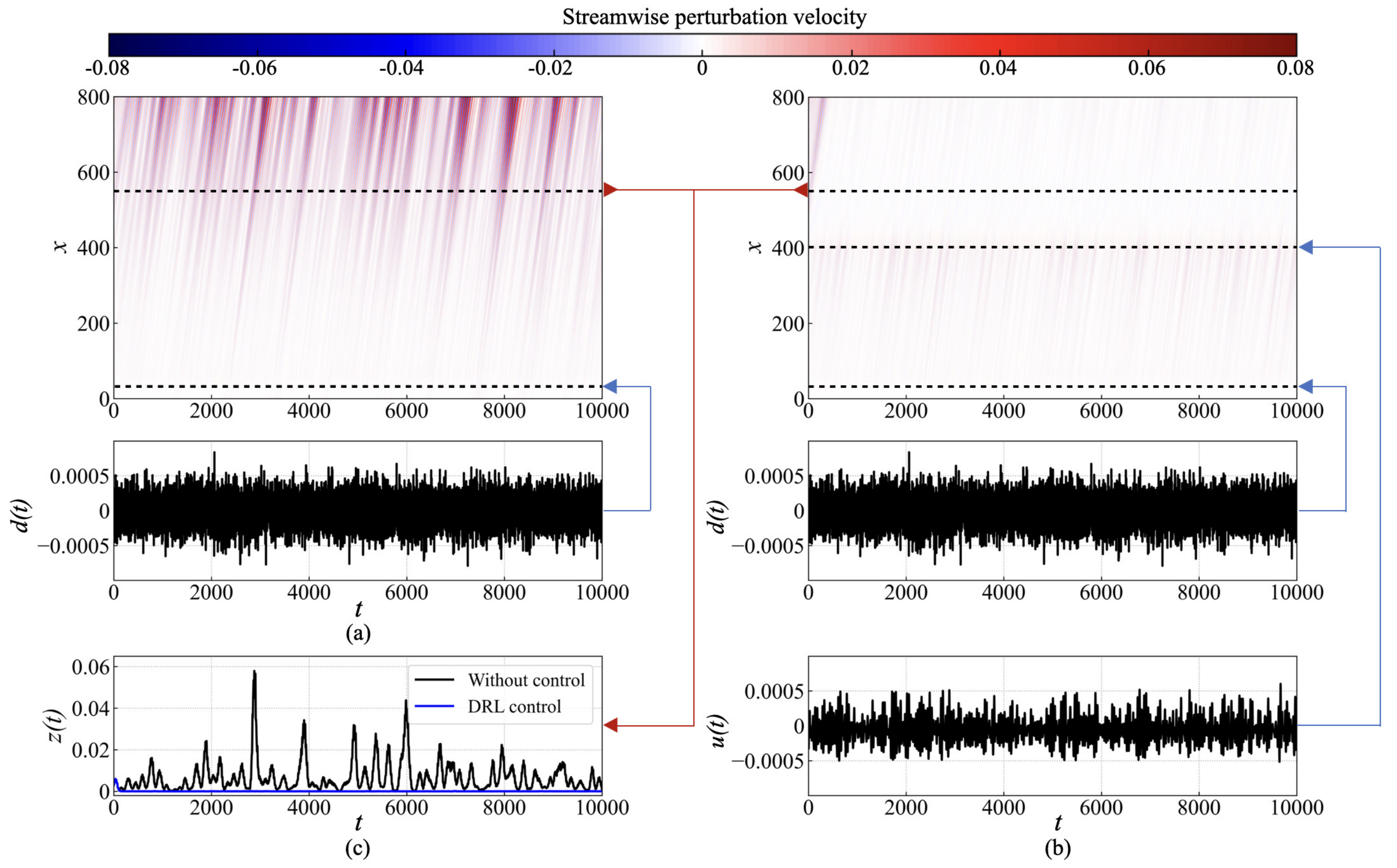}}
  \caption{(a) Contour of streamwise perturbation velocity at $y=1$ induced by an upstream disturbance $d(t)$ with standard deviation $\delta_{d(t)} = 0.0002$; (b) contour of streamwise perturbation velocity with both disturbance input $d(t)$ and control input $u(t)$; (c) output signal $z(t)$ comparison between the uncontrolled and controlled cases, i.e., the localised perturbation energy monitored around (550,1); (d) control performance comparison in terms of the RMS value of streamwise velocity at $y=1$. The blue arrows represent inputs and red arrows output.}
\label{figa1}
\end{figure}

\begin{figure}
  \centerline{\includegraphics[width=1.0\textwidth]{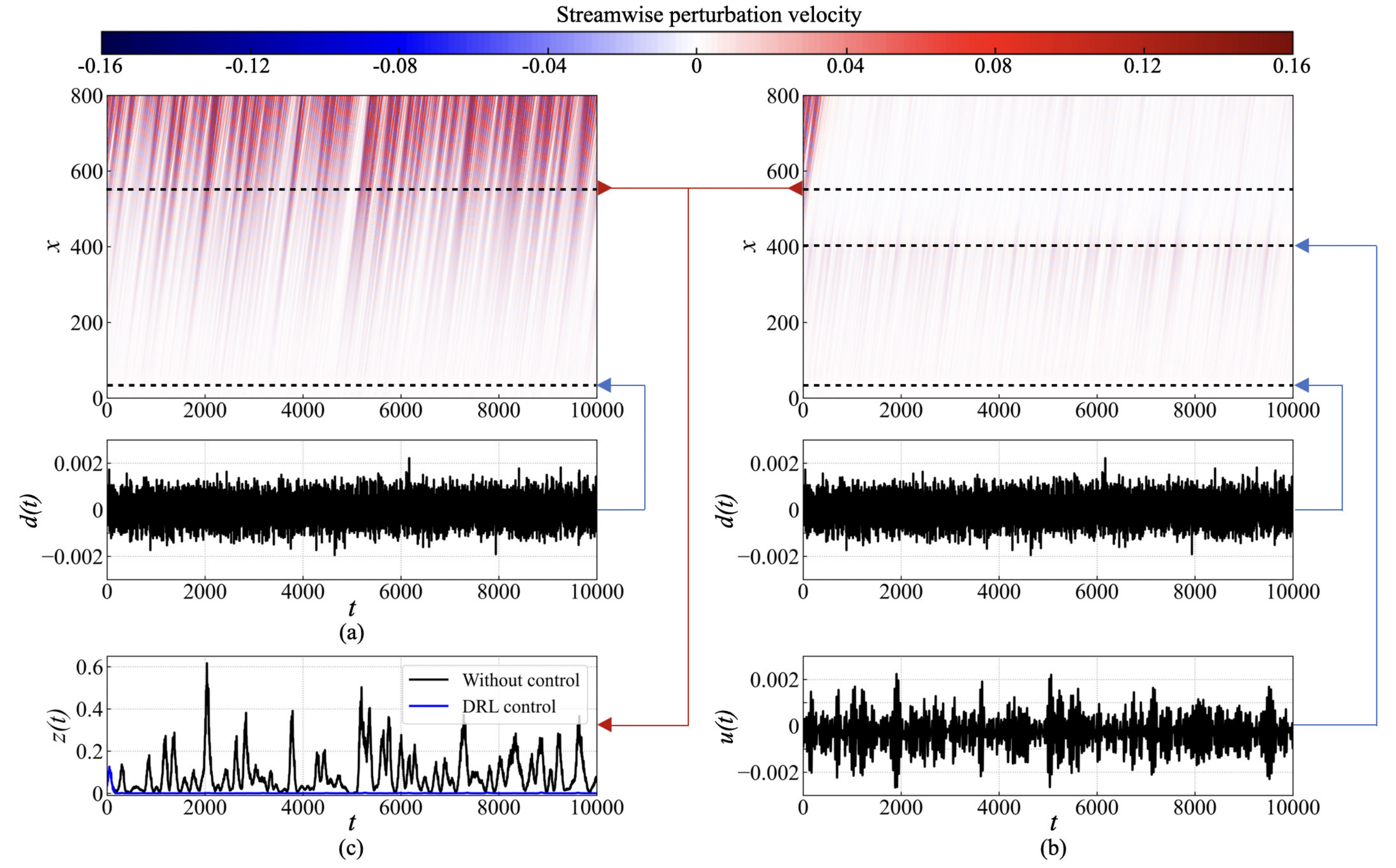}}
  \caption{(a) Contour of streamwise perturbation velocity at $y=1$ induced by an upstream disturbance $d(t)$ with standard deviation $\delta_{d(t)} = 0.0005$; (b) contour of streamwise perturbation velocity with both disturbance input $d(t)$ and control input $u(t)$; (c) output signal $z(t)$ comparison between the uncontrolled and controlled cases, i.e., the localised perturbation energy monitored around (550,1); (d) control performance comparison in terms of the RMS value of streamwise velocity at $y=1$. The blue arrows represent inputs and red arrows output.}
\label{figa2}
\end{figure}

\begin{figure}
  \centerline{\includegraphics[width=1.0\textwidth]{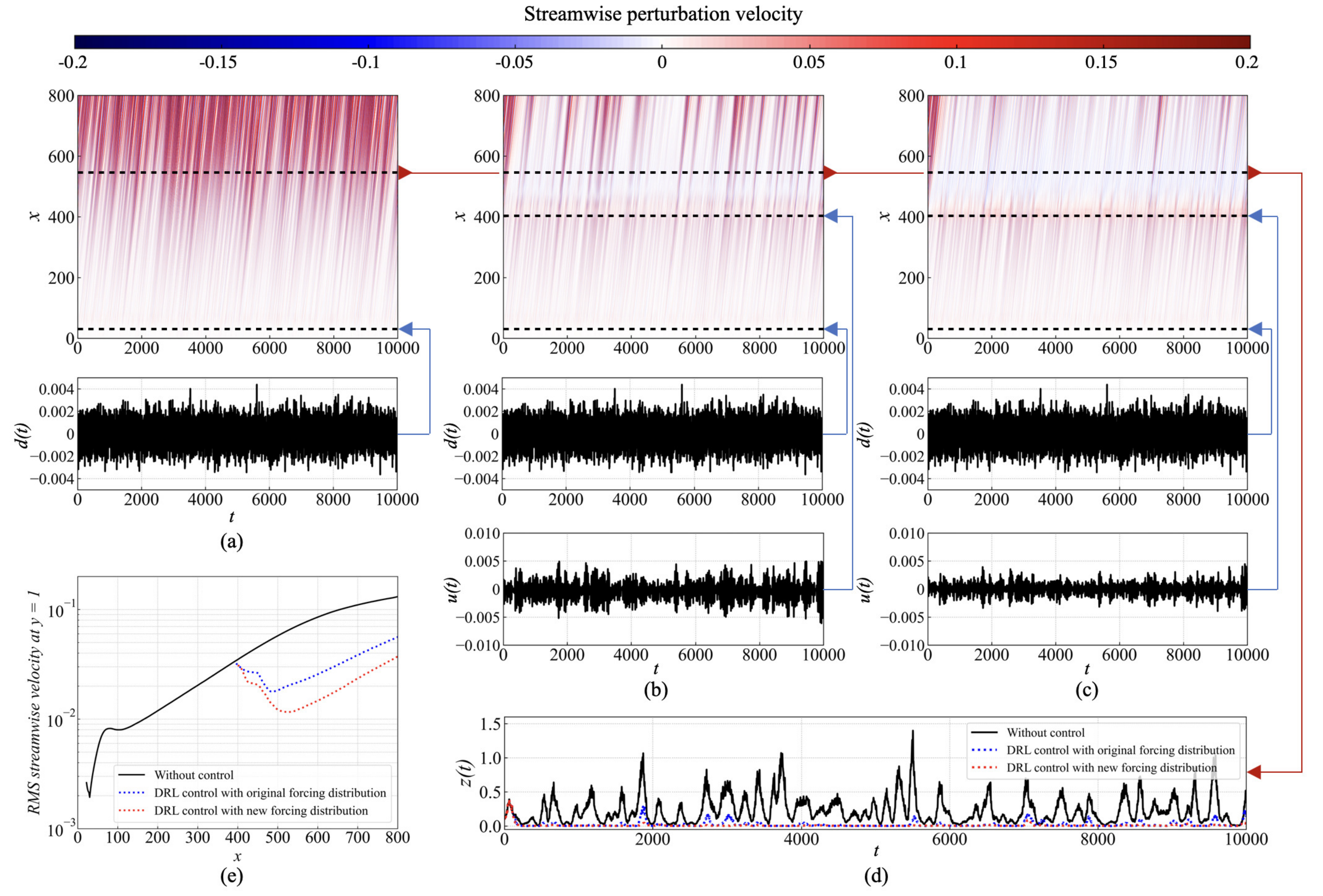}}
  \caption{(a) Contour of streamwise perturbation velocity at $y=1$ induced by an upstream disturbance $d(t)$ with standard deviation $\delta_{d(t)} = 0.001$; (b) contour of streamwise perturbation velocity controlled by $u(t)$ with the original forcing spatial distribution; (c) contour of streamwise perturbation velocity controlled by $u(t)$ with the new forcing spatial distribution; (d) output signal $z(t)$ comparison among the uncontrolled and controlled cases, i.e., the localised perturbation energy monitored around (550,1); (e) control performance comparison in terms of the RMS value of streamwise velocity at $y=1$. The blue arrows represent inputs and red arrows output.}
\label{figa3}
\end{figure}

The performance degradation can be understood as follows. First, the optimised sensor placement applied here is obtained from the linearised KS system, which may be sub-optimal in the nonlinear conditions. Second, with the increase of disturbance level, such a control method via a localised volumetric forcing may not be effective due to its confined region of influence compared to the extended region influenced by the increasing disturbance input. This argument is verified by performing an additional DRL training for a spatially-enlarged localised forcing and then comparing its control performance with the original one. The original control forcing is localised around point (400,1) with 2-D Gaussian supports with $\delta_x = 4$ and $\delta_y = 1/4$ and the new forcing is also localised around (400,1) but with a slightly-enlarged spatial region of influence with $\delta_x = 5$ and $\delta_y = 1/3$. All the other training parameters are the same as the original one and the resultant control performance is shown in figure \ref{figa3}(c) and the red dotted curves in figure \ref{figa3}(d,e). It is demonstrated that the control performance can be improved by slightly enlarging the spatial region of the localised forcing, which can be seen by comparing the downstream contour in figure \ref{figa3}(b,c) where the perturbation is better suppressed with the new forcing distribution, and from the recorded output signal $z(t)$ in figure \ref{figa3}(d), and also from the RMS value of streamwise velocity along $y=1$ plotted in figure \ref{figa3}(e).}

\begin{figure}
\centering  
\subfigure[Disturbance level $\sigma_{d(t)} = 0.0002$]{
\includegraphics[width=1.0\textwidth]{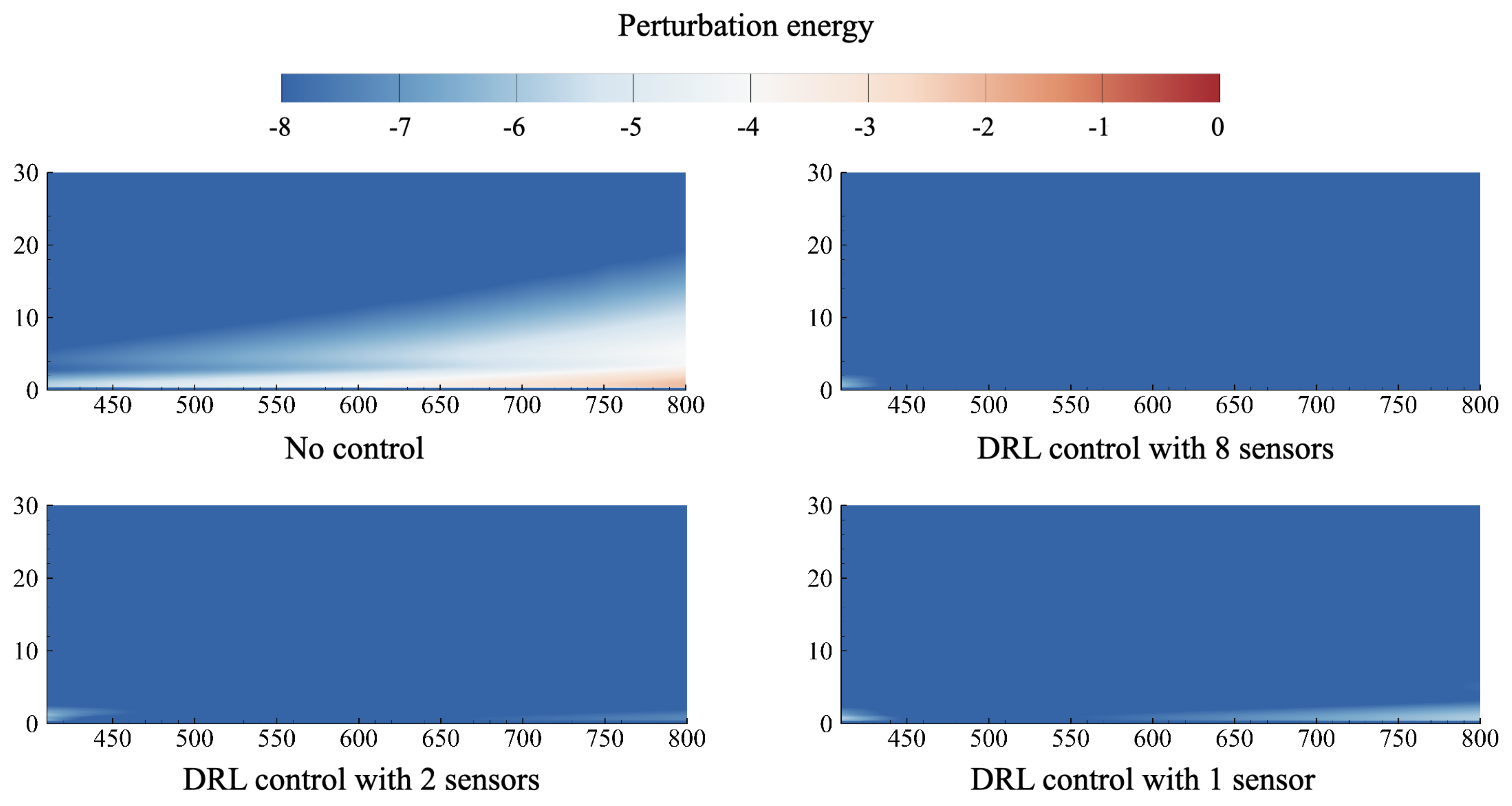}
}
\quad
\subfigure[Disturbance level $\sigma_{d(t)} = 0.0005$]{
\includegraphics[width=1.0\textwidth]{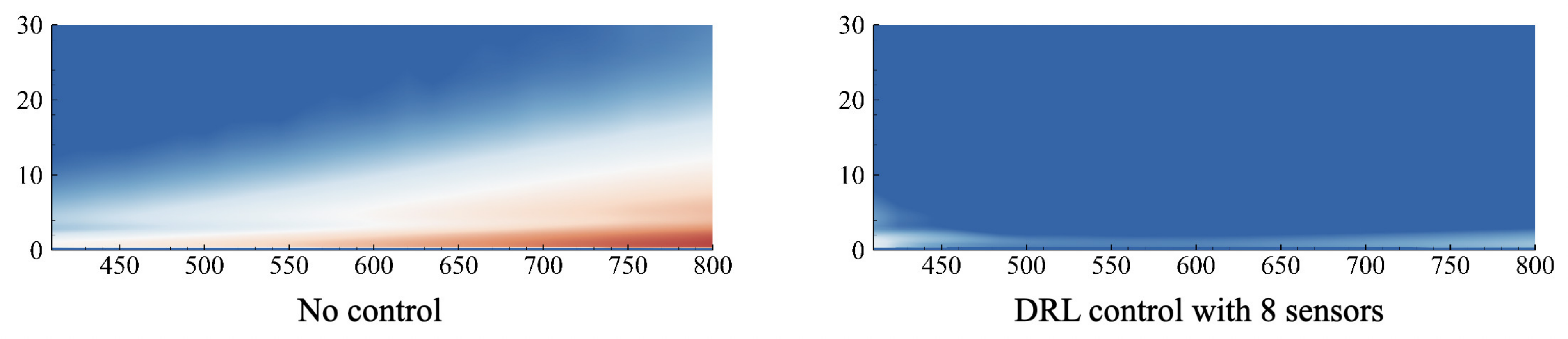}
}
\quad
\subfigure[Disturbance level $\sigma_{d(t)} = 0.001$]{
\includegraphics[width=1.0\textwidth]{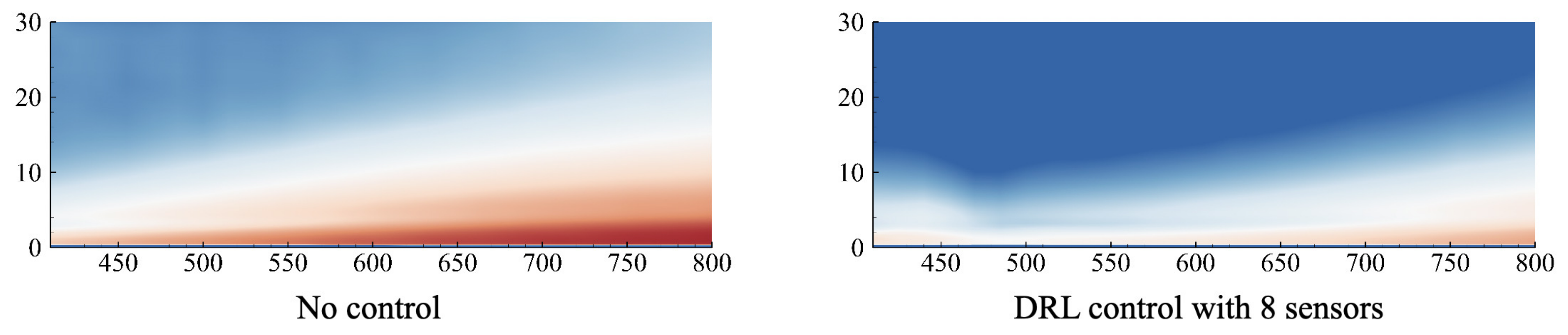}
}
\quad
\caption{Contours of the time-averaged perturbation energy field downstream the actuator with different amplitudes $\sigma_{d(t)}$ of disturbance input. In each panel, the comparison of perturbation energy field with and without control is presented. The colour bar displays the logarithm of the normalised perturbation energy.} 
\label{fig14}
\end{figure}

All the results shown in figures \ref{fig13}-\ref{figa3} are related to the perturbation reduction along a single straight line. Next, we plot the time-averaged perturbation energy field downstream the actuator with different amplitudes of the disturbance input in figure \ref{fig14}(a-c). For the clarity of comparison, all the raw data of perturbation energy field are post-processed by first normalisation using its maximum value and then taking the logarithm. Thus, the right bound of the colour code shown in figure \ref{fig14} is zero, which corresponds to the region having the maximum perturbation energy among the considered cases. As shown in figure \ref{fig14}(a), when no control is implemented, the perturbation velocity keeps increasing as convected downstream and develops from the boundary layer region into the outer region. In contrast, when DRL-based control is activated, the downstream perturbation is significantly suppressed. The effect of the number of sensors is also illustrated here. For DRL control with only one sensor, some small perturbations can still be observed along $y=1$, while for DRL control with two sensors, some small perturbations appear at the outlet. For DRL control with 8 sensors, the downstream perturbations are almost completely suppressed. 

Moreover, the effect of the amplitude of disturbance input is illustrated by the comparison among the first rows of figure \ref{fig14}(a,b,c), using the same eight-sensor placement for consistency. With the increase of disturbance level, the region with high perturbations is obviously enlarged and thus the DRL-based control performance deteriorates as explained before. A quantitative comparison is summarised in Table \ref{table1}, where the time-averaged global perturbation energy $E$, i.e., $E = \overline{e(t)}$ where $e(t)$ is defined in Eq. \ref{eq4-aa}, downstream the actuator is reported and compared. It is found that when the disturbance level is low, DRL-based control is remarkably efficient with an over 96\% reduction of the downstream perturbation energy. With the disturbance level increasing to 0.001, DRL-based control is not that efficient but still effective with a 89.68\% reduction of the downstream perturbation energy.

\begin{table}
  \begin{center}
\def~{\hphantom{0}}
  \begin{tabular}{ccccc}
      $\ Disturbance \ amplitude \ $ & $\ Sensor \ number \ $ & $ \ Perturbation \ energy \ E \ $ & $ \ Relative \ reduction \ $\\[1pt]
             & no control & 0.08886 & -\\
             & 1 & 0.00353 & 96.03\%\\
             & 2 & 0.00232 & 97.39\%\\
     0.0002  & 4 & 0.00078 & 99.12\%\\
             & 6 & 0.00066 & 99.26\%\\
             & 8 & 0.00045 & 99.49\%\\
             & 10 & 0.00056 & 99.37\%\\
       \hline
             & no control & 0.65188 & -\\
     0.0005  & 8 & 0.00452 & 99.31\%\\
       \hline
             & no control & 1.79398 & -\\
     0.001   & 8 & 0.18512 & 89.68\%\\
  \end{tabular}
  \caption{DRL-based control performance quantified by the reduction of perturbation energy.}
  \label{table1}
  \end{center}
\end{table}

\subsection{Interpretation of the learnt control policy}\label{Physics}
\xuda{In this section, we make an attempt to understand the DRL-based flow control from a physical point of view. We stress that the convectively-unstable flow system under control is a selective frequency amplifier, subject to upstream random noise. To make this point clear, we perform the stability analysis of the 2-D Blasius flow and discuss its relevance to the control.

We numerically solve the Orr-Sommerfeld equation with the solution form $\psi(x, y, t)=\varphi(y) \mathrm{e}^{\mathrm{i}(\alpha x-\omega t)}$, where $\varphi(y)$ is the complex amplitude; $\alpha$ is the wavenumber and $\omega = \omega_r + \mathrm{i}\omega_i$ with $\omega_r$ being the angular frequency and $\omega_i$ the exponential growth rate; $c = \omega/\alpha = c_r + \mathrm{i}c_i$ where $c_r$ is the phase velocity of the wave and the sign of $c_i$ determines whether the wave is stable or unstable. The limiting case $c_i = 0$ yields the neutral stability curves as shown in figure \ref{figcc}(a) for $Re-\omega_r$, non-dimensionalised by the free-stream velocity $U_\infty$ and the local displacement thickness $\delta^*$ of the boundary layer. It is shown that the critical $Re$ is 519.4 and for $Re > 519.4$, a narrow range of frequencies are amplified. For our control case whose inlet $Re = 1000$, the local displacement thickness of the boundary layer at the location of control $x = 400$ is about $\delta^* = 1.474$ and thus the local $Re_c = 1474$. Moreover, sensors used to collect state are also positioned around this location. Thus, we can extract the corresponding stability curve at this section, denoted by the black dashed line in figure \ref{figcc}(a), and plot it in a $\omega_r-c_i$ plane in figure \ref{figcc}(b). It can be seen from figure \ref{figcc}(b) that the unstable frequency range at this section is about (0.027, 0.079), denoted by two blue points. To make the control effective, this frequency range should be well resolved by state observations in DRL. In the current case, we adopt a time step of 0.05 unit times for temporal integration in 2-D simulations. The action is constant for 30 time steps and there is only one observation collected by 8 sensors at the end of each 30 time steps which is returned as state. Thus, the control frequency is the same as the state sampling frequency, namely, $2\pi/(30\times 0.05) = 4.189$, which is much larger than the aforementioned amplified frequency, thus satisfying Nyquist criteria and avoiding the aliasing effect.

\begin{figure}
\centering  
\subfigure[Neutral curve: $Re-\omega_r$]{
\includegraphics[width=0.473\textwidth]{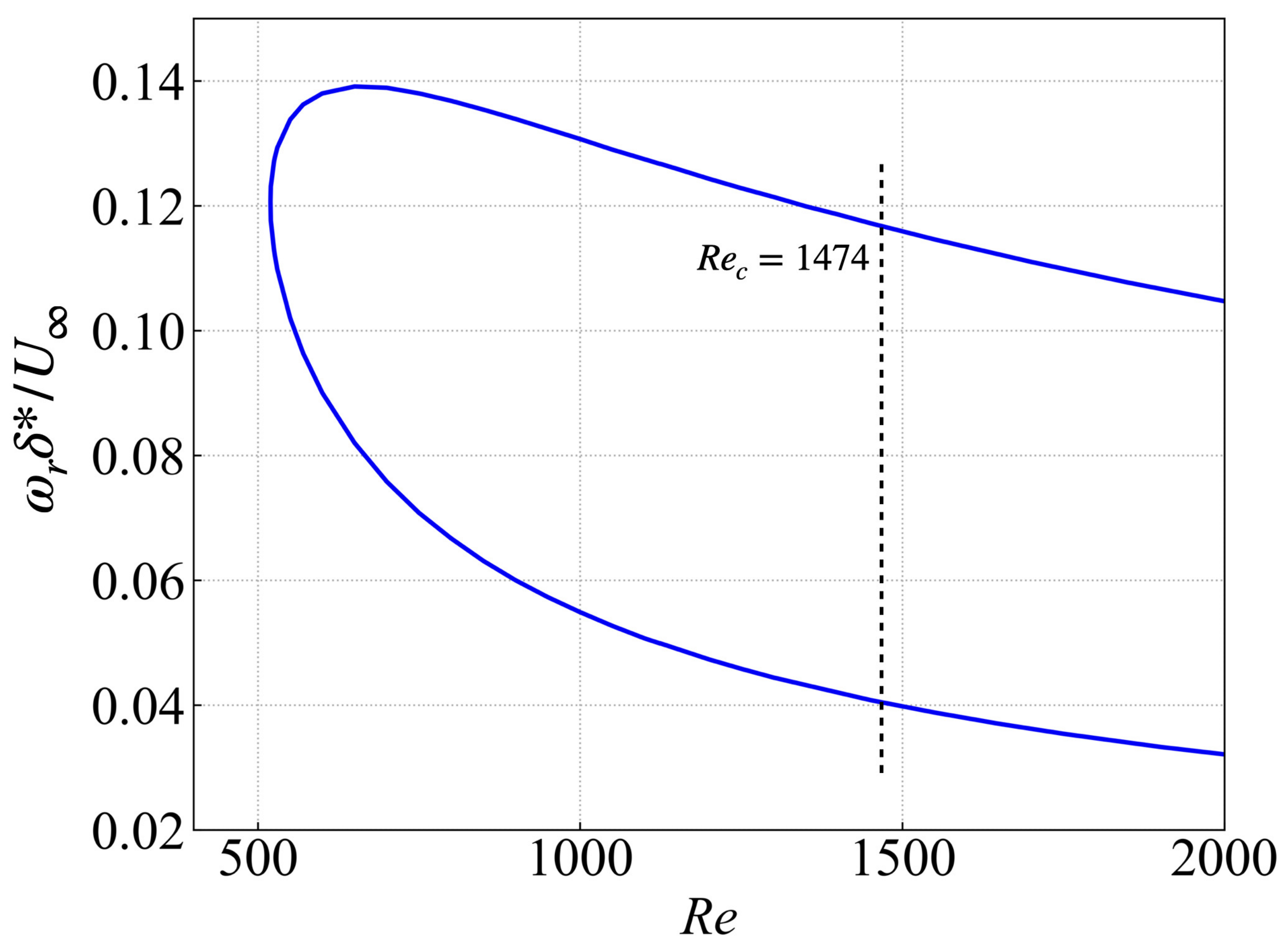}
}
\quad
\subfigure[Curve for $Re_c = 1474$: $\omega_r-c_i$]{
\includegraphics[width=0.463\textwidth]{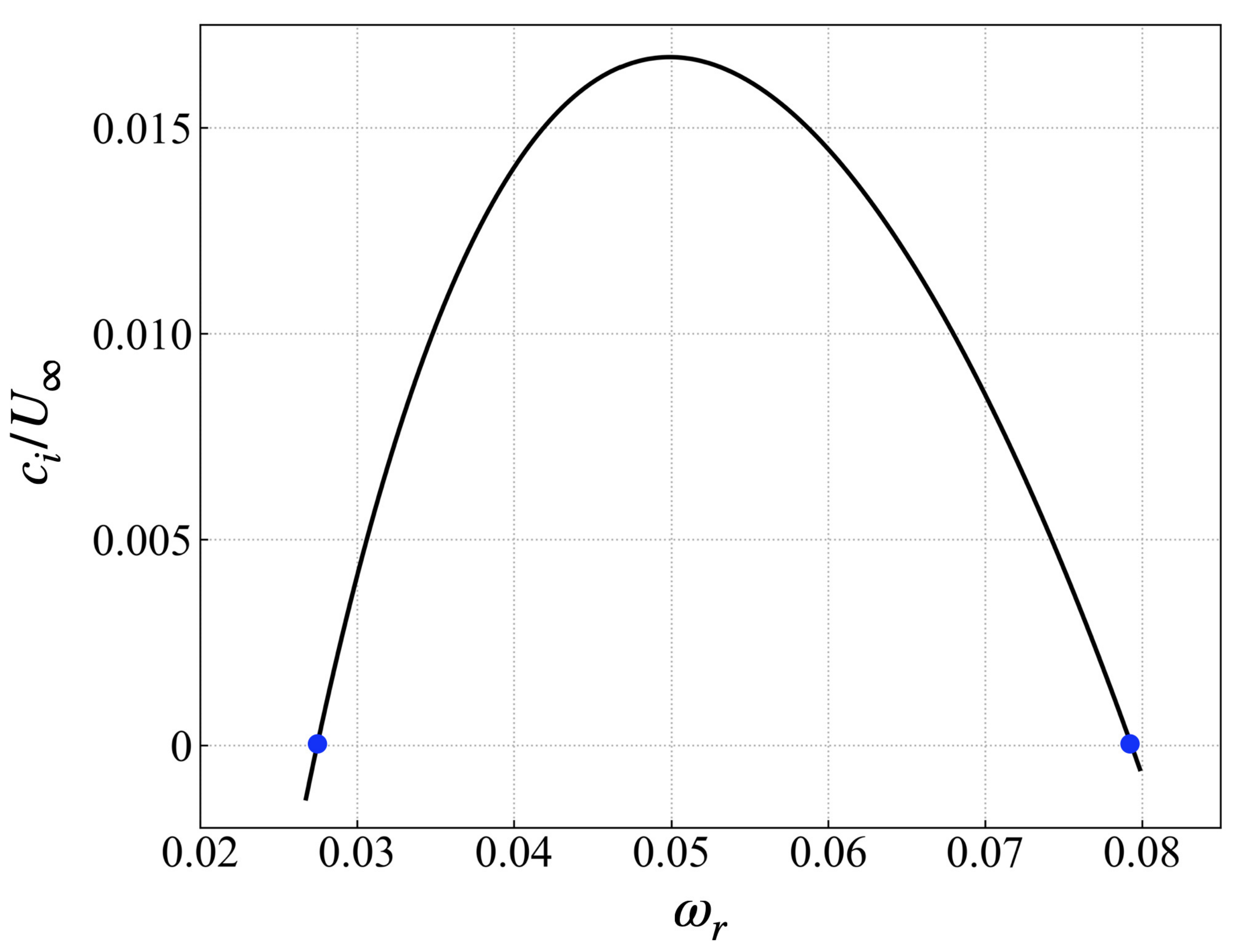}
}
\quad
\caption{Stability curves for Blasius flow by solving the Orr-Sommerfeld equation.}
\label{figcc}
\end{figure}


\begin{figure}
\centering  
\subfigure[$t_1 = 1000$]{
\includegraphics[width=0.9\textwidth]{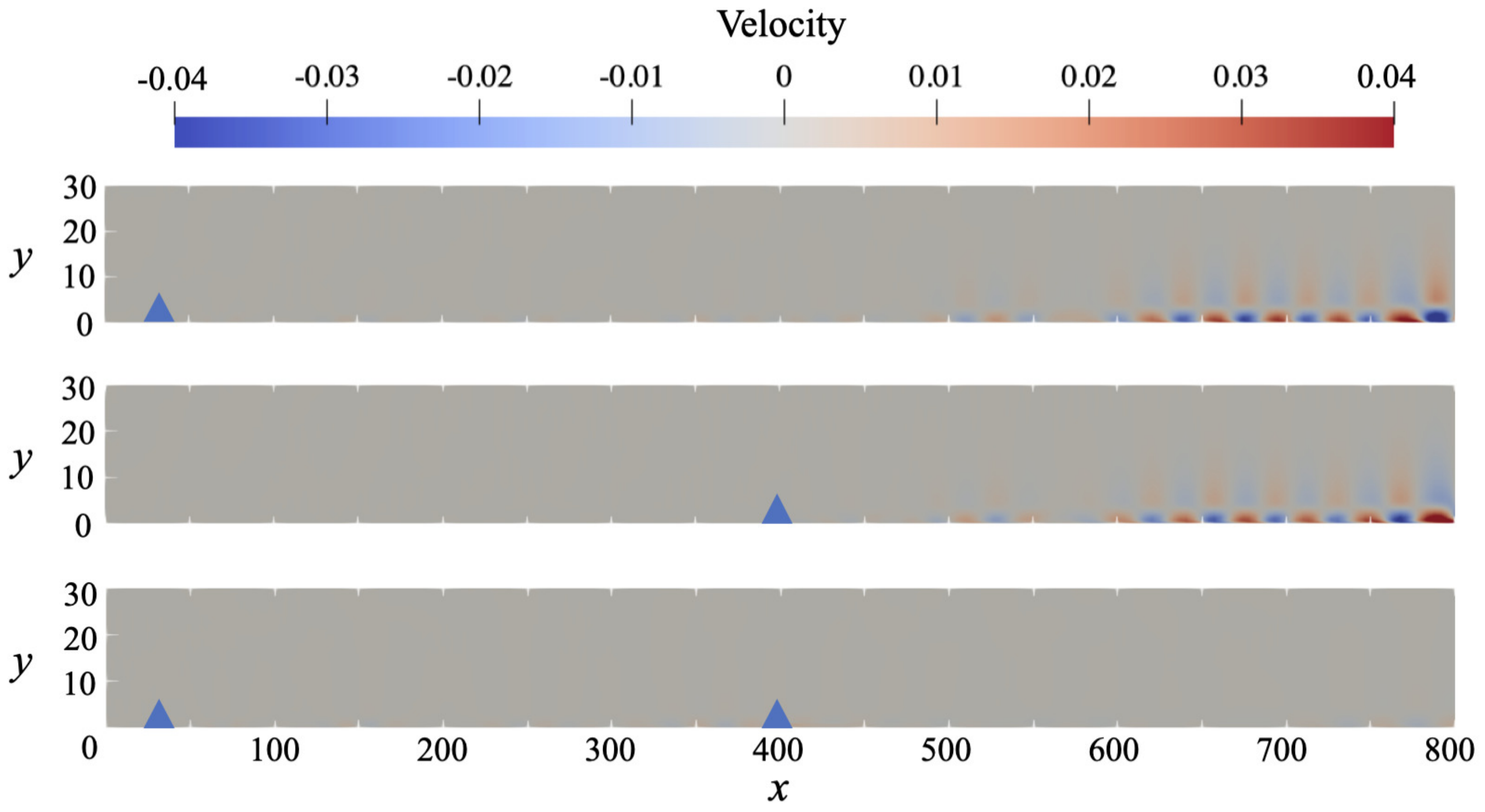}
}
\quad
\subfigure[$t_2 = 3100$]{
\includegraphics[width=0.9\textwidth]{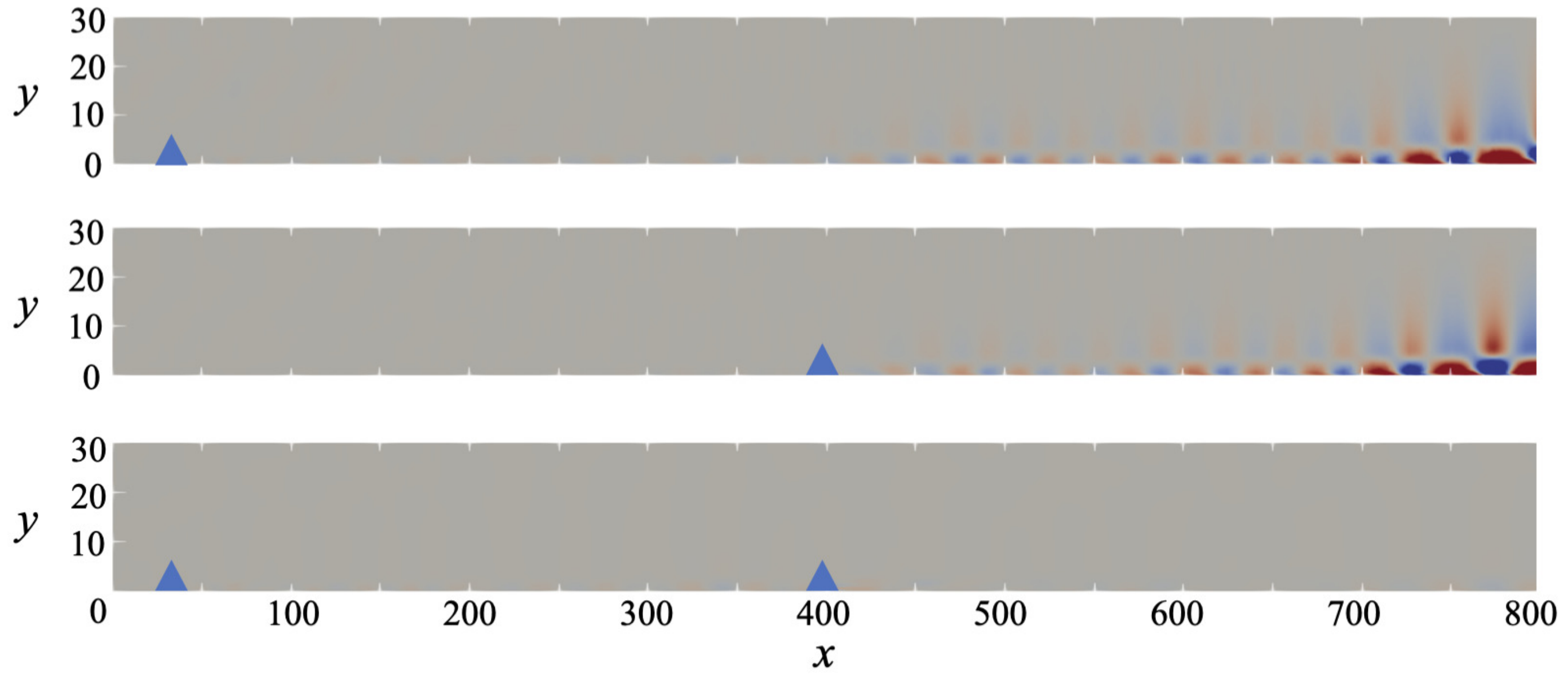}
}
\quad
\subfigure[$t_3 = 5600$]{
\includegraphics[width=0.9\textwidth]{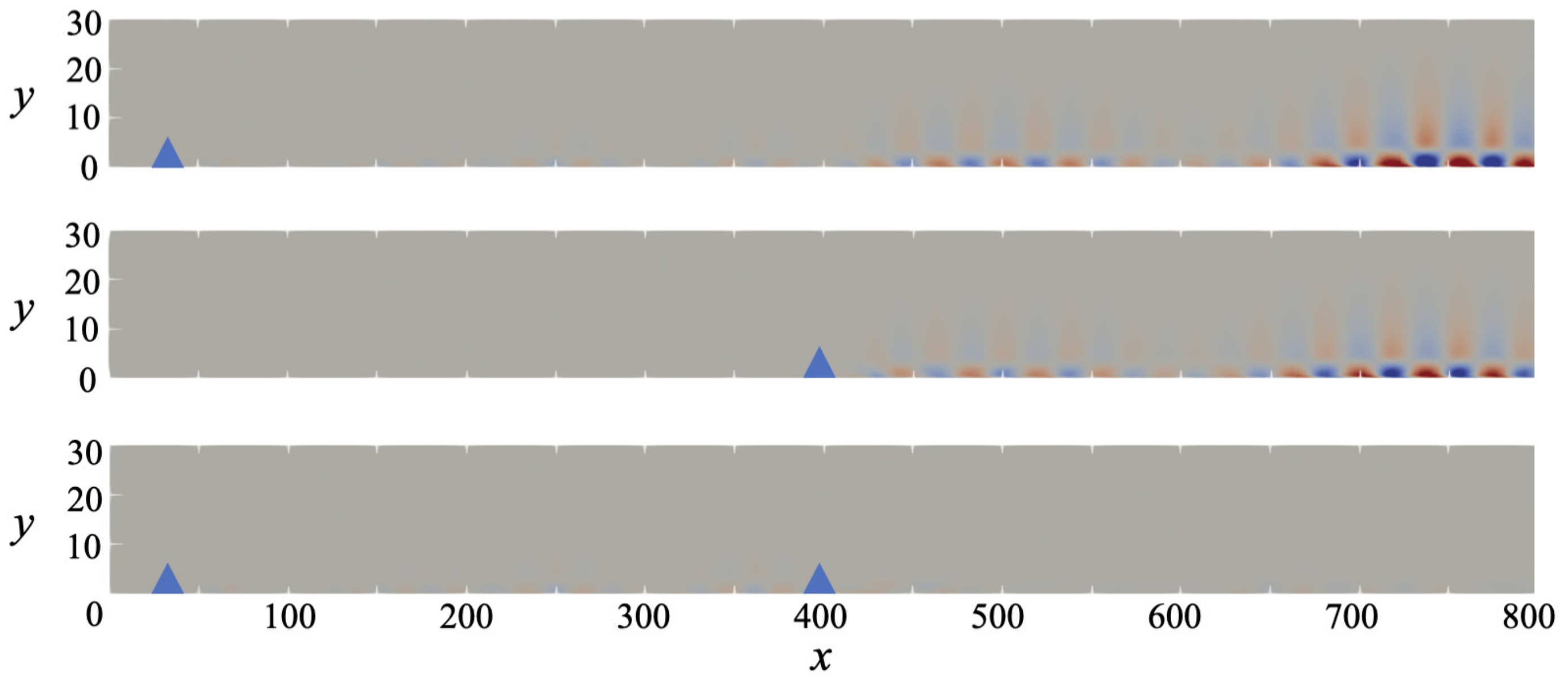}
}
\quad
\caption{Instantaneous snapshots of streamwise perturbation velocity at three different time instants $t_1$, $t_2$ and $t_3$. In each panel, the top row represents the velocity field with only the upstream disturbance input; the middle one represents the velocity field with only the corresponding control input; the bottom one represents the velocity field with both the disturbance input and control input turning on. The position of disturbance input and control input is denoted by the blue triangle.}
\label{fig15}
\end{figure}

}
Moreover, we make an attempt to understand the learnt control policy through an analysis of the macroscopic flow features induced by the control action, since it is difficult to interpret the policy from the weights of neural networks due to the black box feature of ML algorithms \citep{schmidhuber2015deep,rabault2019artificial}. We plot the instantaneous snapshots of streamwise perturbation velocity at three different time instants $t_1$, $t_2$ and $t_3$ in figure \ref{fig15}(a-c), respectively, corresponding to the three spikes as denoted in figure \ref{fig13}(b). In each panel, the top row corresponds to the black curve in figure \ref{fig13}(b), which represents the situation with only the upstream noise input at $x_d = 35$, while the bottom row corresponds to the blue curve in figure \ref{fig13}(b), which represents the situation with both upstream noise input at $x_d = 35$ and DRL-based control implemented at $x_u = 400$. During the control process, we record the action sequence produced by the agent and then perform a new simulation with only the recorded control actions input and obtain the resultant field as shown by the middle row. It is observed in all the three panels that the specific wave induced by the control action (middle row) is almost of the same magnitude but the opposite phase with that generated by the disturbance input (top row), which imposes a destructive inference to the incoming TS wave and thus the downstream perturbations are well suppressed, as shown in the bottom row of each panel. This is exactly the opposition control mechanism as has been discussed in e.g. \cite{choi1994active,grundmann2008active,brennan2021tollmien,sonoda2022reinforcement,Alejandro2022DRL}. To summarise, DRL control is effective in controlling the convective instability evolving in boundary layer flows, using the optimised sensor placement obtained from the KS system. The learnt control policy based on the `sticky action choice' acts as opposition control to cancel the incoming perturbation TS wave.


\section{Conclusions} \label{Conclusions}


In this work, we have studied and evaluated the performance of DRL in controlling the perturbative dynamics in both the 1-D linearised KS system and the more realistic 2-D flat-plate boundary layer flows. The former is  a particular variant of the original KS equation and commonly used to model the perturbation in flat-plate boundary layer flows. \xuda{It is known that traditional model-based controllers usually struggle with convectively-unstable flows subjected to unknown external disturbance, since it is difficult to assume an accurate flow model in a real noise environment \citep{dergham2011accurate,herve2012physics,belson2013feedback}.} Thus, the main objective of this work is to test if the model-free DRL-based flow control strategy can suppress the randomly-perturbed convective instability in these flows and investigate the optimal sensor placement issue in the context of DRL flow control.

Due to the convective instability of boundary layer flows, without control, the amplitude of perturbation grows exponentially along the streamwise direction. After the DRL control is activated, the perturbation downstream the actuator can be significantly suppressed as the perturbation amplitude at the monitoring point reduces significantly. We have also demonstrated the robustness of the learnt control policy to two types of noises, i.e., measurement noise and external noise. The former is used to mimic the realistic control scenarios as the sensors are usually subjected to noise, while the latter is used to test whether the policy learnt from a certain environment can be generalised to other unseen noise conditions. The robustness property of DRL results from the state normalisation operation in the training process and also \xuda{the closed-loop nature of the control system.} 

We have also investigated the optimised placement of different numbers of sensors in the DRL-based flow control using the gradient-free PSO algorithm. We find that one sensor placed upstream the actuator is not enough to generate a good control performance because, this way, the sensor cannot perceive the consequence of the action after it is taken due to the convective nature of the flow. \xuda{In addition, due to the implementation of sticky actions, a single upstream sensor can only inform the agent of a single upcoming parallel slash in a packet of slashes that crosses the actuator over the duration that the action is ``stuck''. As the slashes are generated by random noise, it is not possible for the agent to choose an action that accounts for the packet of slashes based on a single measurement. Thus, a proper number of sensors} placed both upstream and downstream the actuator can provide a more complete measurement of the current flow environment in this model-free method. But more sensors do not necessarily lead to a better performance since the information provided by the additional sensors may be redundant. In our study, a specific eight-sensor placement has been found to be the optimal with the best control of the KS equation. We also investigate the dynamics of the nonlinear KS system and find that, due to the nonlinear effect, the perturbation may no longer grow exponentially along the streamwise direction but saturate. The control policy learnt from the linear equation can be applied directly to control the weakly nonlinear case with an effective performance, although the controlled result is not as good as the policy trained in the real nonlinear condition. Moreover, we have also embedded the information on flow stability, i.e., the leading growth rate evaluated by DMD, into the reward function to penalise the instability, and found that the control performance can be further improved. 

All the above results pertain to the DRL-control of the 1-D KS system. As a further demonstration, we apply the optimised sensor placement from the 1-D KS equation to the control of 2-D Blasius boundary layer flows of $Re=1000$ subjected to a random upstream disturbance input. When the disturbance level is relatively low, DRL-based control is remarkably efficient to reduce the downstream perturbation energy and the effect of the number of sensors on the control performance is very similar to that in 1-D KS system. With the disturbance level increasing to 0.001, DRL-based control is less efficient. This performance degradation may be related to the fact that the \xuda{adopted sensor placement} is obtained from the linear KS system and \xuda{also the confined region of influence of the localised control forcing}. In addition, we can explain the learnt DRL policy by post-processing the flow data and find that the DRL-based control can be interpreted as opposition control, an approach that has been studied in boundary layer flow controls. 

In the end, we would like to discuss the limitations of this work and future directions that can be followed to improve the model-free DRL-based flow control. A more consistent study to determine the optimal sensor placement in the 2-D boundary layer flows would couple sensor optimisation with DRL training directly in the 2-D flows. This was our initial attempt; however, we quickly realised that the computational cost, i.e., PSO executed in the 2-D NS equations with a reasonable resolution, is impractically high and thus resorted to the 1-D KS model of the perturbed boundary layer flows. With more computational resources, future work can consider solving for the optimal sensor placement in a more consistent manner. \xuda{It should be noted that some researchers have begun to use a data-driven reduced-order model in place of the real environment in DRL control to mitigate the problem of high computational cost \citep{ha2018recurrent,zeng2022data} and this may also help circumvent the computational problem that we are facing here.}  Another issue lies in the time delay between the control starting time and the time when its impact on the downstream flow field can be perceived. This delay is due to the convective nature of the studied flow. Our experience is that it is important to remove such delay and match the effect of the action and the corresponding flow state/reward in the DRL framework. In academic flows, the control delay time may be obtained based on our knowledge of the flow system, as described in Appendix \ref{A}. This may, however, be unobtainable in real-world applications. Thus, coming up with a systematic way to eliminate the delay is important. The final issue is to embed flow physics/symmetry properties of the dynamical system in the DRL-based control. This will help to steer the black-box exploration of DRL in a physically correct direction and improve the sample efficiency. Our attempt only deals with it in a specific case. More systematic studies along this direction will be necessary and important in furthering our understanding of the DRL control. 
\\\\
\noindent \textbf{Acknowledgements.} We acknowledge the computational resources provided by the National Supercomputing Centre of Singapore.

\noindent \textbf{Funding.} D.X. is supported by a PhD scholarship from the Ministry of Education, Singapore.

\noindent \textbf{Declaration of Interest.} The authors report no conflict of interest.

\begin{appendix}
\section{\xuda{Time delay in DRL-based flow control}}\label{A}
\xuda{In this section, we explain the time delay issue as mentioned in Sec. \ref{RL} and how we circumvent this issue by correlating the reward with the right state-action tuple.

During DDPG training process, transition data $(s,a,r,s')$ at each step are first stored into the experience memory, where $s,a,r$ and $s'$ are state, action, reward and next state, respectively, and then mini-batch samples are taken from the memory to update the parameters of neural network. In the current DRL framework, reward $r$ is related to the perturbation amplitude/energy monitored at a downstream location, so its instantaneous value is not a corresponding response to the current action at an upstream position, until the impact of action has been convected to the downstream location. To eliminate this mismatch, the key is to identify the corresponding time delay $t_D$. Given a constant $t_D$, we first store $s, a, s'$ in three independent buffers and after $t>t_D$, we start to extract $(s, a, s')$ at $t-t_D$ from their respective buffers which correspond to the data $t_D$ time ago. That is, $(s, a, s')$ at $t-t_D$ is combined with reward $r$ at $t$ and then stored into the experience memory together. In this way, we remove the time delay and help the DRL agent perceive the dynamics of environment in a time-matched way.

Next, we explain how we choose the time delay $t_D$ which is defined as the time difference between the control starting time and the time when downstream monitoring point responds. Before training, we first run simulations using the current control framework with both random noise input and random control input from a zero initial condition in the KS equation and from laminar base flow in the boundary layer flow. As shown in figure \ref{figA1}(a) and figure \ref{figA2}(a), reward signal remains zero before the first control action reaches the downstream location and then turns negative (since the reward is defined as the negative value of downstream perturbation amplitude for KS system and the negative value of downstream perturbation energy for boundary layer case). Thus, the time at this turning point is chosen as the time delay $t_D$, i.e., $t_D=25$ for KS system and $t_D=120$ for boundary layer case. 

\begin{figure}
  \centering  
  \subfigure[\ Reward signal]{
  \includegraphics[width=0.462\textwidth]{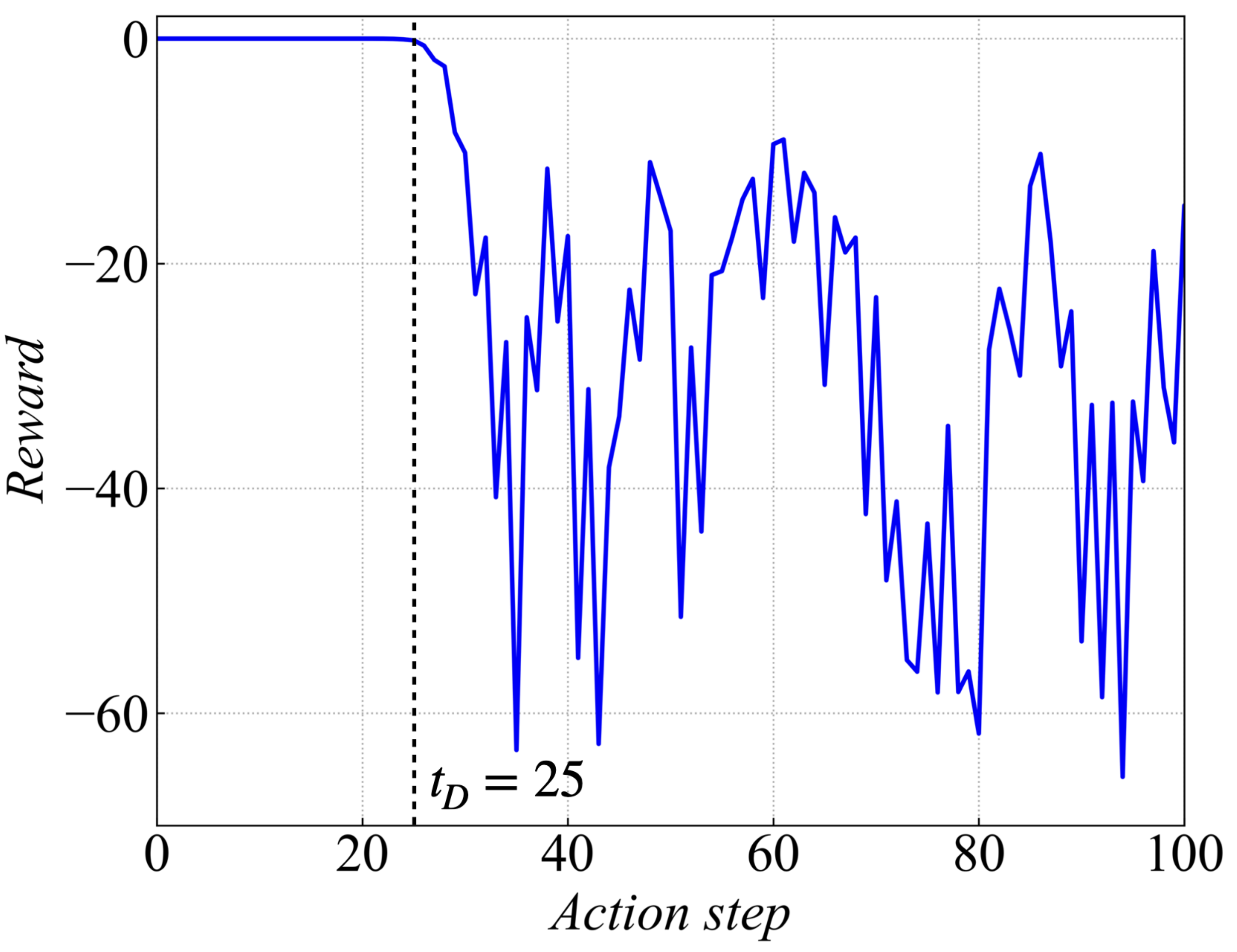}
  }
  \quad
  \subfigure[\ DRL-based control performance]{
  \includegraphics[width=0.472\textwidth]{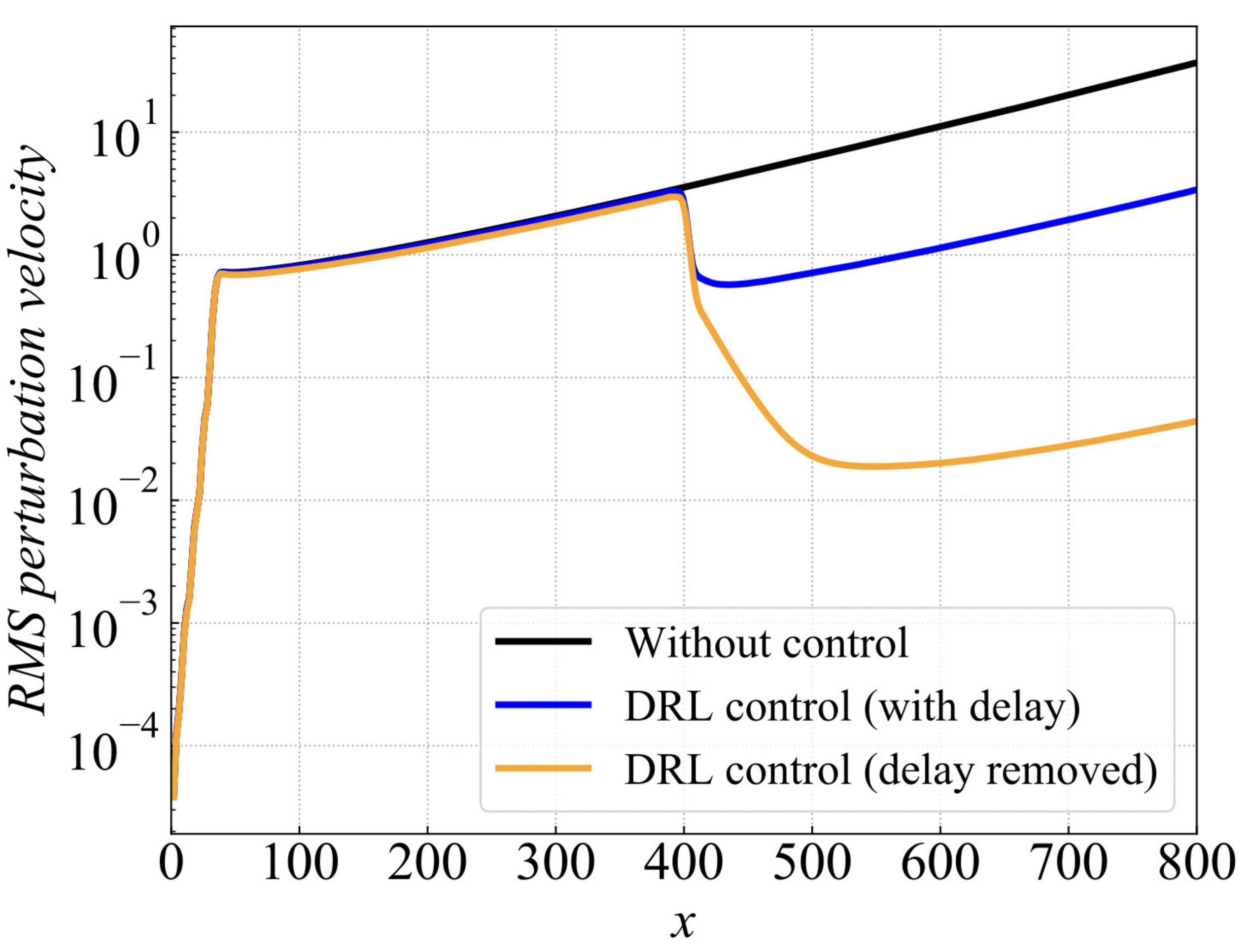}
  }
  \quad
  \caption{Time delay removal for DRL-based control of 1-D KS system. In panel (a), time delay $t_D$ is identified at the turning point when reward turns negative. In panel (b), the RMS value of perturbation along the 1-D domain is plotted for control cases with and without time delay.}
  \label{figA1}
\end{figure}

\begin{figure}
  \centering  
  \subfigure[\ Reward signal]{
  \includegraphics[width=0.467\textwidth]{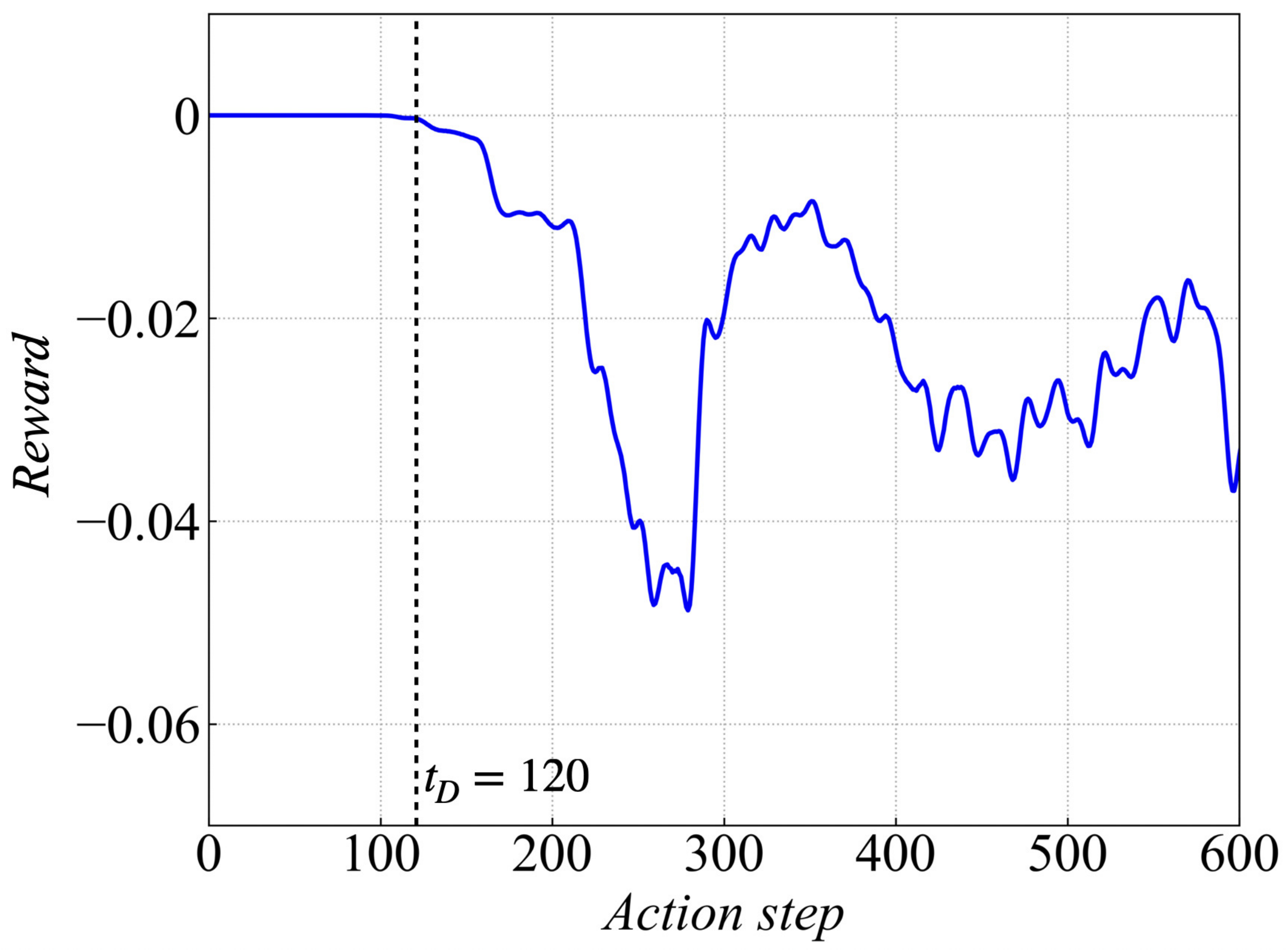}
  }
  \quad
  \subfigure[\ DRL-based control performance]{
  \includegraphics[width=0.465\textwidth]{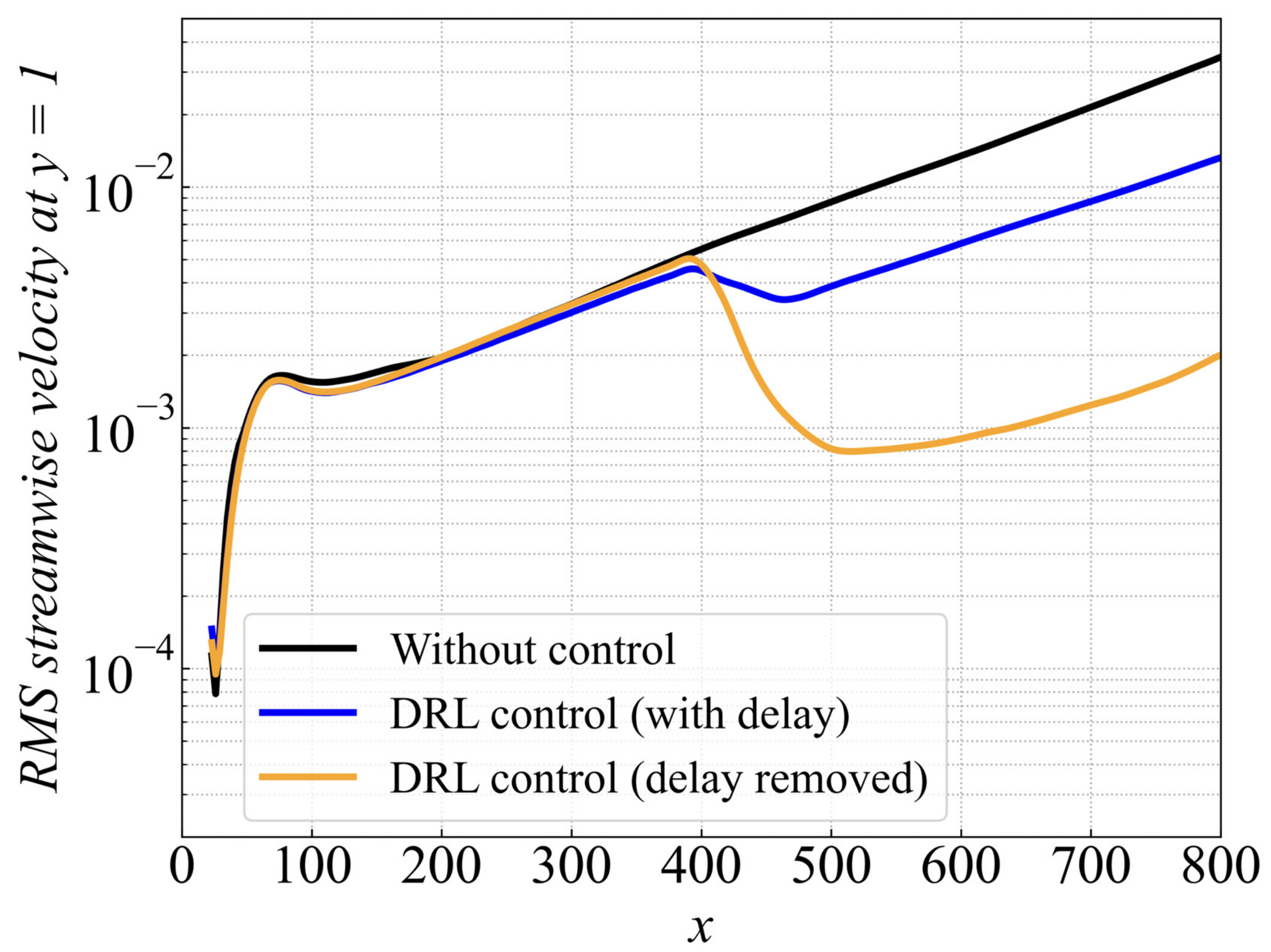}
  }
  \quad
  \caption{Time delay removal for DRL-based control of 2-D boundary layer flow. In panel (a), time delay $t_D$ is identified at the turning point when reward turns negative. In panel (b), the RMS value of streamwise velocity along $y=1$ is plotted for control cases with and without time delay.}
  \label{figA2}
\end{figure}

We also present the comparison of DRL control performance with and without this modification, as shown in figure \ref{figA1}(b) for KS system and figure \ref{figA2}(b) for boundary layer case. For consistency, all the cases adopt the same optimised eight-sensor placement. It is shown that with time delay removed, DRL control performance is significantly improved. Especially for boundary layer flow control, with delay removed, the relative reduction of downstream perturbation energy is 99.49\% (from 0.08886 to 0.00045) while if the time delay is not treated, the perturbation energy reduction is only 87.74\% (from 0.08886 to 0.01089). The control results are visualised in figure \ref{figA3}.

\begin{figure}
  \centerline{\includegraphics[scale=0.36]{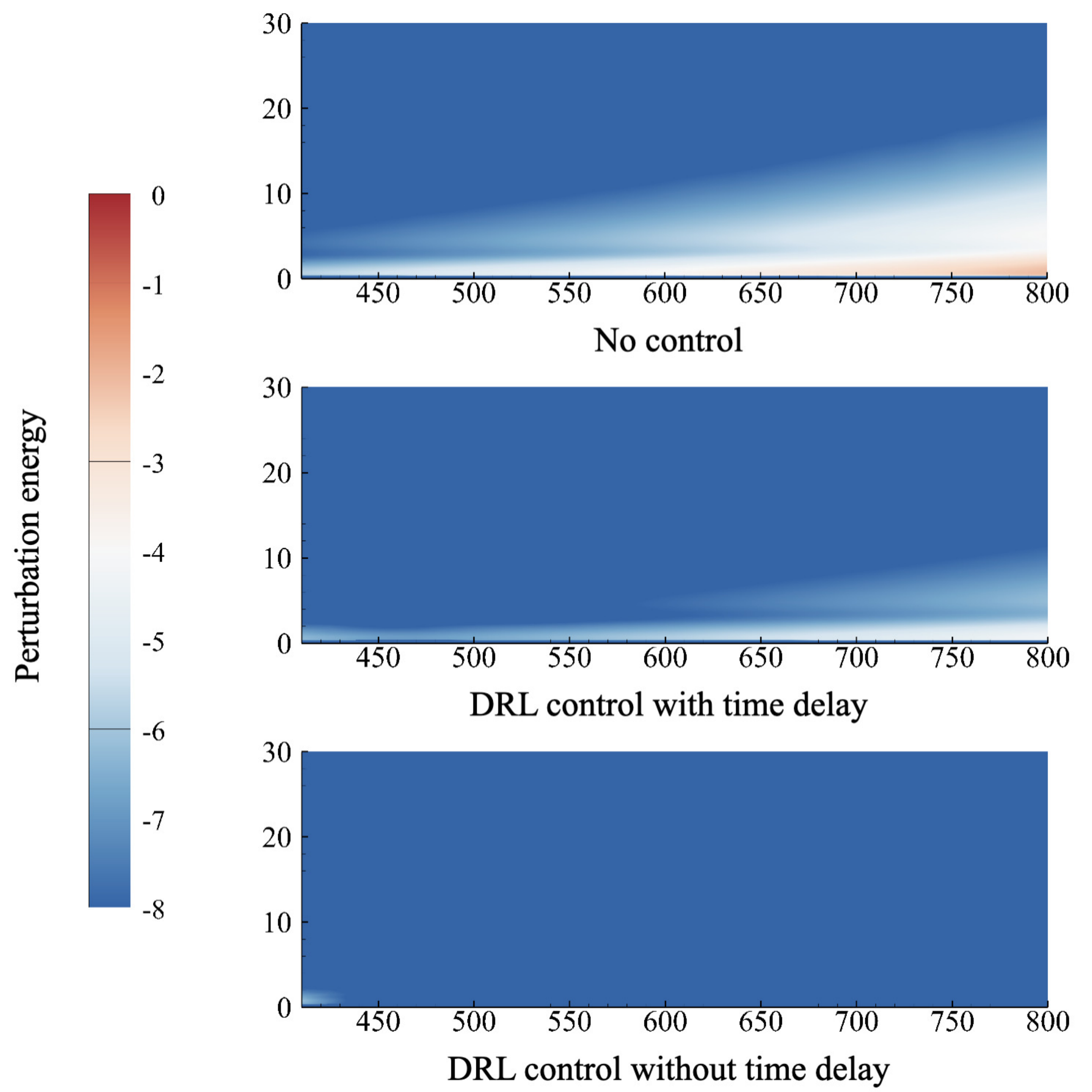}}
  \caption{Contours of the time-averaged perturbation energy field downstream the actuator with upstream noise input $\sigma_{d(t)} = 0.0002$. The top row represents the energy field without control; the middle one displays DRL-based control without regard to time delay and the bottom one with time delay removed.}
\label{figA3}
\end{figure}

Finally, we provide a more detailed description of our DRL framework for the reproducibility of the current work. The environment set-up has been described in Sec. \ref{problemformulation} and the numerical simulations of the two systems are given in Sec. \ref{NS} and Sec. \ref{DNS}, respectively. For the interaction between the agent and the environment, the control action is updated every 30 time steps and there is only one state observation collected by the optimised sensors at the end of each 30 time steps. Thus, the control frequency is the same as the state sampling frequency, which has to be larger than twice the characteristic frequency of the problem to satisfy Nyquist criteria. For hyperparameters adopted in the training process, we define a training episode composed of 120 action steps for the 1-D KS system and 600 action steps for the 2-D boundary layer flow. The experience memory size is chosen to be 10000 for KS system and 15000 for Blasius case with a mini-batch size of 32. Moreover, both actor and critic networks have two hidden layers each with 200 neurons using ReLU as the activation function. Adam optimiser is adopted to update network parameters and the learning rate is selected as 0.001 for both actor and critic networks. In terms of the computational time spent, the training in KS system is quite fast due to the simplicity of this reduced model and it takes about 20 minutes for full convergence, while for training in Blasius case, it takes about 12 hours to learn the optimal control policy. The sensor placement optimisation implemented in KS system is quite time-consuming since the particle searching process is conducted in serial in the \textit{Pyswarm} package we adopted here and each search corresponds to a complete DRL training in KS system. It takes about two weeks to obtain the final optimised sensor placement in our case and how to accelerate the optimisation process is also one of our future research directions. All of the above computations are performed on 12 cores of Intel Xeon(R) Bronze 3104.}

\section{Linear quadratic regulator (LQR) control}\label{B}
LQR is a classical model-based control method which assumes the full knowledge of the field $\boldsymbol{v}$(\textit{t}) to calculate the control signal \textit{u}(t) as below
\begin{equation}
u(t)=\boldsymbol{K}(t) \boldsymbol{v}(t)
\end{equation}
where $\boldsymbol{K}$(\textit{t}) is the feedback gain which will be calculated by solving a Riccati equation. The objective of LQR controller is to seek a control signal \textit{u}(\textit{t}) which minimises the cost function \textit{L} in a quadratic form considering both the sensor output \textit{z}(\textit{t}) and the control effort \textit{u}(\textit{t}) in some time interval $t \in[0, T]$
\begin{equation}
\begin{aligned}
{L}(\boldsymbol{v}(u), u)=& \frac{1}{2} \int_{0}^{T}\left(\boldsymbol{v}^{H} \boldsymbol{W}_{\boldsymbol{v}} \boldsymbol{v}+u^{H} w_{u} u\right) d t+\int_{0}^{T} \boldsymbol{p}^{H}\left(\dot{\boldsymbol{v}}-\boldsymbol{A} \boldsymbol{v}-\boldsymbol{B}_{u} u\right) d t
\end{aligned}
\label{eq-A2}
\end{equation}
where $\boldsymbol{W}_{\boldsymbol{v}}=\boldsymbol{C}_{z}^{H} w_{z} \boldsymbol{C}_{z}$ and $z(t) = \boldsymbol{C}_{z} \boldsymbol{v}(t)$. Both $w_z$ and $w_u$ are weighting factors and selected as 1 here except in the cases we will change them. The second term on the right hand side is due to the dynamic constraint, \textit{i.e.} $\dot{\boldsymbol{v}}(t)=\boldsymbol{A} \boldsymbol{v}(t)+\boldsymbol{B}_{u} u(t)$, and $\boldsymbol{p}$ is a Lagrangian multiplier which is also the adjoint state corresponding to the direct state $\boldsymbol{v}$.

For a linear time-invariant system, the most straightforward way to compute the optimal control signal \textit{u}(\textit{t}) is to utilise the optimal condition, \textit{i.e.} $\partial L /\partial u$ = 0, which gives
\begin{equation}
u(t)=-w_{u}^{-1} \boldsymbol{B}_{u}^{H} \boldsymbol{p}(t).
\end{equation}
Assuming a linear relation between direct state and adjoint state, \textit{i.e.} $\boldsymbol{p}(t)=\boldsymbol{X}(t) \boldsymbol{v}(t)$,  we can obtain the feedback gain $\boldsymbol{K}$(\textit{t}) given by
\begin{equation}
\boldsymbol{K}(t)=-w_{u}^{-1} \boldsymbol{B}_{u}^{H} \boldsymbol{X}(t)
\end{equation}
where the matrix $\boldsymbol{X}(t)$ is the solution of a differential Riccati equation \citep{lewis2012optimal}. When $\boldsymbol{A}$ is stable and time \textit{t} approaches infinity, we can obtain $\boldsymbol{X}(t)$ by solving the following algebraic Riccati equation
\begin{equation}
\boldsymbol{0}=\boldsymbol{A}^{H} \boldsymbol{X}+\boldsymbol{X} \boldsymbol{A}-\boldsymbol{X} \boldsymbol{B}_{u} w_{u}^{-1} \boldsymbol{B}_{u}^{H} \boldsymbol{X}+\boldsymbol{W}_{\boldsymbol{v}}
\end{equation}

The above equations describe the principle of LQR controller. Its advantage is that the feedback gain is constant and thus needs to be computed only once. As a typical example of model-based control methods, LQR controller is compared with our model-free DRL-based controller to control the 1-D linearised KS equation in the Sec. \ref{LQR}.

\section{Stability-enhanced reward design}\label{Reward}
In this section, we would like to analyse the convective nature of the 1-D KS equation and propose to embed flow physical information in the reward design. In the previous sections, the reward is defined as the negative RMS value of $z(t)$ which represents the downstream perturbation level. In this section, we discuss how the reward function in DRL-based control can be improved to incorporate information on the flow instability of the 1-D KS system to improve the control performance. 

\begin{figure}
\centering  
\subfigure[]{
\includegraphics[width=0.471\textwidth]{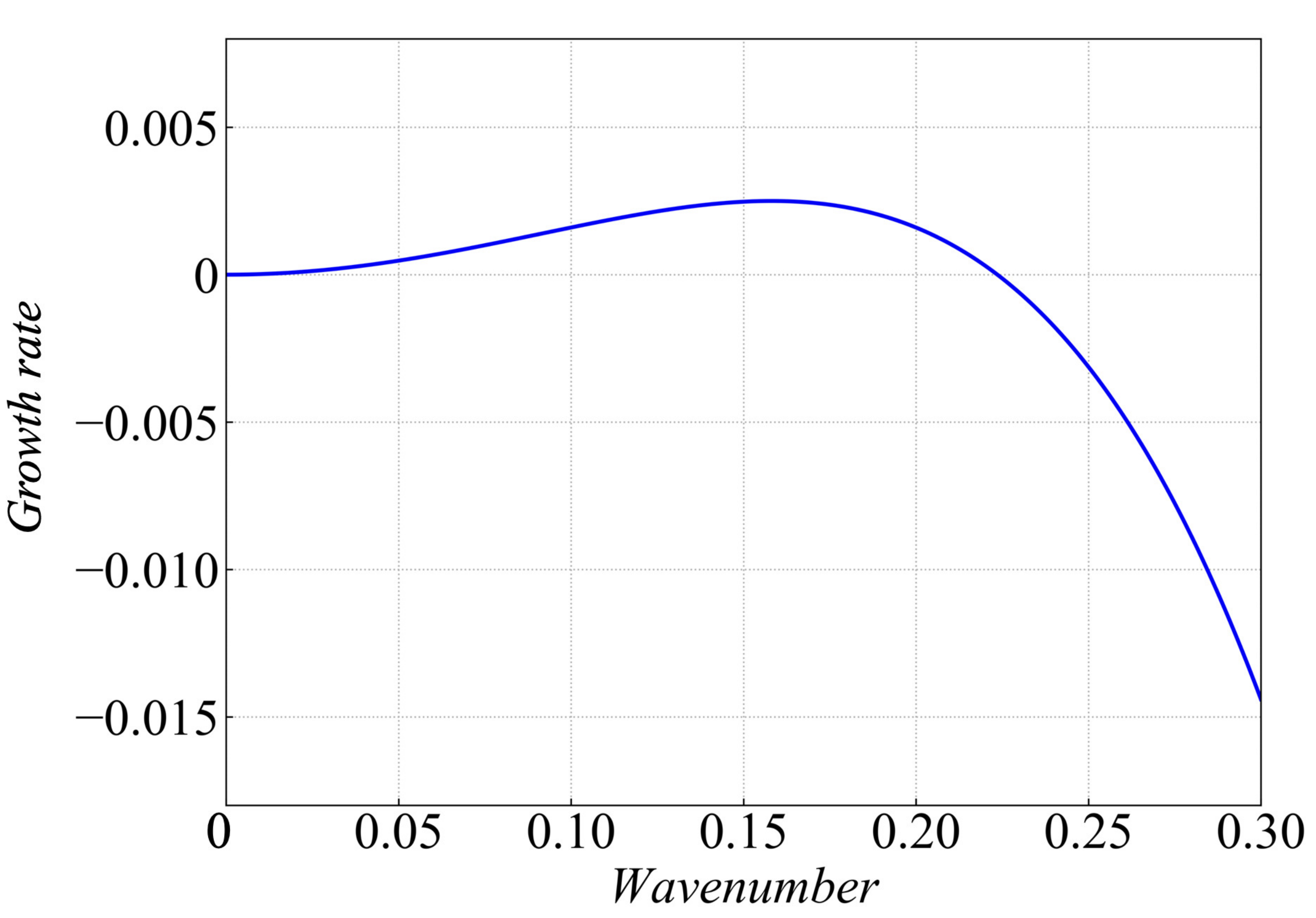}
}
\quad
\subfigure[]{
\includegraphics[width=0.461\textwidth]{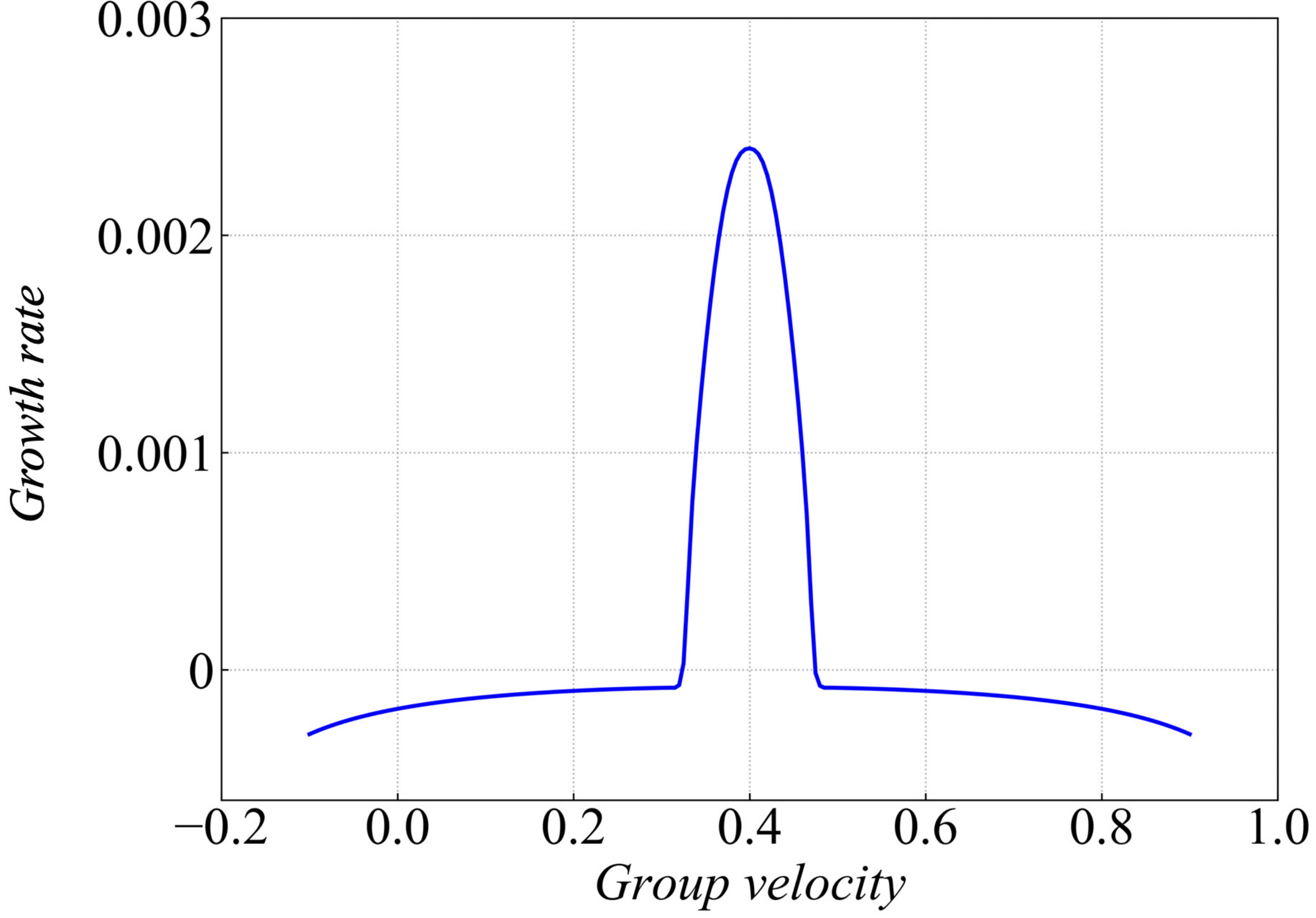}
}
\quad
\caption{Convective instability of the 1-D linearised KS system. In panel (a), relation between wavenumber $\alpha$ and their growth rate  $\omega_i$ is plotted. In panel (b), relation between group velocity $v_g$ and the corresponding growth rate $\sigma$ is plotted.}
\label{fig16}
\end{figure}

We first analyze the stability properties of the 1-D linearised KS system without the external forcing term, i.e., Eq. \ref{eq2-4} with $f(x,t) = 0$. We assume the travelling wave-like solutions in the form of 
\begin{equation}
v^{\prime}=\hat{v} e^{i(\alpha x-\omega t)}
\label{eq4-2}
\end{equation}
where $\alpha \in \mathbb{R}$ is the wavenumber and $\omega=\omega_{r}+i \omega_{i} \in \mathbb{C}$ with $\omega_{r}$ representing the frequency and $\omega_{i}$ the exponential growth rate. Inserting Eq. \ref{eq4-2} into Eq. \ref{eq2-4} yields the following dispersion relation between $\omega$ and $\alpha$
\begin{equation}
\omega=V \alpha+\mathrm{i}\left(\frac{\mathcal{P}}{\mathcal{R}} \alpha^{2}-\frac{1}{\mathcal{R}} \alpha^{4}\right).
\label{eq4-3}
\end{equation}
The relation between $\omega_i$ and $\alpha$ is presented in figure \ref{fig16}(a). It is shown that only a certain range of wavelengths are unstable and will be amplified in the final state, see again figure \ref{fig4}. For the convectively-unstable flows, the spatiotemporal stability analysis is more relevant, which accounts for the wave development in both space and time.

The spatiotemporal instability of the convective KS system subjected to an impulse disturbance is conducted, following the post-processing method in \cite{brancher1997absolute}; our code adapts the one used in \cite{feng2022nonlinear}. To unambiguously extract the amplitude and the phase of a wavepacket, the Hillbert transform is applied to the perturbation velocity as below
\begin{equation}
v^{\prime}(x, t)=\hat{v}(x, t) e^{i \psi(x, t)}
\label{eq4-4}
\end{equation}
where $\hat{v}(x, t)$ represents the complex-valued amplitude function and $\psi(x, t)$ is the phase of the wavepacket. Then, we calculate the amplitude function $\hat{\mathcal{E}}(x, t)$, which is the norm of $\hat{v}(x, t)$, at a (asymptotically) large time
\begin{equation}
\hat{\mathcal{E}}(x, t) \propto t^{-1 / 2} e^{\sigma\left(v_{g}\right) t} \quad \text {with} \quad v_{g}=\left(x-x_{0}\right) / t=\text{const}, t \rightarrow \infty
\label{eq4-5}
\end{equation}
where $x_0$ is the initial location of the impulse disturbance and $\sigma(v_g)$ refers to the dominant growth rate along the ray $v_g$. See \cite{huerre1985absolute} for the derivation of the above equation. Based on two snapshots extracted at instants $t_1$ and $t_2$, the spatiotemporal growth rate can be calculated as
\begin{equation}
\sigma\left(v_{g}\right) \approx \frac{\ln \left[\hat{\mathcal{E}}\left(v_{g} t_{2}, t_{2}\right) / \hat{\mathcal{E}}\left(v_{g} t_{1}, t_{1}\right)\right]}{t_{2}-t_{1}}+\sigma_{0}\left(t_{1}, t_{2}\right), \quad \sigma_{0}\left(t_{1}, t_{2}\right)=\frac{\ln \left(t_{2} / t_{1}\right)}{2\left(t_{2}-t_{1}\right)}
\label{eq4-6}
\end{equation}
where $\sigma_0(t_1, t_2)$ is a finite-time correction for the growth rate due to the $t^{-1/2}$ term in Eq. \ref{eq4-5} \citep{delbende1998nonlinear}.

Following this method, we can calculate the spatiotemporal growth rate $\sigma$ under various group velocities $v_g$ and plot it in figure \ref{fig16}(b). It is shown that the maximum growth rate $\sigma = 2.42\times 10^{-3}$ is obtained at $v_g = 0.4$, which is close to the maximum growth rate $\omega = 2.5\times 10^{-3}$ obtained in figure \ref{fig16}(a) at $\alpha = 0.158$. In addition, the absolute growth rate $\sigma(v_g = 0)$ is negative as shown in \ref{fig16}(b), which demonstrates the nature of convective instability in the 1-D linearised KS system. 

Now we discuss the new reward design in the DRL-based control of the linearised KS system. When the control is turned on and the external forcing term $f(x, t)$ in Eq. \ref{eq2-4} is varying, it is difficult to apply the conventional linear stability analysis. In this case we use the traditional dynamic mode decomposition (DMD, c.f. \cite{rowley2009spectral,schmid2010dynamic}) to extract the leading growth rate and embed it into the reward function design. Here two notable points regarding the DMD used in the current work are mentioned. First, the snapshots are extracted in a moving coordinate system with the velocity of 0.4 which is the group velocity leading to the maximum growth rate as calculated above (see figure \ref{fig16}(b)). If the snapshots were extracted based on a stationary coordinate, we are not able to obtain the correct growth rate to describe the instability of the convective system. Second, the snapshots are extracted within the range of (410, 500) which is near downstream the actuator position, since this specific range is affected by the control action most directly and can reflect the effects of control. 

We embed the physical information on flow instability into reward function in the following way, similar to that in \cite{li2022reinforcement},
\begin{equation}
\text { Reward } = -|z(t)|_{rms}\times e^{cg}
\label{eq4-7}
\end{equation}
where the first term $-|z(t)|_{rms}$ on the right hand side represents the initial reward and the second term $e^{cg}$ is related to the flow instability with $g$ being the leading growth rate evaluated by DMD and $c$ being a constant of 50 (the specific value of $c$ is not essential and we choose 50 in order to compensate the small value of $g$). When the control has a stabilising effect, i.e., $g < 0$ and $e^{cg} < 1.0$,  a larger reward is given to encourage such control attempts. On the contrary, when the control is destabilising, i.e., $g > 0$ and $e^{cg} > 1.0$, the second term acts as a penalty to reduce the probability of the occurrence of such actions. Therefore, the new reward function penalises flow instability. Loosely speaking, one can also take the logarithm of the reward function to understand that the growth rate acts as an additional regulation term in the objective function.

\begin{figure}
\centering  
\subfigure[1 sensor]{
\includegraphics[width=0.465\textwidth]{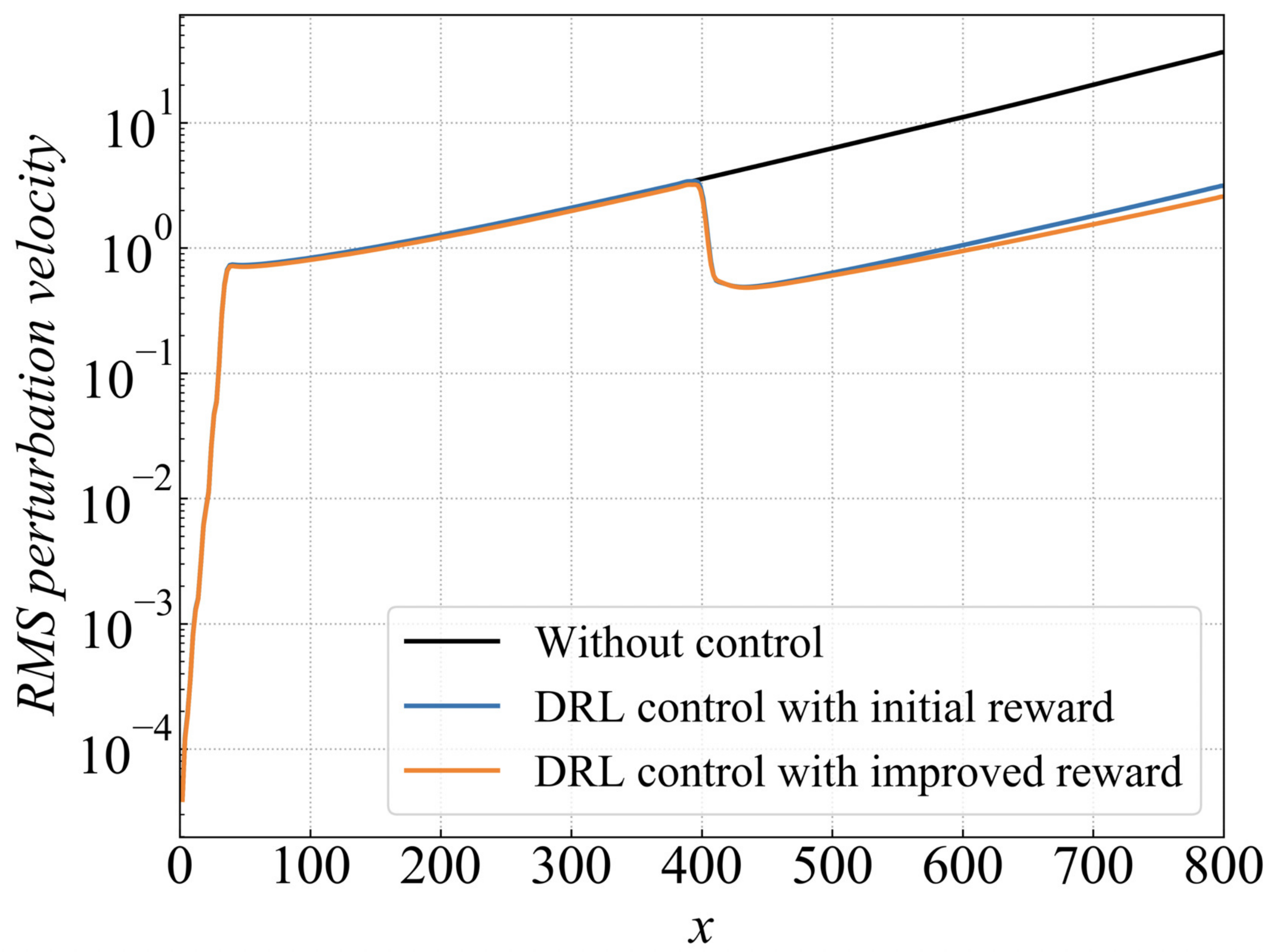}
}
\quad
\subfigure[2 sensors]{
\includegraphics[width=0.465\textwidth]{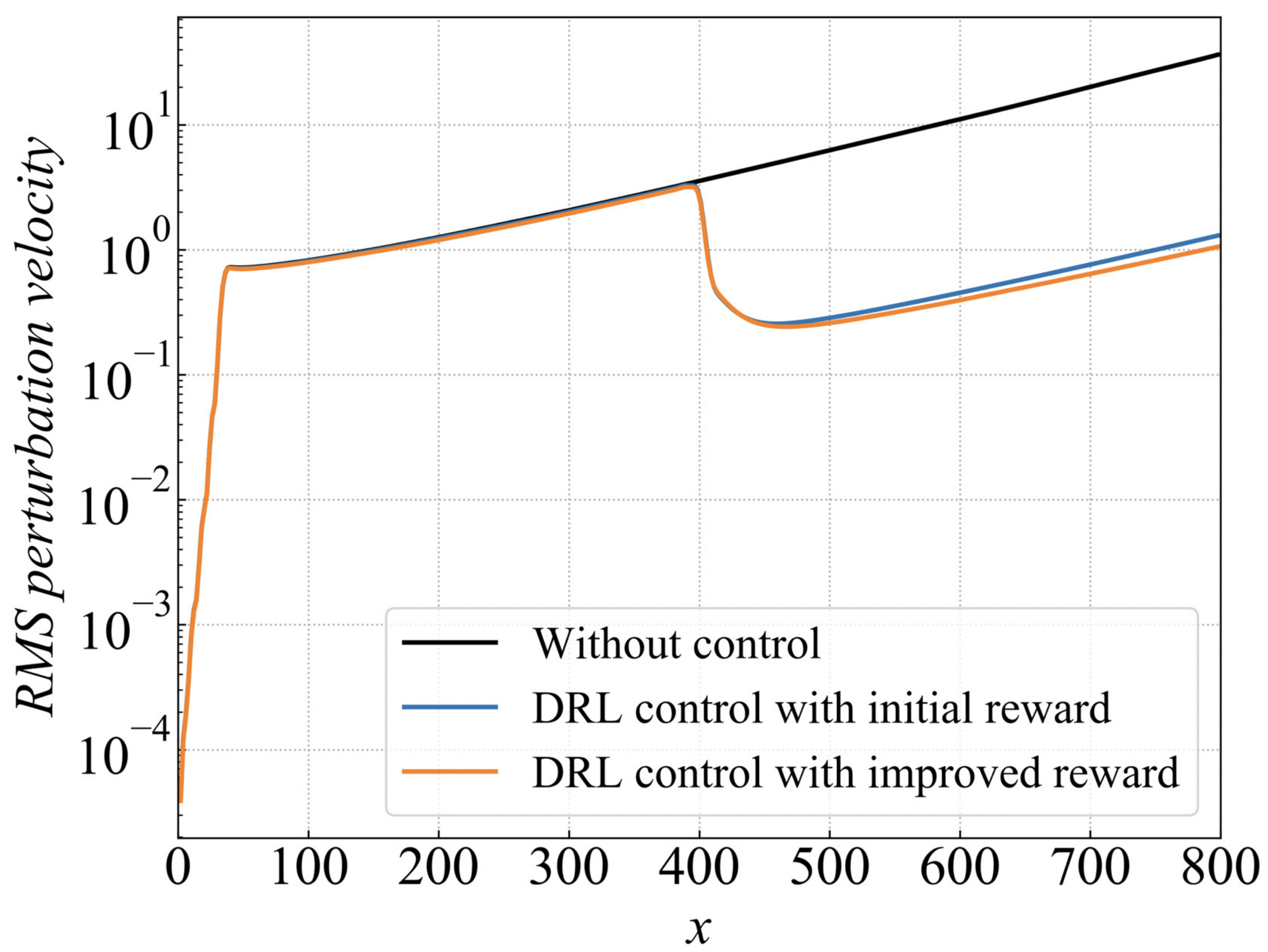}
}
\quad
\subfigure[4 sensors]{
\includegraphics[width=0.465\textwidth]{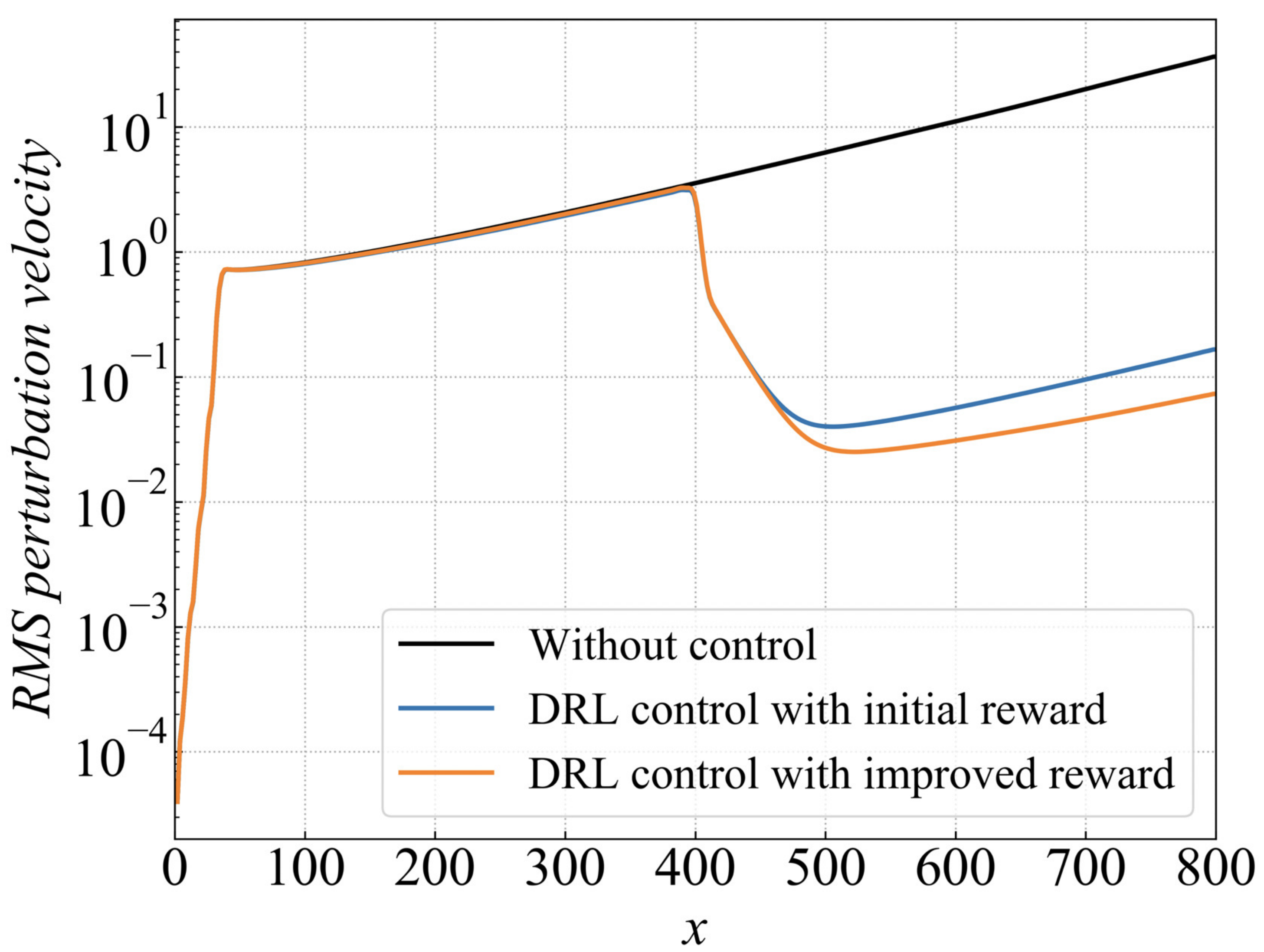}
}
\quad
\subfigure[6 sensors]{
\includegraphics[width=0.465\textwidth]{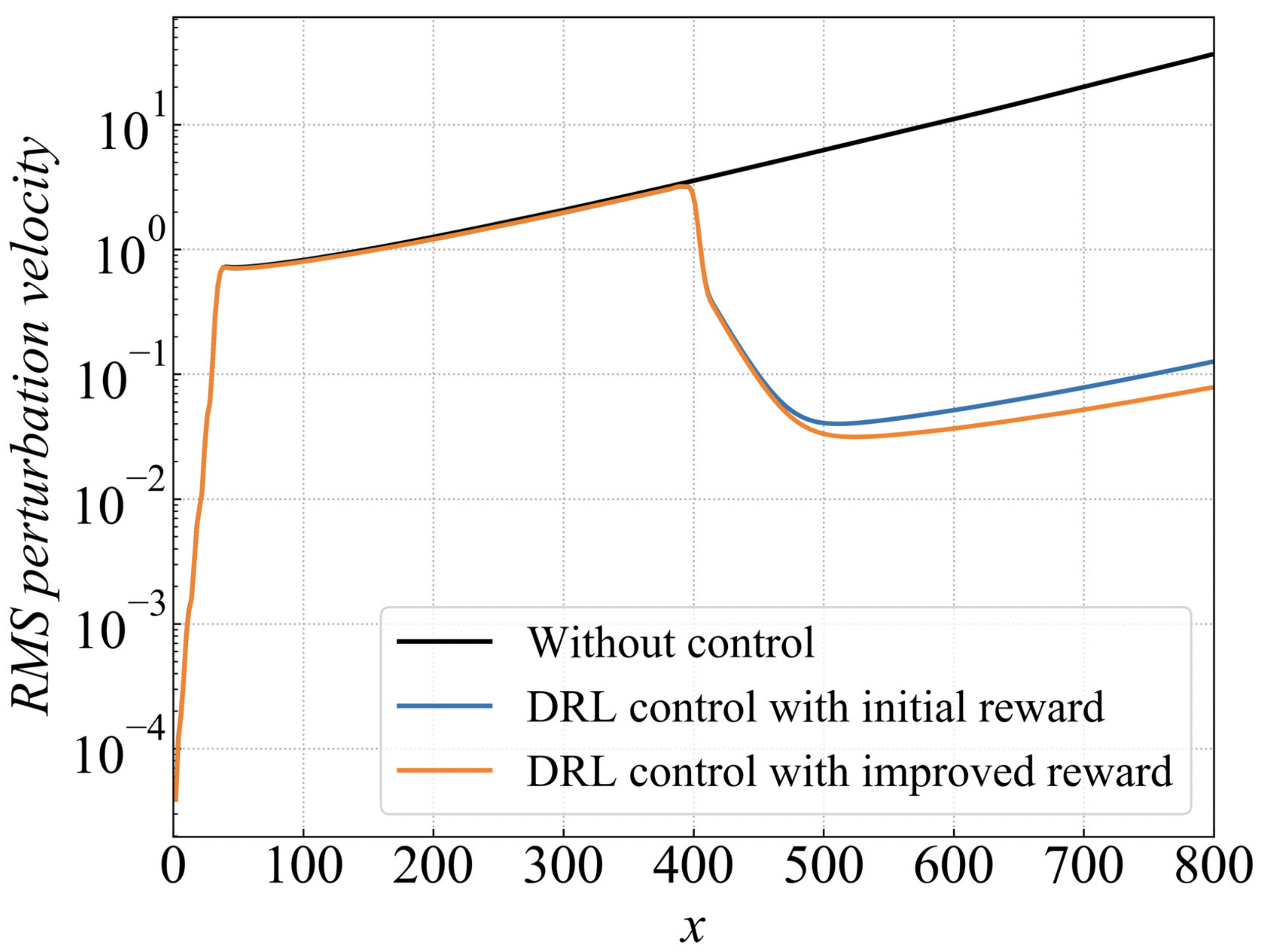}
}
\quad
\subfigure[8 sensors]{
\includegraphics[width=0.465\textwidth]{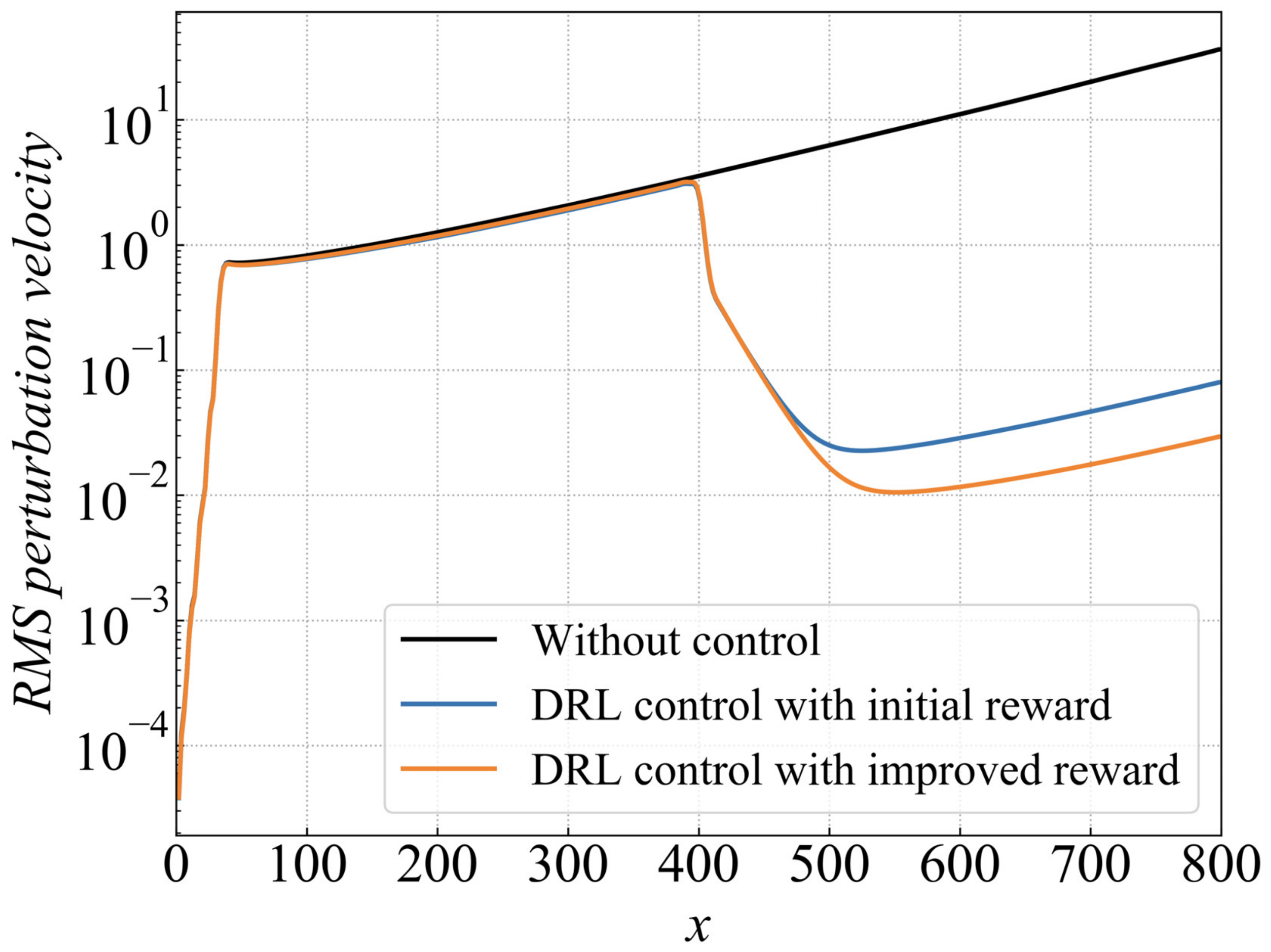}
}
\quad
\subfigure[10 sensors]{
\includegraphics[width=0.465\textwidth]{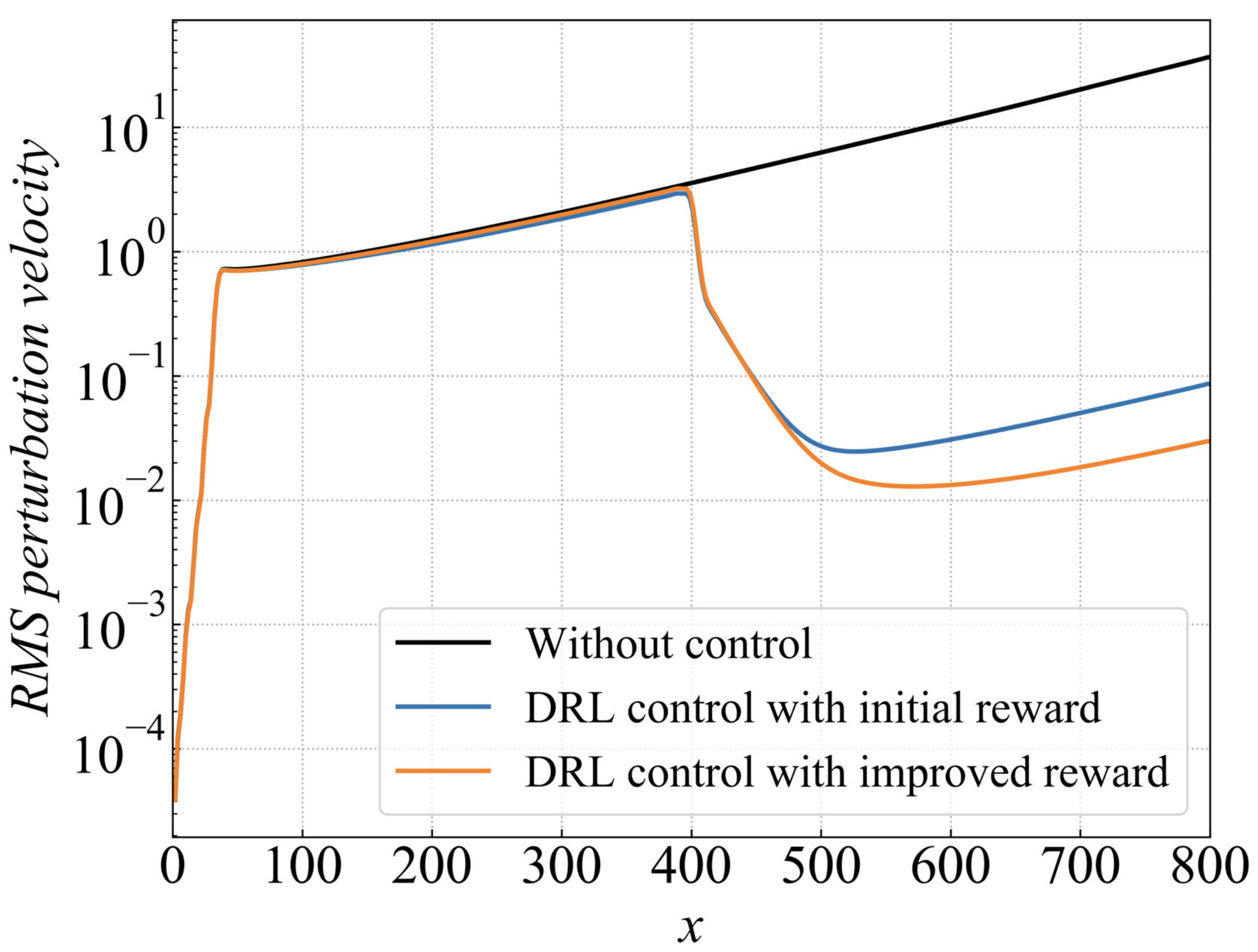}
}
\quad
\caption{DRL-based control performance comparison between using the initial reward and the stability-enhanced reward with different numbers of sensors. The RMS value of perturbation velocity along the 1-D domain is plotted.}
\label{fig17}
\end{figure}

\begin{figure}
\centerline{\includegraphics[width=0.9\textwidth]{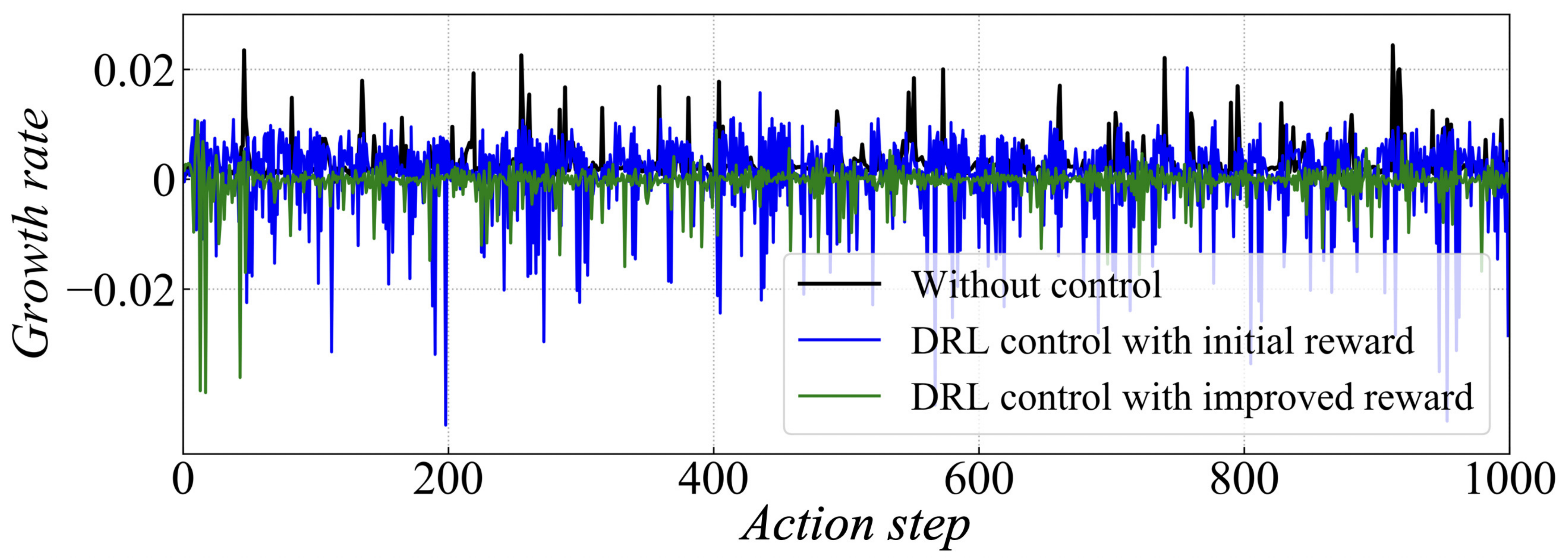}}
\caption{Time-variation curve of the leading growth rate evaluated by DMD during the control process, using both the initial reward and the stability-improved reward. The curve for the uncontrolled process is also plotted for comparison.}
\label{fig18}
\end{figure}

Comparisons on the DRL-based control performance between using the initial reward and the stability-improved reward are presented in figure \ref{fig17}(a)-(f) for different numbers of sensors, respectively. It is shown that with 1 or 2 sensors, the control performance enhancement by using the stability-enhanced reward is limited. This may be due to the fact that too few sensors cannot provide a complete measurement of the noisy environment, preventing the new reward to demonstrate its effectiveness. As the number of sensors increases, the new reward begins to show its advantages in the sense that the perturbation downstream the actuator is further reduced, as shown in figure \ref{fig17}(c)-(f). The manifestation of the performance improvement is revealed in figure \ref{fig18}, where the growth rate calculated by DMD is presented (here we use 8 sensors). When there is no control, the growth rate as a function of time  is always positive which indicates the flow itself is convectively unstable. With the initial reward, the large spikes of positive growth rate decrease, and there are some instants with a negative growth rate, leading to the reduction of downstream perturbations. However, the instants with a positive growth rate are still frequent. With the help of stability-enhanced reward, such instants with a relatively large positive growth rate are penalised and thus become fewer. The average growth rate during the control process is decreased and thus a better control performance is achieved. 

\section{Dynamics of nonlinear KS equation and its control}\label{Nonlinear}
All the previous results are related to the control of the linearised KS equation. However, when the perturbation amplitude increases above a certain level, the nonlinear effect cannot be neglected any more. In this section, we investigate the dynamics of the nonlinear KS equation as described by Eq. \ref{eq2-3}, together with boundary conditions Eqs. \ref{eq2-5}, and its control issue.

\begin{figure}
\centering  
\subfigure[]{
\includegraphics[width=0.459\textwidth]{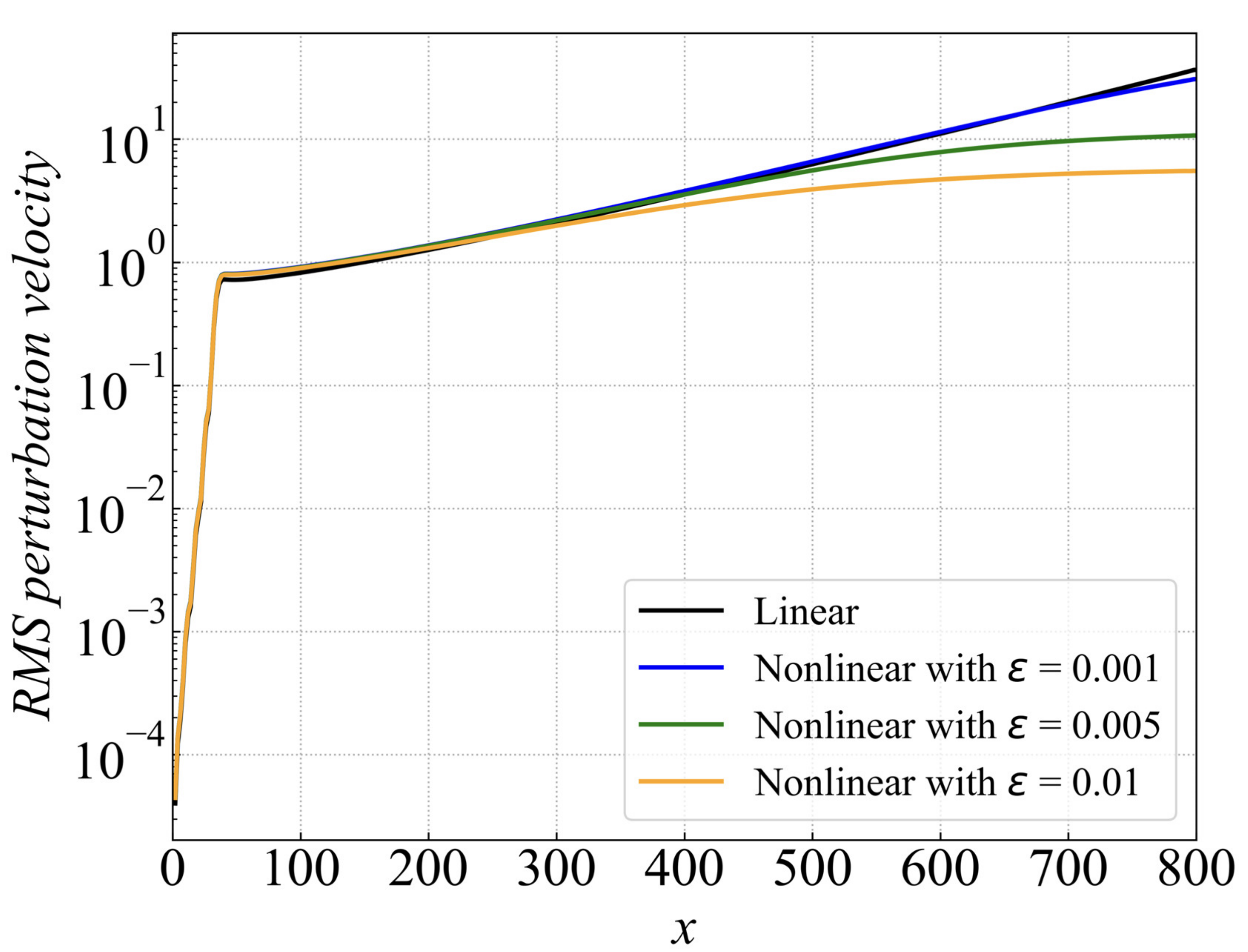}
}
\quad
\subfigure[]{
\includegraphics[width=0.473\textwidth]{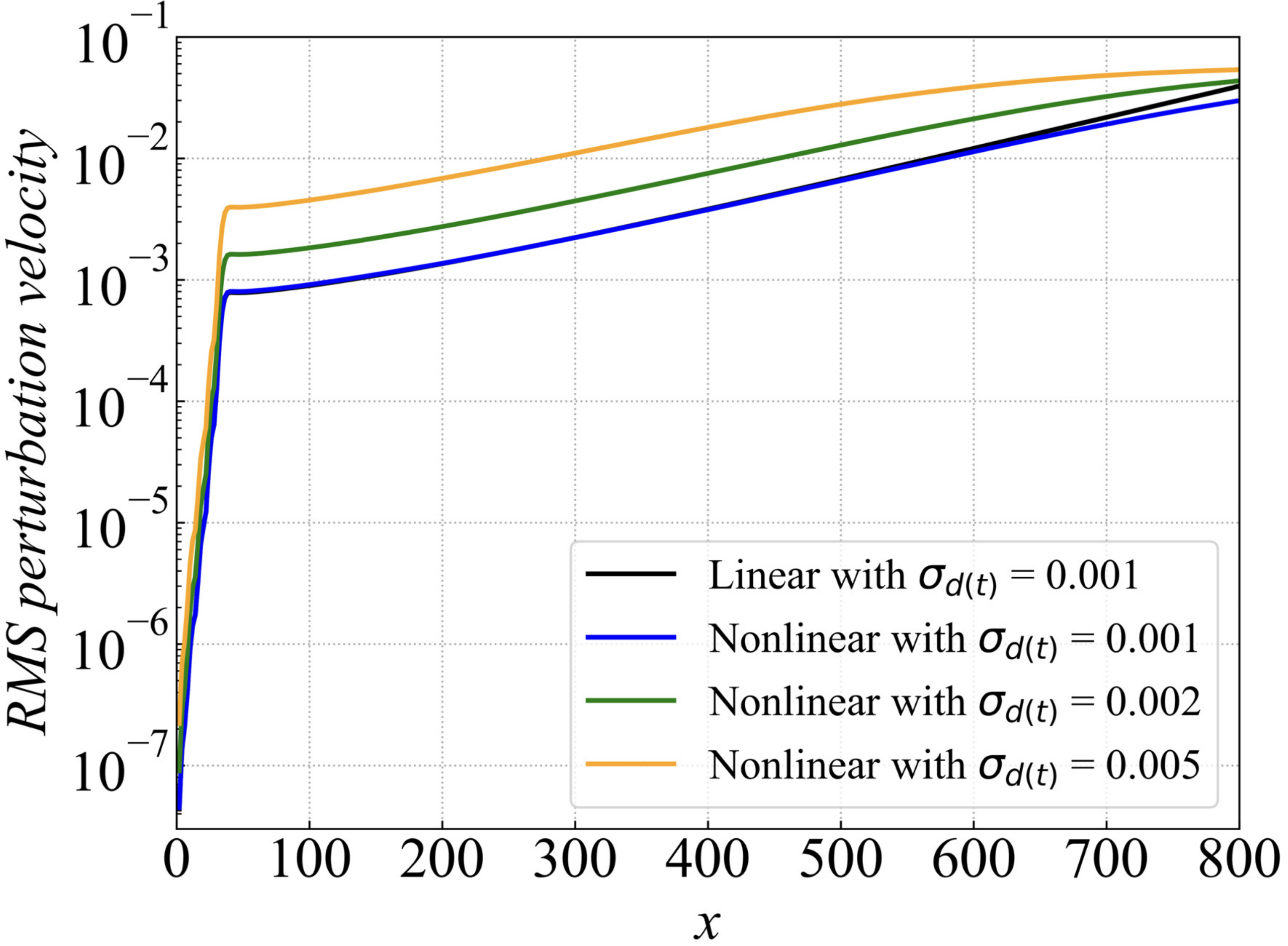}
}
\quad
\caption{Dynamics of the nonlinear KS system when subjected to an upstream random noise. In panel (a), the RMS value of perturbation velocity along the 1-D domain is plotted for KS equation excited by an upstream noise of $\sigma_{d(t)}=1$, with different values of $\varepsilon$. In panel (b), the RMS value of perturbation velocity along the 1-D domain is plotted for KS equation excited by an upstream noise of different standard deviations $\sigma_{d(t)}$, with $\varepsilon$ fixed at 1.}
\label{fig11}
\end{figure}

The intensity of nonlinearity in Eq. \ref{eq2-3} depends on the value of $\varepsilon$. Here we first select three different values of $\varepsilon = 0.001$, 0.005 and 0.01, and simulate the dynamics of the weakly nonlinear KS system subjected to an upstream random noise of a unit variance at $x_d = 35$, using the numerical method described in Sec. \ref{NS}. All the other parameters are identical with those for the linearised system. The corresponding results are presented in figure \ref{fig11}(a), where the RMS value of perturbation velocity along the 1-D domain is plotted. It is shown that with the increase of $\varepsilon$, the nonlinear effect becomes more evident in the sense that the perturbation is no longer growing exponentially along the $x$ direction, in contrast to the linear system as shown by the black curve. Besides, the nonlinear effect can also be demonstrated by varying the external noise level. As shown in figure \ref{fig11}(b), we fix $\varepsilon = 1$ but increase the standard deviation of the upstream noise from $\sigma_{d(t)} = 0.001$ to $\sigma_{d(t)} = 0.005$ and the nonlinear effect strengthens.

\begin{figure}
\centering  
\subfigure[]{
\includegraphics[width=0.458\textwidth]{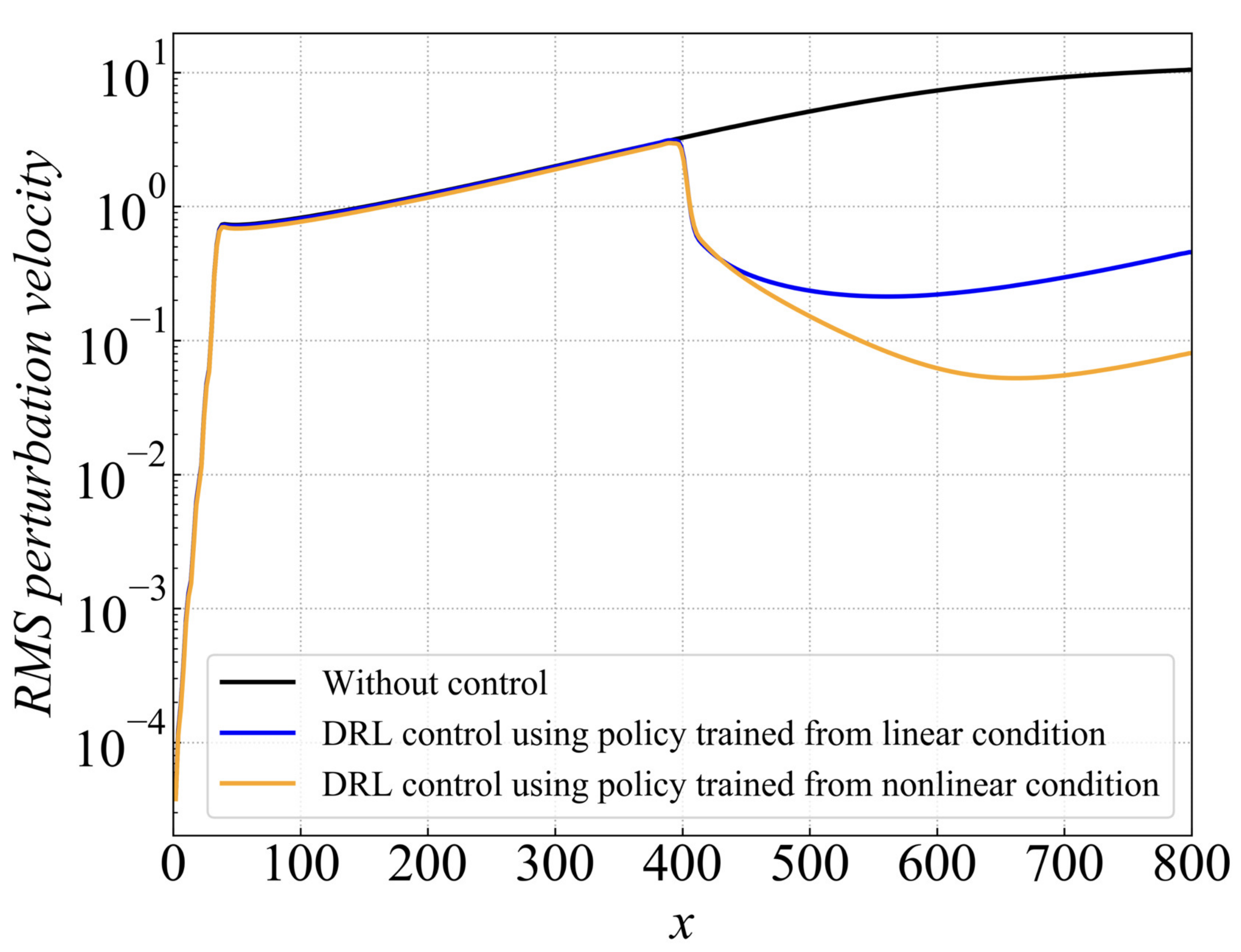}
}
\quad
\subfigure[]{
\includegraphics[width=0.474\textwidth]{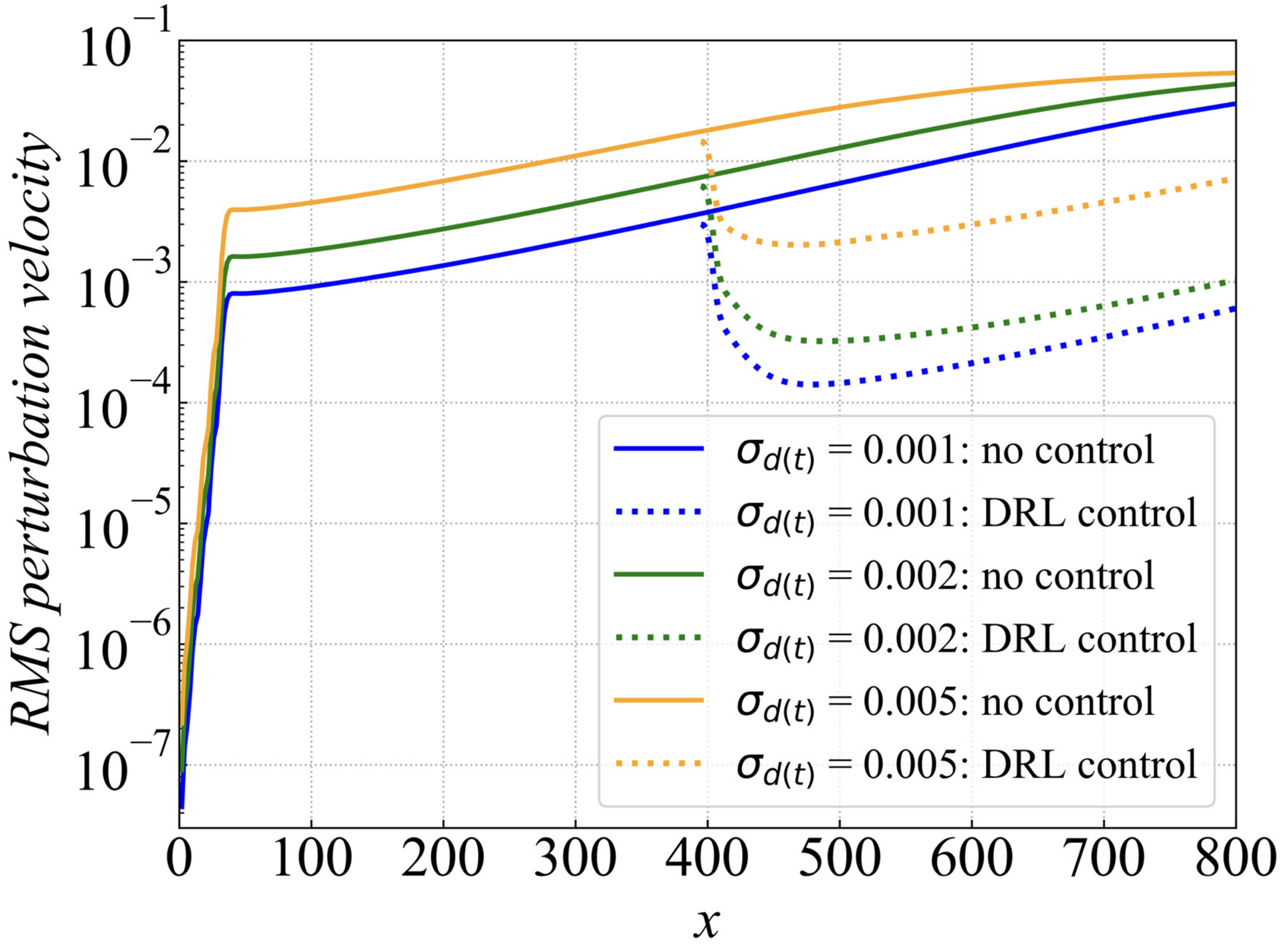}
}
\quad
\caption{DRL-based control of the nonlinear KS system. In panel (a), the RMS value of perturbation velocity is plotted for the uncontrolled case and the controlled cases using both the old policy (trained from the linear condition) and the new policy (retrained in the nonlinear condition with $\varepsilon = 0.005$ and $\sigma_{d(t)} = 1$). In panel (b), the RMS value of perturbation velocity is plotted for the uncontrolled and controlled cases of the KS system excited by an external noise of different levels, and all the control policies applied are trained from the nonlinear condition with $\varepsilon = 1$.}
\label{fig12}
\end{figure}

Then we investigate the DRL-based control of the nonlinear KS system. The control policy which was learnt from the linear condition is directly applied to the nonlinear case with $\varepsilon = 0.005$ and $\sigma_{d(t)} = 1$. It is found that this policy is still effective in suppressing the downstream perturbation, as shown by the blue curve in figure \ref{fig12}(a) where the amplitude at $x_z = 700$ is reduced from about 10 to 0.3, although the performance is not as good as the new policy trained in the nonlinear condition with $\varepsilon = 0.005$ (orange curve). \xuda{Similar results can also be found in \cite{al2015output}, where the controller designed in the linearised KS equation can be used to stabilise the nonlinear KS equation.} In figure \ref{fig12}(b), we compare the DRL-based control performance with the increase of the external noise level under the nonlinear condition with $\varepsilon=1$. It is shown that as the noise level increases, the control performance degrades in terms of the downstream perturbation reduction. This may partly be due to the fact that the sensor placement applied here is obtained directly from the linearised KS system, which may be sub-optimal in the nonlinear condition, and similar results are also found in 2-D boundary layer flows, to be detailed in Sec. \ref{DRL-BL}.

\end{appendix}

\bibliographystyle{jfm}
\bibliography{DRL_KS_BL}

\end{document}